\documentclass[12pt]{article}
\usepackage{amsfonts,amsthm,amsmath,amssymb,upgreek}
\usepackage{hyperref}
\usepackage{cite}
\usepackage[paper=letterpaper,margin=1in]{geometry}
\usepackage{graphicx}
\usepackage{units}
\usepackage{color}

\def\t{\tau}
\def\N{{\mathcal N}}
\def\a'{\alpha'}

\def\F{{\mathcal F}}

\def\la{\langle}
\def\ra{\rangle}
\def\th{\t h}
\def\hat{\widehat}
\renewcommand\bar{\overline}
\def\uphi{{\upphi}}

\newcommand{\be}{\begin{equation}}
\newcommand{\ee}{\end{equation}}
\newcommand{\bea}{\begin{eqnarray}}
\newcommand{\eea}{\end{eqnarray}}
\newcommand{\ba}{\begin{array}}
\newcommand{\ea}{\end{array}}
\newcommand{\bee}{\begin{enumerate}}
\newcommand{\eee}{\end{enumerate}}
\newcommand{\bi}{\begin{itemize}}
\newcommand{\ei}{\end{itemize}}
\newcommand{\bc}{\begin{center}}
\newcommand{\ec}{\end{center}}
\newcommand{\bfig}{\begin{figure}}
\newcommand{\efig}{\end{figure}}

\def\O{{\mathcal O}}
\def\c{{\sf c}}

\def\CC{{\Bbb C}}

\def\C{{\mathcal C}}
\def\V{{\mathcal V}}
\def\Z{{\Bbb Z}}
\def\CC{{\Bbb C}}
\def\R{{\Bbb R}}
\def\sgn{{\mathrm{sgn}}}
\def\Re{{\mathrm{Re}}}
\def\t{\widetilde}
\def\h{\widehat}
\def\b{\overline}
\def\bz{\b z}
\def\x{{\mathrm x}}
\def\X{{\mathrm X}}
\def\y{{\mathrm y}}

\def\p{{\mathrm p}}
\def\th{{\widetilde h}}
\def\N{{\mathcal N}}
\def\A{{\mathcal A}}
\def\B{{\mathcal B}}
\def\q{{\h q}}

\def\U{{\mathcal U}}
\def\ret{{R}}

\numberwithin{equation}{section}

\begin{document}
\thispagestyle{empty}

\begin{center}

\bf{{\LARGE More On Supersymmetric And 2d \vskip.5cm Analogs of the SYK Model}\\

\vspace{1cm}}

  \begin{center}

 \bf {Jeff Murugan,$^{a,b}$ Douglas Stanford,$^b$  and Edward Witten$^b$}\\
  \bigskip \rm
  
\bigskip
 $^a$Laboratory for Quantum Gravity and Strings, Department of Mathematics \\ and Applied Mathematics, University of Cape Town, South Africa  \\
 \medskip
   $^b$Institute for Advanced Study, Princeton NJ USA 08540

\rm
  \end{center}

\vspace{2cm}
{\bf Abstract}
\end{center}
\begin{quotation}
\noindent
In this paper, we explore supersymmetric and 2d analogs of the SYK model. We begin by working out a basis of (super)conformal eigenfunctions appropriate for expanding a four-point function. We use this to clarify some details of the 1d supersymmetric SYK model. We then introduce new bosonic and supersymmetric analogs of SYK in two dimensions. These theories consist of $N$ fields interacting with random $q$-field interactions. Although models built entirely from bosons appear to be problematic, we find a supersymmetric model that flows to a large $N$ CFT with interaction strength of order one. We derive an integral formula for the four-point function at order $1/N$, and use it to compute the central charge, chaos exponent and some anomalous dimensions. We describe a problem that arises if one tries to find a 2d SYK-like CFT with a continuous global symmetry.
\end{quotation}

\setcounter{page}{0}
\setcounter{tocdepth}{2}
\setcounter{footnote}{0}
\newpage
 \tableofcontents
\section{Introduction}

The SYK model \cite{Sachdev:1992fk,KitaevTalks} is a strongly interacting but solvable quantum mechanics system, described by $N$ Majorana fermions interacting with random $q$-fermion couplings:
\be\label{SYKH}
H = i^{\frac{q}{2}}J_{i_1i_2\cdots i_q}\psi_{i_1}\psi_{i_2}\cdots\psi_{i_q}.
\ee
Here we sum over the indices and $J$ is a random but fixed tensor. At large $N$, a summable set of Feynman diagrams dominates, but the effective strength of interaction does not become small. In a sense, this model finds a sweet spot between intractable systems with large $N$ matrix degrees of freedom and solvable but weakly interacting systems built from large $N$ vectors. Although the holographic dual of this theory remains mysterious, it has been shown that at low temperature the SYK model is dominated by a universal ``Schwarzian'' sector \cite{KitaevTalks} (see also \cite{Maldacena:2016hyu}) that also describes dilaton gravity theories in $AdS_2$ \cite{Almheiri:2014cka,Jensen:2016pah,Maldacena:2016upp,Engelsoy:2016xyb}. This makes SYK useful for studies of $AdS_2$ gravity.

Many interesting generalizations of the original SYK model have been studied, including models with complex fermions \cite{Sachdev:2015efa,Davison:2016ngz}, higher-dimensional lattices \cite{Gu:2016oyy}, global symmetry \cite{Gross:2016kjj}, extra quadratic fermions \cite{Banerjee:2016ncu}, and supersymmetry \cite{Fu:2016vas}. Some progress has been made towards higher dimensional continuum theories \cite{Berkooz:2016cvq,Turiaci:2017zwd,Berkooz:2017efq}. Also, models have been proposed \cite{Witten:2016iux,KT} that eliminate the random couplings $J_{i_1\cdots i_q}$ in favor of a specific interaction tensor that leads to the same behavior at large $N$, including the first $1/N$ correction.

In this paper we study a new class of two-dimensional field theories in the spirit of SYK. Previous work has attempted to construct such models using fermion fields. This is tricky in higher than one dimension, for the following reason. In one dimension, a free fermion with a canonical kinetic term has dimension zero, so the four fermi interaction in (\ref{SYKH}) is relevant and the model flows in the IR to a strongly interacting SYK phase. By contrast, in two dimensions a canonical free fermion has dimension $1/2$, which makes a four-fermion interaction marginal (and marginally irrelevant \cite{Berkooz:2017efq} or relevant \cite{GN}) and higher $q$ interactions irrelevant. This complicates the effort to get SYK-like physics. 

An obvious idea, in order to make the interaction relevant, would be to use bosons instead of fermions as the fundamental variables. In two dimensions a free canonical boson has dimension zero, so a random $q$-boson term will be relevant. One could imagine studying an action of the form
\be\label{naivebose}
I = \int d^2 \x \left[\frac{1}{2}(\nabla \phi_i)^2 +  J_{i_1i_2\cdots i_q}\phi_{i_1}\phi_{i_2}\cdots\phi_{i_q}\right].
\ee
Unfortunately, the potential will generically have negative directions, and the model will not be well-defined. Still, we will find it useful to study the theory (\ref{naivebose}) as a formal warmup. Another possibility would be to take an interaction with a somewhat random but positive potential,
\be\label{betterbose}
I = \int d^2 \x \left[\frac{1}{2}(\nabla \phi_i)^2+ \sum_a\left(C^a_{i_1i_2\cdots i_{\frac{q}{2}}}\phi_{i_1}\phi_{i_2}\cdots\phi_{i_{\frac{q}{2}}}\right)^2\right],
\ee
where $C$ is a random tensor. This model is well-defined, but we will not be able to show that it flows to an SYK-like fixed point.

The most promising model that we find is a supersymmetric model involving both bosons and fermions, organized into superfields $\uphi$:
\be\label{introsuperaction}I=\int d^2\x d^2\theta \Bigg[\frac{1}{2} D_{\b\theta}\uphi_i D_\theta\uphi_i +iC_{i_1i_2\cdots i_{\h q}}\uphi_{i_1}
 \uphi_{i_2}\cdots \uphi_{i_{\h q}}  \Bigg] .\ee
We will maintain explicit supersymmetry by studying this model directly in terms of the superfields. However, one can also integrate out an auxiliary field, and write the theory in terms of component fields, $N$ bosons $\phi_i$ and $N$ pairs of chiral fermions $\psi_i,\b\psi_i$. The above action then has standard kinetic terms for these fields, plus two types of interaction term. We have a purely bosonic interaction term similar to the one in (\ref{betterbose}), and also an interaction coupling two fermions to $\q-2$ bosons: $C_{i_1i_2\cdots i_\q}\psi_{i_1}\b\psi_{i_2}\phi_{i_3}\cdots\phi_{i_\q}$. At large $N$, this model can be studied in a straightforward way. At long distances, it appears to flow to a conformal field theory. In particular, the emergent reparametrization invariance that led to conformal symmetry breaking in one-dimensional SYK is harmless here. It leads instead to the finite and conformally-invariant contribution of the stress tensor, with central charge $c = \frac{3N}{2}(1 - \frac{2}{\hat{q}})$.

This makes the model simple to analyze, since we preserve conformal symmetry. But it is not entirely good news: in one-dimensional SYK, the reparametrization mode dominated in the IR, leading to a close connection with dilaton gravity. It also led to  saturation \cite{kitaevfirsttalk}  of the chaos bound \cite{Maldacena:2015waa}. In two dimensions, the morally similar stress tensor contribution does not dominate. Instead, it is simply part of a Regge trajectory of higher spin operators with order one anomalous dimensions  that
all  contribute on the same footing. Because of this, one would not expect the model to saturate the chaos bound, and indeed it does not. The chaos exponent is small for large $\hat{q}$, and the largest physical value is $\lambda_L \approx 0.5824 \frac{2\pi}{\beta}$ for $\hat{q} = 3$. So, unlike in the one-dimensional case, there does not seem to be any sense in which gravity could be dominant in the holographic dual of the theory (\ref{introsuperaction}). Still, we hope that it might be possible to understand some aspects of holography and shed further light on the dual of SYK using this model.

Setting holography aside, we hope that these models will also be interesting to study as new candidate CFTs that are interacting but tractable. One feature is that the large $N$ solution presents the four-point function in the form of an integral over the principal series representations of the conformal group, with a contour that can be deformed to give a conventional OPE expansion. The coefficient function has a simple expression in terms of gamma functions. It can be analytically continued in spin in order to describe the Regge region, as anticipated in \cite{Cornalba:2007fs,Costa:2012cb,Caron-Huot:2017vep}. It might be useful in exploring the formalism of \cite{Caron-Huot:2017vep} to have this example of an interacting theory in which the coefficient function is known exactly (at order $1/N$) as a function of dimension and spin.

Instead of the random interaction in (\ref{introsuperaction}), one can instead consider interactions with a particular fixed tensor as in \cite{Witten:2016iux,KT}. However, there seems to be an obstacle to getting a conformal field theory in these cases. The problem is that these models have an exact global symmetry, and the naive analysis of the 
four-point function for such models leads to a divergence associated to the would-be symmetry currents. We have not understood this completely, but we believe this divergence implies that the large $N$ model does not really find a critical point.

The paper is organized as follows. {\bf In section \ref{supercasimir}} we review superconformal symmetry in one dimension. We discuss the super-Casimir and cross ratios. We derive the $\mathcal{N} = 1$ superconformal blocks as eigenfunctions of the super-casimir operator, following \cite{Fitzpatrick:2014oza}.

{\bf In section \ref{shadow}} we discuss the shadow representation \cite{Ferrara:1972uq,Dolan:2000ut,Dolan:2011dv,SimmonsDuffin:2012uy}, which is a tool for generating conformal or superconformal blocks appropriate for representing the four-point function. We review these functions and how they can be used to write the four-point correlator in the one-dimensional SYK model. We do this for the nonsupersymmetric case following \cite{Maldacena:2016hyu} and also for the supersymmetric case, following \cite{Fu:2016vas}. We work in the superfield formalism to maintain explicit supersymmetry, and we work out some details not discussed in \cite{Fu:2016vas}, including an explicit integral formula for the four-point function. This involves a complete set of one-dimensional superconformal blocks, which includes a continuum and a discrete set. As in the nonsupersymmetric case, the integration contour over the continuum can be deformed, cancelling the discrete set and giving a conventional OPE expansion.

{\bf In section \ref{twod}} we move to two dimensions, using the shadow representation to work out a complete set of conformal eigenfunctions, with weights
\be
h = \frac{1+\ell}{2}+is, \hspace{20pt}\widetilde{h} = \frac{1-\ell}{2}+is
\ee
where the spin $\ell$ is integer and $s$ is real. The completeness relation involves an integral over $s$ and a sum over $\ell$. We show how the four-point function in an SYK-like model can be written as an integral over these conformal eigenfunctions with a weighting factor (determined by the ladder kernel $k(h,\widetilde{h})$ computed in a later section) that implements the sum over the ladder diagrams. Although we have the SYK application in mind, the considerations of this section and the next are quite general, since they essentially just involve working out the resolution of the identity in the basis of conformal eigenfunctions. A similar representation would also be possible for other conformal field theories. 

To get an OPE form for the correlator, the integration contour over $s$ can be deformed. We explain how some spurious poles cancel in making this contour deformation, leaving only the expected OPE expansion, with operator dimensions determined by the conditions $k(h,\widetilde{h}) = 1$ and $h - \th = \ell =\text{integer}$.

{\bf In section \ref{twoscft}} we repeat the previous section but for supersymmetric models in two dimensions, setting up the completeness relation for superconformal eigenfunctions. The important difference from the bosonic model is that the conformal eigenfunctions have discrete labels that indicate whether the exchanged operator is
in a multiplet with a  fermionic or bosonic primary from a holomorphic or antiholomorphic point of view. 
To get a complete set of states, we have to sum over these labels in addition to $\ell,s$.

{\bf In section \ref{bosemod}} we move beyond kinematics and begin discussing dynamics of the two-dimensional bosonic SYK models (\ref{naivebose}) and (\ref{betterbose}). This section is somewhat formal, because as we explain, the models either do not exist or do not appear to flow to conformal phases (although our analysis of this point for the model (\ref{betterbose}) is not conclusive). However, we compute the two-point functions and the ladder kernel $k(h,\widetilde{h})$ formally for all $q$ and discuss some features such as the contribution of the stress tensor.

{\bf In section \ref{supermod}} we discuss dynamical aspects of the supersymmetric theory (\ref{introsuperaction}). First, we compute the two-point function, finding the superconformal form
\be\label{2ptintro}
\langle \uphi_i(1)\uphi_j(2)\rangle = \frac{b\,\delta_{ij}}{|\x_{12} - \theta_1\theta_2|^{2\Delta}},
\ee
with $\Delta = 1/\hat{q}$. We then compute the eigenvalues of the ladder kernel. For each value of $h,\widetilde{h}$, there are four eigenvalues $k^{BB},k^{FB},k^{BF},k^{FF}$, corresponding to the four choices of bosonic or fermionic primaries for the holomorphic and antiholomorphic sectors. We compute each of these and insert the expressions in the formula for the four-point function derived in section \ref{twoscft}. 


In addition to the stress tensor, many other operators appear in the OPE. We can think of these schematically as dressed versions of bilinear operators $\uphi_i \mathcal{D}\uphi_i$ where $\mathcal{D}$ is a differential or superdifferential operator. In these expressions a single index $i$ is summed over, so we refer to these as ``single-sum'' operators. They acquire order one anomalous dimensions at the critical point. The particular dimensions are determined by solving $k(h,\widetilde{h}) = 1$ where $k$ is one of the  ladder kernels 
of the model. The anomalous dimensions become small at large $\hat{q}$, and the central charge also approaches the free value, suggesting that the large $\hat{q}$ fixed point is weakly coupled.

We also discuss the symmetries of the model. Naively, in going to the low energy theory, we can drop the kinetic term in (\ref{introsuperaction}). The theory would then appear to be invariant under general diffeomorphisms of $\x,\overline{\x},\theta,\overline{\theta}$, together with a multiplicative transformation of $\uphi$. However, typical diffeomorphisms will change the UV behavior of the correlator (\ref{2ptintro}), and if we act with such a transformation, we will leave the space of configurations for which it was allowable to drop the gradient terms. The correct low energy symmetry group consists of the diffeomorphisms that preserve the short distance behavior of (\ref{2ptintro}), namely the $\mathcal{N} = 1$ superconformal transformations. This symmetry group is spontaneously broken to a finite-dimensional subgroup by the two-point function (\ref{2ptintro}). In the one-dimensional SYK model, the Goldstone modes associated to a similar breaking were normalizable zero-modes that led to a divergence that spoiled conformal symmetry. In the two-dimensional case, the zero-modes are not normalizable, so they are not on the defining contour of the functional integral and do not lead to a divergence.

However, we do find normalizable zero-modes in a related class of models where the UV theory has a global $U(1)$ symmetry. In this case the IR theory has a local $U(1)$ symmetry that is spontaneously broken to a global $U(1)$ by the vacuum solution. The integral over the action of the broken symmetry generators leads to a divergence. Mathematically, this divergence appears as a double pole on the integration contour at $\ell = 1,s = 0$. We offer a possible interpretation of this divergence as indicating the presence of an operator in the low energy effective action that has a nonzero beta function and prevents models with $U(1)$ symmetry from finding a true critical point.

{\bf In section \ref{ret}} we discuss the behavior of out-of-time-order (OTO) correlators in real time. We review how to set up perturbation theory on a folded time contour in order to get a diagrammatic expansion. We then review an approach introduced by Kitaev \cite{kitaevfirsttalk} that defines a ``retarded'' ladder kernel whose eigenfunctions with unit eigenvalue give the allowed growth exponents for the OTO correlator. We review the computation of the retarded kernel for the one-dimensional SYK model, and then extend this to the two-dimensional models, both bosonic and supersymmetric. Using these computations, we learn that the models do not saturate the chaos bound, but instead have chaos exponents that are less than the bound by order one factors. In other words, for these models, the Regge intercept is at a spin somewhere between one and two.

{\bf In section \ref{sec:anacon}} we show how the behavior in the chaos limit can also be obtained by analytic continuation of the four-point formula derived in section \ref{supermod}. This follows the approach anticipated in \cite{Cornalba:2007fs,Costa:2012cb} in the discussion of the Regge limit. The only subtlety here has to do with picking the right analytic continuation of the kernels $k(h,\widetilde{h})$ and the conformal eigenfunctions as a function of spin.

{\bf In the discussion}, we mention some expectations for the theories at finite $N$, suggest an extension to three dimensions, and comment briefly on the possible holographic dual.

Several details are explained in appendices.

\section{Superconformal Symmetry In One Dimension}\label{supercasimir}
In this section, we give an introduction to global superconformal symmetry in one dimension. We work out the super-Casimir operator and its associated inner product, describe the cross ratio invariants for a configuration of four points, and write the superconformal blocks \cite{Fitzpatrick:2014oza}.

The global conformal group in one dimension is $SL(2,\mathbb{R})$. The global superconformal group extends this to $OSp(1|2)$ by the addition of two fermionic generators. The $osp(1|2)$ super-algebra is defined through
\begin{eqnarray}
  \left[L_m,L_n\right] = (m-n)L_{m+n},\qquad\left\{G_{r},G_{s}\right\} = 2 L_{r+s}, \qquad 
  \left[L_{m},G_{\pm1/2}\right] = \left(\frac{m}{2}\mp \frac{1}{2}\right)G_{m\pm 1/2}\,,\nonumber
\end{eqnarray}
where $r,s = \pm1/2$ and $m,n = -1,0,1$.  This algebra acts on functions $\Phi(t,\theta)$ on a 1-dimensional superspace by\footnote{We have chosen sign conventions
so that these formulas are a specialization of a standard realization of the super-Virasoro algebra: \begin{eqnarray}
  G_{r} = t^{r+1/2}\left(\partial_{\theta}-\theta\partial_{t}\right),\qquad L_{m}=-\left(t^{m+1}\partial_{t}+\frac{m+1}{2}t^{m}\theta
  \partial_{\theta}\right).
\end{eqnarray}}
\begin{eqnarray} \label{algv}
   L_{0} &=&  - t\partial_{t} - \frac{1}{2}\theta\partial_{\theta}-\Delta \hspace{50pt}G_{-1/2} = \partial_{\theta} - \theta\partial_{t},\nonumber\\
   L_{-1} &=& -\partial_{t}\hspace{119pt}G_{+1/2} = t\partial_{\theta} - t\theta \partial_{t} - 2\Delta \theta\\
   L_{+1} &=& -t^{2}\partial_{t} - t\theta\partial_{\theta}- 2\Delta t
   .\nonumber 
\end{eqnarray}
Here $\Delta$ is the conformal dimension of the field $\Phi$. The quadratic Casimir of $osp(1|2)$ is
\begin{eqnarray}
  C = L_{0}^{2} - \frac{1}{2}\left\{L_{-1}, L_{+1}\right\} - 
  \frac{1}{4}\left[G_{-1/2},G_{+1/2}\right].
\end{eqnarray}
For the one-particle realization (\ref{algv}) of the algebra, we find 
\begin{eqnarray}\label{onepc}
  C = \Delta^{2} - \frac{1}{2}\Delta.
\end{eqnarray}
The corresponding bosonic result is $\Delta^2-\Delta$.

We can now construct the  Casimir  $\C_{12}$ of a two-particle system with coordinates $(t_{1},\theta_{1})$ and $(t_{2},\theta_{2})$:\footnote{The notation here is, for example, that $L_{0}^{(1)}$ refers to the $L_{0}$ generator of particle 1.}
\begin{eqnarray}
  \C_{12} &=& \left( L_{0}^{(1)}+L_{0}^{(2)} \right)^{2} - 
  \frac{1}{2}\left\{ (L_{-1}^{(1)}+L_{-1}^{(2)}), (L_{+1}^{(1)}+L_{+1}^{(2)})\right\}
  - \frac{1}{4}\left[ (G_{-1/2}^{(1)}+G_{-1/2}^{(2)}), (G_{+1/2}^{(1)}+G_{+1/2}^{(2)})\right]
  \nonumber\\
  &=& 2\left(\Delta^{2} - \frac{1}{2}\Delta\right) + 2L_{0}^{(1)}L_{0}^{(2)} - L_{-1}^{(1)}L_{+1}^{(2)} - 
  L_{+1}^{(1)}L_{-1}^{(2)} - \frac{1}{2} G_{-1/2}^{(1)}G_{+1/2}^{(2)} + 
  \frac{1}{2} G_{+1/2}^{(1)}G_{-1/2}^{(2)}\,.
\end{eqnarray}
In going to the second line we collected terms proportional to the individual casimirs $C^{(1)}$ and $C^{(2)}$ and then used (\ref{onepc}), assuming both particles have dimension $\Delta$. Using (\ref{algv}), this becomes an explicit differential operator, which for $\Delta = 0$ is
\begin{align}\label{zelo}
  \C_{12}\Phi(t_{1},\theta_{1},t_{2},\theta_{2}) =& -\left(t_{12} - 
  \frac{1}{4}\theta_{1}\theta_{2}\right)^{2}\partial_{1}\partial_{2}\Phi+ 
  \frac{1}{2}\left(t_{12}-\theta_{1}\theta_{2}\right)\partial_{\theta_{1}}
  \partial_{\theta_{2}}\Phi \nonumber\\ 
  &+t_{12}\left(\theta_2-\frac{1}{2}\theta_{1}\right)\partial_{\theta_{2}}
  \partial_{1}\Phi- t_{12}\left(\theta_{1}-\frac{1}{2}\theta_{2}\right)\partial_{\theta_{1}}\partial_{2}\Phi\,,
\end{align}
where $t_{12} = t_1-t_2$. For general $\Delta$, we conjugate by a factor of $(t_{12}-\theta_1\theta_2)^{2\Delta}$.

\subsection{The Casimir in $OSp(1|2)$-invariant coordinates}
Now we consider a four-particle system with coordinates $t_1,\theta_1;\dots; t_4,\theta_4$.  The normalized four-point function in a superconformally-invariant
theory will be a superconformally-invariant function of these variables.  For four fermionic (or bosonic) operators, the four-point function is also Grassmann-even.
A function with these properties can only depend on the Grassmann-even invariants \cite{Qiu:1986if} which we parametrize as
\begin{align}\label{zelf}
  \chi&\equiv\frac{\langle 1,2\rangle \langle 3,4\rangle}{\langle 1,3\rangle\langle 2,4\rangle} -\zeta,\\
  \zeta&\equiv\frac{\langle 1,2\rangle\langle 3,4\rangle+\langle 2,3\rangle\langle 1,4\rangle 
  + \langle 3,1\rangle\langle 2,4\rangle}{\langle 1,3\rangle\langle 2,4\rangle},\nonumber
\end{align}
with $\langle i,j\rangle\equiv t_{i}-t_{j}-\theta_{i}\theta_{j}$.  Here $\chi$ is a supersymmetric analog of the usual cross ratio of four points on the real line, and $\zeta$ is nilpotent, $\zeta^2 = 0$. One could have defined $\chi$ without the $-\zeta$ term, but this parametrization turns out to be more convenient.
To see why these are the only even invariants, observe that $OSp(1|2)$, which has bosonic dimension 3 and fermionic dimension 2, can be used to fix 3 bosonic coordinates and 2 fermionic ones, say
\be\label{zelof}t_1=0,\,t_3=1,t_4=\infty,~~\theta_3=\theta_4=0.\ee  This leaves one even modulus $t_2$ and two odd moduli $\theta_1$ and $\theta_2$.  But the four-point function that we are studying, since it is bosonic, can only be a function of
$t_2$ and $\theta_1\theta_2$, or equivalently of
\be\label{chival}\chi=t_2,~~\zeta = \theta_1\theta_2.\ee
Here we see easily that $\zeta^2=0$, since $\zeta$ is a bilinear in two fermionic variables.

As a result, we need only understand how the super Casimir acts on functions of the form $\Phi(\chi,\zeta) = F(\chi)+\zeta G(\chi)$. The computation  of $\C_{12}$ as an operator acting on $\Phi(\chi,\zeta)$ is facilitated by first acting with $\C_{12}$ on a general function 
of all variables $t_1,\theta_1,\dots, t_4,\theta_4$.   It is convenient to use $OSp(1|2)$ to fix $t_3=1,t_4=\infty$ and $\theta_3=\theta_4=0$, but since we will be taking derivatives with respect to both $t_1,t_2$, it is important not to fix either of these coordinates. Then we have
\be\label{chizeta12}
\chi = \frac{t_1-t_2}{t_1-1},\hspace{20pt}\zeta = \frac{\theta_1\theta_2}{1-t_1}.
\ee
We substitute these expressions into a general function $ \Phi(\chi,\zeta)=F(\chi)+\zeta G(\chi)$, apply the Casimir using the representation as a differential operator in (\ref{zelo}), and then reorganize the result in terms of $\chi,\zeta$. (At this point, one can set $t_1=0$ and hence $\chi=t_2$, $\zeta=\theta_1\theta_2$.) After some work, one finds
\begin{eqnarray}
  \C_{12}\left(\begin{array}{c}
                                                                                                     F(\chi)\\
                                                                                                     G(\chi)
                                                                                                  \end{array}
                                                                                       \right)
  = \mathcal{D}\left(\begin{array}{c}
                                                                                                     F(\chi)\\
                                                                                                     G(\chi)
                                                                                                  \end{array}
                                                                                       \right),
\label{Casimir-system}
\end{eqnarray}
with the $2\times2$ matrix differential operator
\begin{eqnarray}\label{matdiffop}
\mathcal{D} = \left(
                     \begin{array}{cc}
                       \chi^{2}(1{-}\chi)\partial_{\chi}^{2} 
                       - \chi^{2}\partial_{\chi} & \frac{\chi}{2}\\
                       \frac{\chi}{2}(1{-}\chi)\partial_{\chi}^{2} 
                       - \frac{\chi}{2}\partial_{\chi} & 
                       \chi^{2}(1{-}\chi)\partial_{\chi}^{2} 
                       +\chi(2{-}3\chi)\partial_{\chi} - \chi+\frac{1}{2}
                     \end{array}
                     \right)\,.
\end{eqnarray}

\subsection{The Measure}\label{measure}
This operator is not  Hermitian with respect to any positive-definite inner product. To understand in what sense it is Hermitian, let us revisit the purely bosonic case as discussed in \cite{Maldacena:2016hyu}. There, setting $\Delta = 0$, the Casimir acts on a function $F(t_{1}, t_{2})$ as
\begin{eqnarray}
  C_{12}\,F(t_{1}, t_{2}) &=& \left(-K_{1}P_{2}-P_{1}K_{2}+2D_{1}D_{2}\right)
  F(t_{1}, t_{2})\nonumber\\
  &=&-(t_{1}-t_{2})^{2}\,\frac{\partial^{2}}{\partial t_{1}\partial t_{2}}F(t_{1},t_{2})\,.
\end{eqnarray}
It is clear that this operator is Hermitian with respect to the inner product
\begin{eqnarray}\label{bosan}
  \langle F| G\rangle = \int\frac{dt_{1}dt_{2}}{(t_{1}-t_{2})^{2}}F^{*}G.
\end{eqnarray}
For a general function of four variables $t_1,\dots,t_4$, we could use the conformally-invariant inner product
\be \label{longerone} \langle F| G\rangle = \int\frac{dt_{1}dt_{2}}{(t_{1}-t_{2})^{2}}\frac{dt_{3}dt_{4}}{(t_{3}-t_{4})^{2}}F^{*}G, \ee
and $C_{12}$ is still hermitian acting in this space.  Now we can specialize to $SL(2,\R)$-invariant functions, that is functions of the ordinary
cross-ratio $\chi=\frac{t_{12}t_{34}}{t_{13}t_{24}}$.  The integral in (\ref{longerone}) is
divergent in the case that $F$ and $G$ are $SL(2,\R)$-invariant.  In the usual way, to remove this divergence, one fixes the action
of $SL(2,\R)$ by setting any three of $t_1,\dots, t_4$ to constant values $a,b,c$ and multiplying by $|a-b||b-c||c-a|$.  Fixing $t_1=0$,
$t_3=1$, and $t_4=\infty$, by virtue of which $t_2=\chi$, one arrives at the inner product
\be\label{invone}\langle F|G\rangle =\int_0^2 \frac{d\chi}{\chi^2} F^* G. \ee
Here, anticipating the  symmetry $\chi\to \chi/(\chi-1)$ of  the SYK model (see  below), we 
have restricted the integral to a fundamental
domain
 $0\leq\chi\leq 2$; in general, the integral would run over the whole real line.
 
In the supersymmetric case, the invariant inner product for two particles is 
\be\label{nvone}\langle A|B\rangle =\int \frac{d t_1 d t_2 d \theta_2 d\theta_1}{\langle 1,2\rangle} AB= \int \frac{d t_1 d t_2 d \theta_2 d\theta_1}{t_1-t_2-\theta_1\theta_2} AB. \ee
This will reduce to the bosonic analog (\ref{bosan}) if $A$ and $B$ are independent of the $\theta_1,\theta_2$ and we integrate over
the $\theta$'s.\footnote{We have written the fermionic measure as $\int d\theta_2d\theta_1$ so that $\int d\theta_2d\theta_1\theta_1\theta_2 = 1$. In our conventions, $\int d\theta_1 d\theta_2 \theta_1\theta_2 = -1$.} In contrast to the bosonic case, the supersymmetric inner product is not positive definite and will not lead to a Hilbert
space structure no matter how we proceed.  Accordingly, to include complex conjugation as part of the definition does not appear to
be helpful, and we have defined the inner product in eqn.~(\ref{nvone}) with no complex conjugation, 
as a bilinear inner product rather than a hermitian one.

It is not difficult to see that the Casimir operator as defined in (\ref{zelo}) is Hermitian with respect to the indefinite inner product defined
in eqn.~(\ref{nvone}). After integrating by parts one can show
\begin{align}\label{zother}
  \langle A(t_{1},\theta_{1},t_{2},\theta_{2}) | & \C_{12}B(t_{1},\theta_{1},t_{2},\theta_{2}) \rangle = \int 
  \frac{dt_{1}dt_{2}d\theta_{2}d\theta_{1}}{t_{1}-t_{2}-\theta_{1}\theta_{2}}\,A \left(\C_{12}B\right)
  \cr=& \int  dt_{1}dt_{2}d\theta_{2}d\theta_{1} \left(-\partial_{1}A + \partial_{2}A +\partial_{1}A - \partial_{2}A\right)B+ 
  \int \frac{dt_{1}dt_{2}d\theta_{2}d\theta_{1}}{t_{1}-t_{2}-\theta_{1}\theta_{2}}\,\left(\C_{12}A\right)B 
  \cr =& \langle  \C_{12}A(t_{1},\theta_{1},t_{2},\theta_{2}) |B(t_{1},\theta_{1},t_{2},\theta_{2}) \rangle\,. 
\end{align}
To go over to functions of $OSp(1|2)$ invariants, we first introduce additional variables $t_3,\theta_3$ and $t_4,\theta_4$,
generalizing eqn.~(\ref{longerone}) in the obvious way.  Then if we restrict to the case that $A$ and $B$ are $OSp(1|2)$-invariant,
we can gauge fix the $OSp(1|2)$ action by setting, for example, $t_1,t_3,t_4=0,1,\infty$, $\theta_3=\theta_4=0$.  The inner product then becomes
\be\label{zem}\langle A|B\rangle =-\int\frac{d \chi d\theta_2d\theta_1} {\chi+\zeta}A(\chi,\theta_1,\theta_2)B(\chi,\theta_1,\theta_2).\ee
The Casimir is hermitian with respect to this inner product, since it was hermitian as an operator acting on the full set of
variables $t_1,\dots,\theta_4$, and as it is $OSp(1|2)$-invariant, this does not change when we restrict to $OSp(1|2)$-invariant functions.

In these coordinates, the nilpotent invariant $\zeta$ is simply $\zeta=\theta_1\theta_2$.  For the case that $A$ and $B$ are
functions of $\chi$ and $\zeta$ only, it is convenient to abbreviate $d\theta_2 d\theta_1$ as $d\zeta$, with the rule
$\int d\zeta (a+b\zeta)=b$, for constants $a,b$.
Then we can write
\be\label{uzem}\langle A|B\rangle =-\int\frac{d \chi d\zeta} {\chi+\zeta}A(\chi,\zeta)B(\chi,\zeta).\ee

The ordinary SYK model has an important symmetry under which
 the ordinary cross ratio transforms as  $\chi\to \chi/(\chi-1)$.  This has the effect of interchanging particles 1 and 2 (or 3 and 4).
The same exchange is also a symmetry of the supersymmetric SYK model.  A look back to eqn.~(\ref{zelf}) or more simply (\ref{chizeta12}) reveals that the $1\leftrightarrow 2$
exchange acts on $\chi$ and $\zeta$ by
\be\label{nchisym}\chi\to \frac{\chi}{\chi-1},~~\zeta\to\frac{\zeta}{\chi-1}. \ee
For functions with this symmetry, we can restrict the range of integration for $\chi$ in (\ref{uzem}) to run from 0 to 2, and take the inner product to be
\be\label{iluzem}\langle A|B\rangle =-\int_0^2\frac{d \chi d\zeta} {\chi+\zeta}A(\chi,\zeta)B(\chi,\zeta).\ee
The $2\times 2$ matrix operator in \eqref{matdiffop} is Hermitian with respect to this inner product in that, for $A(\chi,\zeta)$ and $B(\chi,\zeta)$ as above, $\langle A|\mathcal{D}B\rangle = \langle \mathcal{D}A|B\rangle$.
Now let us discuss the invariance of $\mathcal{D}$ under the discrete symmetry (\ref{nchisym}).   For
$\Phi(\chi,\zeta)=F(\chi)+\zeta G(\chi)$ to be invariant under that symmetry means that $F$ is invariant but $G$ transforms
to $(\chi-1)G$.  
Using the transformations of the derivatives under the discrete symmetry
\begin{eqnarray}
 \partial_{\chi}\to -(\chi-1)^{2}\partial_{\chi}\,\quad\mathrm{and}\quad
 \partial_{\chi}^{2} \to (\chi-1)^{4} \partial_{\chi}^{2} +
 2(\chi-1)^{3}\partial_{\chi},
\end{eqnarray}
it is not difficult to show that the $2\times 2$ matrix operator $\mathcal{D}$ is invariant under the symmetry in the sense that
if $\Phi$ is invariant then $\mathcal{D}\Phi$ is also invariant.

\subsection{Solving the super-Casimir Differential Equation}\label{Solving}
Returning to the system of equations (\ref{Casimir-system}), we would like to solve the eigenvalue problem
\begin{eqnarray} \label{oldfish}
 \mathcal{D}\left(\begin{array}{c}
                                         F(\chi)\\
                                         G(\chi)
                                     \end{array}
                            \right) = h\left(h-\frac{1}{2}\right)
                            \left(\begin{array}{c}
                                         F(\chi)\\
                                         G(\chi)
                                     \end{array}
                            \right)\,,
\end{eqnarray}
since the Casimir for two particles coupled to a dimension-$h$ primary is $h\left(h-\frac{1}{2}\right)$. 
As noted in \cite{Fitzpatrick:2014oza}, this system of coupled second order equations can be solved by setting  $G(\chi) = \frac{h}{\chi}F(\chi)$, 
leading to
\begin{eqnarray}\label{boseq}
  \chi^{2}(1-\chi)\partial_{\chi}^{2}F - \chi^{2}\partial_{\chi}F - h(h-1)F=0\,.
\end{eqnarray}
This is simply the condition for $F$ to be an eigenfunction of the Casimir operator of $SL(2,\R)$, which acts as the differential operator
\be
C_{12} = \chi^2(1-\chi)\partial_\chi^2-\chi^2\partial_\chi.
\ee
Its general solution is the linear combination
\begin{eqnarray}\label{zumbo}
  F(\chi) = c_{1}\,\chi^{h}\,{}_{2}F_{1}(h,h;2h;\chi) + c_{2}\,\chi^{1-h}\,{}_{2}F_{1}(1-h,1-h;2-2h;\chi)
\end{eqnarray}
where $_2F_1$ is a hypergeometric function and $c_1,c_2$ are constants. 

Since the eigenvalue problem (\ref{oldfish}) is invariant under $h\to 1/2-h$, we could just as well set $G(\chi)=\frac{1/2-h}{\chi}F(\chi)$.  In this
case, $F$ has to be an eigenfunction of $C_{12}$ with eigenvalue $(1/2-h)(1/2-h-1)=h^2-1/4$ (and we should replace $h$ by $1/2-h$ in eqn.~(\ref{zumbo})).
The two choices of ansatz for $G$ together with the choice of two constants $c_1$ and $c_2$  in eqn.~(\ref{zumbo}) give
a total of four linearly independent solutions.  This is the right number, as the supersymmetric eigenvalue problem (\ref{oldfish}) is a second
order differential equation for two functions.  From the four linearly independent solutions, we
 will need to select a subset that will form a complete basis for the space of functions satisfying the appropriate boundary conditions. 
 For this we will use the shadow representation.

\section{The Shadow Representation}\label{shadow}


\renewcommand{\thefootnote}{\arabic{footnote}}

\subsection{Overview}

The ``shadow representation''  \cite{Ferrara:1972uq,Dolan:2000ut,Dolan:2011dv,SimmonsDuffin:2012uy} is, for our purposes, 
a way to construct a possible four-point function 
in a conformal field theory with a specified
value of the conformal Casimir operator in a chosen channel.

Suppose, for example, that $\O$ and $\O'$  are conformal primaries of some dimension $\Delta$ in a conformal field theory in $D$ dimensions, and
that we want to understand a connected
four-point function $\langle \O(x_1)\O(x_2)\O'(x_3)\O'(x_4)\rangle$.  This four-point function can be expanded as a linear
combination of eigenfunctions of the conformal Casimir operator\footnote{We write $C_{12}$ for the two particle Casimir of the conformal group, and $\C_{12}$ for its  superanalog.} $C_{12}$ in the $12$ channel.  
However, instead of directly solving the eigenvalue equation associated with the Casimir, it is much easier to  write down an integral representation
of a function that has all of the necessary properties.    To do this, we imagine that $\O$ and $\O'$ are operators in two decoupled CFT's,
and that the first theory has a primary field\footnote{For simplicity we assume for the moment that $\V$ and $\V'$ are bosonic and spinless.}
 $\V$ of some dimension $h$ while the second has a primary field $\V'$ of complementary dimension
$D-h$.   In the product theory, the connected four-point function $\langle \O(x_1)\O(x_2)\O'(x_3)\O'(x_4)\rangle$ simply vanishes.
However, if we perturb the product of the two decoupled theories by
\be\label{perturb}\varepsilon\int d^D y \,\V(y)\V'(y),\ee
 then to first order in $\varepsilon$, we get a connected four-point function
\be\label{connfour}\langle \O(x_1)\O(x_2)\O'(x_3)\O'(x_4)\rangle=\varepsilon\int d^Dy\langle \O(x_1)\O(x_2)\V(y)\rangle\langle\V'(y)\O(x_3)\O(x_4)\rangle.
\ee
The quantity on the right hand side is manifestly single-valued and conformally-covariant as a function of $x_1,x_2,\dots,x_4$.  It is an eigenfunction
of $C_{12}$ with an eigenvalue that depends on $h$.  In fact, for any $y$,
the three-point function $\langle \O(x_1)\O(x_2)\V(y)\rangle$ describes coupling of 
$\O(x_1)$ and $\O(x_2)$ to a primary of dimension $h$, and is an eigenfunction of $C_{12}$ with the corresponding eigenvalue (for $\V$ a spinless
field in a bosonic CFT in $d$ dimensions, the eigenvalue is $h(h-D)$).  Integration over $y$ as in eqn.~(\ref{connfour}) does not affect this statement, so
the right hand side of eqn.~(\ref{connfour}) is a conformally-invariant and single-valued wavefunction that is an eigenstate of $C_{12}$.

This approach and its generalizations for fields with spin is a convenient way to construct 
appropriate basis functions from which the full four-point function of a model like the SYK model
can be reconstructed.   In what follows, after illustrating the method by reviewing some results of \cite{Maldacena:2016hyu} 
for the SYK model in 1 dimension, we apply these ideas to supersymmetric and/or 2-dimensional analogs of the SYK model.

\subsection{The SYK Model in 1 Dimension}\label{syk}

The SYK model in 1 dimension,  with $q$-fold couplings,  has  in the large $N$ limit fermionic primary fields $\psi_i$ of dimension $\Delta=1/q$
and disorder-averaged two-point functions
\be\label{two}\langle \psi_i(t)\psi_{i'}(t')\rangle=\delta_{ii'}\frac{\sgn(t_1-t_2)}{|t_1-t_2|^{2\Delta}} ,\ee
where we have normalized $\psi$ to remove a constant. One wishes to understand the connected four-point function $\langle\psi_i(t_1)\psi_i(t_2)\psi_j(t_3)\psi_j(t_4)\rangle$, $i\not=j$.
It is convenient to normalize the four-point function by dividing by a product of two-point functions.  This gives a function
\be\label{normfn}W(t_1,t_2,t_3,t_4)=\frac{\langle\psi_i(t_1)\psi_i(t_2)\psi_j(t_3)\psi_j(t_4)\rangle}{\langle\psi_i(t_1)\psi_i(t_2)\rangle\,\langle\psi_j(t_3)\psi_j(t_4)\rangle}\ee
 that is conformally-invariant  (rather than conformally covariant).  
 
 In an SYK-like model, $W$ is not exactly the most convenient normalized four-point function.  In that context, one wishes to average each correlation function
 in the numerator or denominator of (\ref{normfn}) over disorder and over the labels $i$ and/or $j$.  One also wants to remove the  contribution
 in the numerator that is disconnected in the 12 channel (that is, the contribution from the identity operator in that channel). We write $\langle ~~\rangle'$ for
 a correlation function that is averaged and partly connected in this sense. Finally, one multiplies by an overall factor of $N$ to get a function that has a large $N$ limit.  
 Thus the natural normalized four-point function in an SYK-like model is actually
 \be\label{normalfn}\F(t_1,t_2,t_3,t_4)=N\frac{\langle\psi_i(t_1)\psi_i(t_2)\psi_j(t_3)\psi_j(t_4)\rangle'}{\langle\psi_i(t_1)\psi_i(t_2)\rangle'\,\langle\psi_j(t_3)\psi_j(t_4)\rangle'}\ee
 In this paper, general remarks on conformal field theory in 1 or 2 dimensions are applicable to either version of the normalized four-point function, but specific
 applications to SYK-like models always refer to $\F$.   
 
  In 1 dimension, 
 the three-point function $\langle\psi_i(t_1)\psi_i(t_2)\V(y)\rangle$,
 where $\psi_i$ is a fermionic primary of dimension $\Delta$ and $\V$ is a suitably normalized primary  of dimension $h$ is
 \be\label{thrept}\langle \psi_i(t_1)\psi_i(t_2)\V(y)\rangle = \frac{\sgn(t_1-t_2)}{|t_1-t_2|^{2\Delta-h}|t_1-y|^h|t_2-y|^h}.\ee
 (Here $\V$ is necessarily bosonic or this correlation function would vanish.)
 Inserting this formula and its analog for $\V'$
 in the shadow representation and dividing by the product of two-point functions, we find that a contribution to the normalized
 four-point function $\F(t_1,t_2,t_3,t_4)$ that is an eigenfunction of $C_{12}$ is a multiple of\footnote{\label{trumf}
 The factors $\sgn(t_1-t_2)$ and $\sgn(t_3-t_4)$
 that come from the three-point functions cancel against similar factors in the two-point functions that are in the denominator of (\ref{normfn}),
 so there are no such factors in the following formula.}
 \be\label{gruff}\Psi_h(t_1,t_2,t_3,t_4)=\frac{1}{2}\int_{-\infty}^\infty \!\!\!\!dy \,\,\frac{|t_1-t_2|^h|t_3-t_4|^{1-h}}{|t_1-y|^h|t_2-y|^h|t_3-y|^{1-h}|t_4-y|^{1-h}}.\ee
 This function was introduced in eqn.~(3.67) of \cite{Maldacena:2016hyu}.    The integral converges
 if $0<\mathrm{Re}\,h <1$.   However, as we discuss below,
 the integral representation can be used to prove that $\Psi_h$ has an analytic continuation as a meromorphic
 function throughout the complex $h$ plane and to  locate its poles. 
 
 Conformal invariance implies that $\Psi_h(t_1,t_2,t_3,t_4)$ is actually a function only of the conformally-invariant cross ratio
 \be\label{crossr}\chi=\frac{(t_1-t_2)(t_3-t_4)}{(t_1-t_3)(t_2-t_4)}. \ee
$SL(2,\R)$ symmetry  can be used to map $t_1,t_2,t_3,t_4$
to $0,\chi,1,\infty$, whereupon we get
\be\label{psih}\Psi_h(\chi)=\frac{1}{2}\int_{-\infty}^\infty \!\!\!\!dy \,\,\frac{ |\chi|^h}{|y|^h|\chi-y|^h|1-y|^{1-h}}.\ee
From this representation, one can immediately deduce two important properties of $\Psi_h(\chi)$ that also have analogs in all of the other models
we will study.  First, by considering the change
of variables $y\to y/(y-1)$, one can deduce from (\ref{psih}) that
\be\label{reflection}\Psi_h(\chi)=\Psi_h(\chi/(\chi-1)). \ee   Note that $y\to y/(y-1)$ exchanges the two points $t_3=1$ and $t_4=\infty$, leaving fixed $t_1=0$.  The symmetry
of the shadow representation under this operation exists because identical operators are inserted at $t_3,t_4$.  Note also that $y\to y/(y-1)$
is orientation-reversing; it is of the form $y\to (ay+b)/(cy+d)$ with $ad-bc=-1$, so it is in $GL(2,\R)$, not $SL(2,\R)$.    The $\chi\to \chi/(\chi-1)$ symmetry
is even more obvious in eqn.~(\ref{gruff}); it reflects the fact that the integrand is invariant under the exchanges $1\leftrightarrow 2$ or $3\leftrightarrow 4$.
Second, we can consider a change of variables that exchanges $t_1,t_2$ with $t_3,t_4$, leaving
$\chi$ invariant.  Assuming that the operators $\psi_i$
and $\psi_j$ have the same dimension, as in the SYK model,
eqn.~(\ref{connfour}) is manifestly invariant under $h\leftrightarrow 1-h$ together with $t_1,t_2\leftrightarrow t_3,t_4$.
This reflects the fact that $\Psi_h$ is an eigenfunction of the Casimir $C_{12}$ with eigenvalue $h(h-1)$, a formula that is invariant under $h\leftrightarrow
1-h$.  
Concretely, with $t_1,t_2,t_3,t_4=0,\chi,1,\infty$, the requisite change of variables in the shadow integral (\ref{psih}) is $y\to \chi/y$ and leads
to 
\be\label{secref}\Psi_h(\chi)=\Psi_{1-h}(\chi). \ee

Because of the symmetry under $\chi\to \chi/(\chi-1)$, one can restrict $\Psi_h$ to  $0\leq \chi\leq 2$.  In understanding the
operator product expansion of the four-point function, it is important to understand the behavior of $\Psi_h$ for small (positive) $\chi$.  
This can be deduced directly from the integral representation.  To begin with, we work in the region $0<\mathrm{Re}\,h<1$ where the integral
converges.  If we further restrict to $\Re(h)<1/2$, then the small $\chi$ behavior of $\Psi_h$ can be found by just naively setting $\chi$ to 0 in the
denominator in eqn.~(\ref{psih}).   Thus we get
\be\label{nefl}\Psi_h(\chi)\sim\frac{\chi^h}{2} \int_{-\infty}^\infty \!\!\!\!dy \,\,\frac{1}{|y|^{2h}|1-y|^{1-h}}. \ee
The integral over any of the three regions  $y\leq 0$, $0\leq y\leq 1$, and $1\leq y$ is a standard representation of an Euler beta function.  Adding
the three contributions and using standard identities, one finds that
\be\label{brefl} \Psi_h(\chi)\sim \chi^h\, \frac{\tan \pi h}{2\tan(\pi h/2)} \frac{\Gamma^2(h)}{\Gamma(2h)},~~ 0<\Re(h)<1/2. \ee
For $1/2<\Re(h)<1$, the integral in (\ref{nefl}) diverges, and the small $\chi$ behavior of $\Psi_h$ cannot be obtained simply by setting
$\chi$ to 0 in the denominator in (\ref{psih}).  Rather, for this range of $h$, the small $\chi$ behavior of the integral comes from the region $|y|\sim \chi$.
To extract the leading contribution, we set $y=\chi z$, after which $|1-y|$ can be replaced by 1 in the denominator in (\ref{psih}) and we get
\be\label{befl}\Psi_h(\chi)\sim\frac{\chi^{1-h}}{2} \int_{-\infty}^\infty \!\!\!\!dz \,\,\frac{1}{|z|^{h}|1-z|^{h}}. \ee
After a further change of variables $z\to 1/z$, one finds that the integral in (\ref{befl}) is the same as that in (\ref{nefl}) but with $h\to 1-h$.
Hence in this region,
\be\label{uwefl} \Psi_h(\chi)\sim \chi^{1-h}\frac{\tan \pi (1-h)}{2\tan(\pi (1-h)/2)} \frac{\Gamma^2(1-h)}{\Gamma(2(1-h))},~~ 1/2<\mathrm{Re}\,h<1. \ee
This could also have been deduced from (\ref{brefl}) and the relation $\Psi_h(\chi)=\Psi_{1-h}(\chi)$.  

The differential equation $C_{12}\Psi_h=h(h-1)\Psi_h$ is
a hypergeometric equation that in the region $0<\chi<1$ has the two linearly independent solutions $F_h$ and $F_{1-h}$, where we define
\be
F_h(\chi)\equiv \chi^h \,_2F_1(h,h,2h,\chi).
\ee
Here $_2F_1$ is  standard notation for a hypergeometric function, and $F_h$ is familiar as the usual $SL(2,\R)$ conformal block.
For $0<\chi<1$,  $\Psi_h$ must be a linear combination of these two functions.  To determine
the coefficients, we just observe that the hypergeometric functions equal 1 at $\chi=0$.  
In the region $\Re\,h<1/2$, the $F_h$ function dominates and its coefficient  can be determined by comparing to (\ref{brefl}),
while for $\Re\,h>1/2$, the $F_{1-h}$ term dominates and its coefficient  can be determined by comparing to (\ref{uwefl}).

We therefore have
\be\label{rrefl} \Psi_h= \t A(h) F_h(\chi)+\t B(h) F_{1-h}(\chi)    , ~~0\leq \chi<1     \ee
with
\be\label{zefl}\t A(h)=\frac{\tan \pi h}{2\tan(\pi h/2)} \frac{\Gamma^2(h)}{\Gamma(2h)},~~~
\t B(h)=\t A(1-h).\ee
In \cite{Maldacena:2016hyu}, this is written $\t A(h)=A(h)\Gamma^2(h)/\Gamma(2h)$, $\t B(h)=B(h)\Gamma^2(1-h)/\Gamma(2-2h)$,
with
\be\label{defl}A(h)=\frac{\tan(\pi h)}{2\tan(\pi h/2)},~~~B(h)=-\frac{1}{2}\tan(\pi h)\tan(\pi h/2). \ee  It is also possible to prove
(\ref{rrefl}) by directly comparing the integral formula (\ref{gruff}) for $\Psi_h$ to the standard integral formula for $_2F_1$.  

In the present case, the expansion (\ref{rrefl}) and standard properties of the hypergeometric functions establish that $\Psi_h(\chi)$, $0\leq\chi<1$,
 can be continued meromorphically in $h$.  However, we will briefly explain how one could deduce this (for all $\chi\not=0,1,\infty$) 
 directly from the shadow representation,
 without reference to the differential equation.  The dangerous regions in the shadow integral (\ref{psih}) are for $y$ near $0, \chi,1$, and $\infty$.
 Remove a small ball around each of these bad points (by a small ball around, say, $y=0$ we mean the set $|y|\leq \varepsilon$ for some
 small $\varepsilon$, and by a small ball around $\infty$ we mean the set $|y|\geq 1/\varepsilon$).  The integral over the complement of the small balls
 is trivially an entire function of $h$.  To understand the integrals over the small balls, we note for example that the integral over the small ball
 near 0 is
 \be\label{wefl}\int_{-\varepsilon}^\varepsilon  \frac{d y}{|y|^h} \,f(y,h,\chi), \ee
 for some smooth function $f(y,h,\chi)$.  If $f(y,h,\chi)$ is a polynomial in $y$, this integral can be performed in closed form and is a meromorphic
 function with a finite number of simple poles at positive odd integer values of $h$.  If $f(y,h,\chi)$ vanishes near $y=0$ to a degree greater than $\mathrm{Re}\,h-1$, then the integral is
 holomorphic in $h$.  In any region of bounded $\Re\,h$, the function $f(y,h,\chi)$ can be written as the sum of a polynomial and a function that vanishes
 to the desired high degree.  So the integral over the ball is meromorphic in $h$, with simple poles at positive odd integers and explicitly calculable
 residues of these poles.    The behavior in the other small balls is similar except that near $y=1$ or $\infty$, $h$ is replaced by $1-h$.
 So $\Psi_h$ is meromorphic with its only singularities being simple poles if $h$ is a positive odd integer or a negative even one.
 
\subsection{The Supersymmetric SYK Model}\label{susysyk}

Now we will adapt this discussion to the supersymmetric SYK model \cite{Fu:2016vas}, still in $D=1$, with $\N=1$ supersymmetry.   A primary field $\V(t,\theta)$
now depends on a fermionic coordinate $\theta$ as well as a bosonic coordinate $t$.  The supersymmetric shadow representation is constructed
with insertion of 
\be\label{double}\int dy\,d\theta \,\V(y,\theta)\V'(y,\theta),\ee
where $\V'$ is another superconformal primary.  Now, however, the measure $dy \,d\theta$ has length dimension $1/2$, rather than 1
as in the bosonic case.  Consequently, if $\V$ has dimension $h$, then $\V'$ must have dimension $1/2-h$.  This is related to the fact that,
as we saw in our discussion of the two-particle Casimir $\C_{12}$, the Casimir for two particles coupled to a primary of dimension $h$ is
$h(h-1/2)$, which is invariant under $h\leftrightarrow 1/2-h$.

Moreover, the measure $dy\,d\theta$ is fermionic, so the product $\V\V'$ must be fermionic to make the integral in eqn.~(\ref{double}) bosonic.
Accordingly, there are two cases: $\V$ may be a bosonic primary and $\V'$ a fermionic one, or vice-versa.  

The supersymmetric SYK model  has fermionic primary fields $\uppsi_i(t,\theta)=\psi_i(t)+\theta b_i(t)$.  If the superspace interactions are of degree $\h q$, with $\hat{q}$ an odd integer,
\be\label{ouble} i^{\frac{\hat{q}-1}{2}}\int dt\,d \theta \,J_{i_1i_2\dots i_{\h q}}\uppsi_{i_1}(t,\theta)\dots \uppsi_{i_{\h q}}(t,\theta), \ee
then the $\uppsi$'s have dimension $\Delta=1/2\h q$, again because the measure $dt\,d\theta$ has length dimension $1/2$.
As in the ordinary SYK model, these primary fields can be normalized to have canonical disorder-averaged two-point functions:
\be\label{twopt}\langle \uppsi_i(t_1,\theta_1)\uppsi_j(t_2,\theta_2)\rangle =\delta_{ij}\frac{\sgn(t_1-t_2)}{|t_1-t_2-\theta_1\theta_2|^{2\Delta}}=\delta_{ij}G(1,2).\ee

A bosonic primary field $\V(y,\theta)$ can be normalized so that its three-point function with one of the $\uppsi_i$ (if not zero) is
\be\label{nnouble}\langle \uppsi_i(t_1,\theta_1) \uppsi_i(t_2,\theta_2)\V(t_3,\theta_3)\rangle= \frac{\sgn(t_1-t_2)}{|\langle 1,2\rangle|^{2\Delta-h}
|\langle 1,3\rangle|^h|\langle 2,3\rangle|^h}, \ee
where we abbreviate $t_i-t_j-\theta_i\theta_j$ as $\langle i,j\rangle$.   If $\V(y,\theta)$ is fermionic, the corresponding formula
is 
\be\label{ffnouble}\langle \uppsi_i(t_1,\theta_1) \uppsi_i(t_2,\theta_2)\V(t_3,\theta_3)\rangle= \frac{1}{|\langle 1,2\rangle|^{2\Delta-h}
|\langle 1,3\rangle|^h|\langle 2,3\rangle|^h} P(1,2,3),\ee
with
\be\label{pdef}P(1,2,3)=\frac{\theta_1(t_2-t_3)+\theta_2(t_3-t_1)+\theta_3(t_1-t_2)-2\theta_1\theta_2\theta_3}
{|\langle 1,2\rangle \langle 2,3\rangle \langle 3,1\rangle|^{1/2}}.  \ee

We now want to study a four-point function of the primary fields $\uppsi_i$, normalized as in eqn.~(\ref{normfn}) by dividing by two-point functions:
\be\label{normfun}W(t_1,\theta_1;\dots; t_4,\theta_4)=\frac{\langle\uppsi_i(t_1,\theta_1)\uppsi_i(t_2,\theta_2)\uppsi_j(t_3,\theta_3)\uppsi_j(t_4,\theta_4)\rangle}{\langle\uppsi_i(t_1,\theta_1)\uppsi_i(t_2,\theta_2)\rangle\,\langle\uppsi_j(t_3,\theta_3)\uppsi_j(t_4,\theta_4)\rangle}\ee
We can write down the shadow representation almost as in eqn.~(\ref{gruff}), but now there are two versions depending on whether $\V$ or $\V'$
is fermionic.   If $\V$ is bosonic, the obvious imitation of eqn.~(\ref{normfn}) gives\footnote{A sign factor is present in the numerator here because, as 
one of the three-point functions (namely the fermionic one (\ref{ffnouble})) lacks such a factor, the cancellation mentioned in footnote \ref{trumf} does not occur.}
\be\label{onec}\Upsilon_h^B(t_1,\theta_1;\dots; t_4,\theta_4)=\frac{1}{2}\int dy\,d\theta_y \frac{|\langle 1,2\rangle|^h|\langle 3,4\rangle|^{1/2-h} \sgn(t_3-t_4)P(3,4,y)}{|\langle 1,y\rangle|^h
|\langle 2,y\rangle|^h|\langle 3,y\rangle|^{1/2-h}|\langle 4,y\rangle|^{1/2-h}},\ee
where $\langle i,y\rangle=t_i-y-\theta_i\theta_y$.   
If instead $\V$ is fermionic, we get
\be\label{zonec} \Upsilon_h^F(t_1,\theta_2;\dots;t_4,\theta_4)=\frac{1}{2}\int dy\,d\theta_y \frac{|\langle 1,2\rangle|^h \sgn(t_1-t_2)P(1,2,y)|\langle 3,4\rangle|^{1/2-h}}{|\langle 1,y\rangle|^h
|\langle 2,y\rangle|^h|\langle 3,y\rangle|^{1/2-h}|\langle 4,y\rangle|^{1/2-h}}.\ee

$\Upsilon_h^B$ and $\Upsilon_h^F$ are functions only of the $OSp(1|2)$ invariants $\chi$ and $\zeta$ that were introduced
in eqns.~(\ref{zelf}).  
Just as in the nonsupersymmetric theory, because the primary fields $\uppsi_i$ and $\uppsi_j$ have the same dimension,  the construction has a symmetry
that exchanges $t_1,\theta_1$ and $t_2,\theta_2$ with $t_3,\theta_3$ and $t_4,\theta_4$.  Both $\chi$ and $\zeta$ are  invariant under
this symmetry.
This symmetry exchanges $\V$ and $\V'$, so
it now exchanges $h$ with $1/2-h$.   But as $\V$ and $\V'$ have opposite statistics, the effect of exchanging them is to also exchange the two
shadow constructions.  So the relation $\Psi_h(\chi)=\Psi_{1-h}(\chi)$ of the bosonic theory is replaced by
\be\label{superan} \Upsilon_h^B(\chi,\zeta)=\Upsilon_{1/2-h}^F(\chi,\zeta). \ee
This means that a complete set of states can be constructed just in terms of $\Upsilon^B_h$, but with a larger fundamental domain than one has
in the bosonic theory.  The relation (\ref{superan}) holds likewise if $\Upsilon^B_h$
and $\Upsilon^F_{1/2-h}$ are regarded as functions of the full set of variables:
\be\label{uperan}\Upsilon_h^B(t_1,\theta_1,\dots,t_4,\theta_4)=\Upsilon_{1/2-h}^F(t_1,\theta_1,\dots,t_4,\theta_4). \ee

Now we will express $\Upsilon_h^B$ explicitly in terms of $\chi$ and $\zeta$ and in fact in terms of the bosonic wavefunction $\Psi_h$.
 With the 
coordinates chosen as in eqn.~(\ref{zelof}), $P_{34y}$ reduces to $-\theta_y/|1-y|^{1/2}$.  Because this gives an explicit factor of $\theta_y$ in the numerator of the shadow integral (\ref{onec}), we can set $\theta_y$ to 0 in the
denominator, so that $\langle i,y\rangle$ reduces to $t_i-y$. We further have $t_2=\chi$, $\zeta=\theta_1\theta_2$,
 $\langle 1,2\rangle =-t_2-\theta_1\theta_2=-(\chi +\zeta)$.  The shadow representation becomes
\be\label{shadowb}\Upsilon^B_h(\chi,\zeta)=\frac{1}{2}\int dy\,d\theta_y  \frac{|\chi+\zeta|^h\,\theta_y}{|y|^h|\chi-y|^h|1-y|^{1-h}}.\ee
Integrating over $\theta_y$ and expanding in powers of $\zeta$,
we get
\be\label{shadowf}\Upsilon^B_h(\chi,\zeta)=\frac{1}{2}\int dy\, \frac{|\chi|^h}{|y|^h|\chi-y|^h|1-y|^{1-h}}\left(1+\frac{h\zeta}{\chi}\right)
=\left(1+\frac{h\zeta}{\chi}\right)\Psi_h(\chi).\ee 

As we have already explained in section \ref{Solving}, this function $\Upsilon^B_h(\chi,\zeta)$ is
 an eigenfunction of the Casimir operator for $OSp(1|2)$, with eigenvalue
$h(h-1/2)$.  As this is invariant under $h\to 1/2-h$, the function $\Upsilon^B_{1/2-h}=\Upsilon^F_h$ is another
eigenfunction with the same eigenvalue.  These are the eigenfunctions that possess the discrete symmetry $1\leftrightarrow 2$
(or $3\leftrightarrow 4$) which acts on $\chi$ and $\zeta$ as 
in eqn.~(\ref{nchisym}).  This is a manifest symmetry of the shadow integral (\ref{zonec}).  The Casimir equation also has two more
eigenfunctions (arising from different choices of the constants in eqn.~(\ref{zumbo})) that are odd under the discrete symmetry.

\subsection{Inner Products}\label{innerprods}

The natural inner product for understanding the four-point function of the ordinary SYK model is\cite{Maldacena:2016hyu}
\be\label{innerp}\bigl(f_1(\chi),f_2(\chi)\bigr)=\int_0^2\frac{d\chi}{\chi^2} f_1(\chi)f_2(\chi).\ee
This formula was explained in the derivation of eqn.~(\ref{invone}) above.\footnote{In \cite{Maldacena:2016hyu}
and also in eqn.~(\ref{invone}) above, $f_1$ is
complex-conjugated in this formula to make a hermitian, rather than bilinear, inner product.  Here we will omit this because,
as explained in section \ref{measure}, in the supersymmetric case, there is little benefit in defining a hermitian rather than bilinear
inner product.
  At any rate the following arguments can be expressed in either language.  Since the states
$\Psi_{2n}$ and $\Psi_{1/2+is}$ that appear in the completeness relation of the bosonic theory are all real, their inner products are not affected by complex-conjugating
one factor.} 

A set of eigenstates of the two-particle Casimir that satisfy a completeness relation for this inner product was described in \cite{Maldacena:2016hyu}.
Because the Casimir is hermitian, there is a complete set of states for which its eigenvalue $h(h-1)$ is real, meaning that $h$ is real or is of the form $1/2+is$ with
real $s$.  In fact, a complete set of states is given by 
 the discrete states $\Psi_{2n}$, with $n$ a positive integer, and the continuum states $\Psi_{1/2+is}$ with $s$ real and positive.
(In view of the relation $\Psi_h=\Psi_{1-h}$, it would be equivalent to consider $\Psi_{1-2n}$ instead of $\Psi_{2n}$ or $\Psi_{1/2-is}$ instead
of $\Psi_{1/2+is}.$)  

The natural inner product of the supersymmetric theory was similarly described in eqn.~(\ref{iluzem}):
\be\label{innprod}\big\langle F(\chi,\zeta),G(\chi,\zeta)\big\rangle=-\int_0^2\frac{d\chi\,d\zeta}{\chi+\zeta}F(\chi,\zeta)G(\chi,\zeta). \ee

We can easily compute inner products of $\Upsilon^B_h$ in terms of
those of $\Psi_h$.   Using the relation $\Upsilon^B_h(\chi,\zeta)=(1+h\zeta/\chi)\Psi_h(\chi)$ and comparing the definitions (\ref{innprod})
and (\ref{innerp}) of the inner products, we get
\be\label{superprod}\langle \Upsilon^B_h,\Upsilon^B_{h'}\rangle =(1-h-h')\bigl(\Psi_h,\Psi_{h'}\bigr).  \ee
To be more exact, this formula is true when the wavefunctions behave well enough near $\chi=0$ that both sides are defined, that is, if
$h=2n$ or $1/2+is$, or the image of one of these under $h\to 1-h$.  

The right hand side of (\ref{superprod}) was computed in \cite{Maldacena:2016hyu}.  For $h=1/2+is$, $h'=1/2+is'$, the result (eqn.~(3.78)
of that paper) is
\be\label{bosprod} \bigl(\Psi_h,\Psi_{h'}\bigr) = \frac{\pi \tan\pi h}{4h-2}2\pi\left(\delta(s-s')+\delta(s+s')\right). \ee
The right
hand side of (\ref{bosprod}) is symmetric in $h$ and $h'$ because the function $\tan(\pi h)/(4h-2)$ is invariant
under $h\to 1-h$.    We have included  a term $\delta(s+s')$ so that the formula holds for either sign of $s$ and $s'$. 
However, when (\ref{bosprod}) is used in (\ref{superprod}), the $\delta(s+s')$ term does not contribute because $h+h'-1=0$ on the support
of this delta function.  Hence we get 
\be\label{superprodb}\langle \Upsilon^B_h,\Upsilon^B_{h'}\rangle=-2\pi\delta(s-s')\frac{\pi\tan\pi h}{2}. \ee
Similarly, according to eqn.~(3.79) of \cite{Maldacena:2016hyu}, for the discrete states $h=2n$, $h'=2n'$, one has
\be\label{discprod}\bigl(\Psi_{2n},\Psi_{2n'}\bigr) = \frac{\delta_{nn'}\pi^2}{4h-2}. \ee
Hence
\be\label{superdisc} \langle \Upsilon^B_{2n},\Upsilon^B_{2n'}\rangle=-\frac{\delta_{nn'}\pi^2}{2}. \ee
However, in the case of the supersymmetric theory, in addition to the discrete states $\Upsilon^B_{2n}$, we have to consider
discrete states $\Upsilon^B_{1-2n}$; the state $\Upsilon^B_{1-2n}$ is different from $\Upsilon^B_{2n}$ (and even has a different value of the
two-particle Casimir $\C_{12}$) but  behaves for $\chi\to 0$ similarly to $\Upsilon^B_{2n}$ because of the relation (\ref{shadowf})
between $\Upsilon^B_h$ and $\Psi_h$.
Using (\ref{superprod}) and the fact that $\Psi_{2n}=\Psi_{1-2n}$, we have
\be\label{nosprod} \langle \Upsilon^B_{1-2n},\Upsilon^B_{1-2n'}\rangle = (1-h-h') \bigl(\Psi_{1-2n},\Psi_{1-2n'}\bigr) =(1-h-h')\bigl(\Psi_{2n},\Psi_{2n'}\bigr)
=\frac{\delta_{nn'}\pi^2}{2}. \ee
The last such relation among the discrete states is
\be\label{osprod}\langle \Upsilon^B_{2n},\Upsilon^B_{1-2n'}\rangle=0, \ee
where the vanishing results from the factor $h+h'-1$ in (\ref{superprod}).

Finally, in either the bosonic theory or the supersymmetric theory, the inner product between a continuum state and a discrete state vanishes.
This actually follows from the fact that the two types of state have different values of the Casimir $\C_{12}$.  

The precise normalization of the discrete state wavefunctions in (\ref{discprod}) played an important role in the derivation of the operator product
expansion in \cite{Maldacena:2016hyu}, leading to a cancellation between discrete state contributions and poles associated to the continuous
spectrum.  Something similar happens in the supersymmetric model, as we will see below.  This phenomenon can be understood in terms of general
facts about Schrodinger-like operators, as is explained in Appendix \ref{normo}.

\subsection{A Complete Set Of States}\label{completeset}

In the nonsupersymmetric theory, since the $\Psi_h$ for $h=2n$ or $h=1/2+is$ are the normalizable or continuum normalizable eigenstates of the hermitian
operator $C_{12}$, they must on general grounds give a basis for the full Hilbert space.  In the supersymmetric theory, we cannot make a similar
argument because $\C_{12}$ is hermitian with respect to an indefinite inner product.  But from the fact that the $\Psi_h$ are a basis
for the bosonic Hilbert space, it follows that the functions $\Upsilon^B_h=(1+h\zeta /\chi)\Psi_h$ are a basis for the space of functions $F(\chi,\zeta)$. 
Indeed, the fact that any function of $\chi$ can be expressed as a linear combination of the $\Psi_h$ implies that any $F(\chi,\zeta)=f(\chi)+\zeta
g(\chi)$ is a linear combination of $\Psi_h(\chi)$ and $(\zeta/\chi)\Psi_h(\chi)$; but $\Psi_h(\chi)$ and $(\zeta/\chi)\Psi_h(\chi)$ can each
be expressed as a linear combination of $\Upsilon^B_h$ and $\Upsilon^B_{1-h}$.

By borrowing formulas from the bosonic theory, we can be more precise about how to express a given function $F$ in terms of the $\Upsilon_h^B$.
The completeness relation for the bosonic theory reads
\be\label{combos}\int_0^\infty \frac{ds}{2\pi} \frac{4h-2}{\pi\tan\pi h} \Psi_h(\chi)\Psi_{h}(\chi')+\sum_{n=1}^\infty\frac{4h-2}{\pi^2}  \Psi_h(\chi)\Psi_h(\chi')
=\chi^2\delta(\chi-\chi')=(\chi')^2\delta(\chi-\chi'). \ee (It is understood here that in the integral $h=1/2+is$, and in the sum $h=2n$.)
The consistency of this with the formulas for the inner products is as follows.  Start with
\be\label{ombos}  \Psi_{h'}(\chi)=\int_0^2{d \chi'}\delta(\chi-\chi') \Psi_{h'}(\chi') \ee
where $h'$ is of the form $1/2+is$ or $2n$.
Using eqn.~(\ref{combos}) to express $\delta(\chi-\chi')$ as a sum over states, we get
\begin{align}\label{wombos}\Psi_{h'}(\chi)=&\int_0^2\frac{d\chi'}{(\chi')^2} \Bigl(\int_0^\infty \frac{ds}{2\pi} \frac{4h-2}{\pi \tan\pi h}\Psi_{h}(\chi)\Psi_h(\chi')
                    +\sum_{n=1}^\infty \frac{4h-2}{\pi^2}\Psi_h(\chi)\Psi_h(\chi')\Bigr) \Psi_{h'}(\chi')\cr
                        =& \int_0^\infty \frac{ds}{2\pi} \frac{4h-2}{\pi \tan\pi h}\Psi_h(\chi) \left(\Psi_h,\Psi_{h'}\right)
                        +\sum_{n=1}^\infty \frac{4h-2}{\pi^2}\Psi_h(\chi)\left(\Psi_h,\Psi_{h'}\right).   \end{align}
This can be confirmed using eqns.  (\ref{bosprod}) and (\ref{discprod}) for the inner products.     This verifies that eqn.~(\ref{combos}) is
the correct form of the completeness relation.        Since this relation holds for any basis function
$\Psi_{h'}$ of the Hilbert space, it actually holds for any function $F(\chi)$:
\be\label{eigendecomp}F(\chi)=\int_0^\infty \frac{ds}{2\pi} \frac{4h-2}{\pi \tan\pi h}\Psi_h(\chi) \bigl(\Psi_h,F\bigr)
                        +\sum_{n=1}^\infty \frac{4h-2}{\pi^2} \Psi_h(\chi)\bigl(\Psi_h,F\bigr). \ee
                        
In the analogous completeness relation in the supersymmetric theory, since there is no symmetry under $h\to 1-h$,
we have to integrate over the whole real $s$ axis, and we have  to sum over discrete states at $h=1-2n$ as well as $h=2n$.
The completeness relation of the supersymmetric theory is\footnote{On the right hand side of this formula,
 $\zeta+\zeta'$ plays the role of $\delta(\zeta-\zeta')$,
since $\int d\zeta' (\zeta+\zeta') f(\zeta')=f(\zeta)$ for any function $f$.}
\begin{align}\label{superrel}-\int_{-\infty}^\infty\frac{ ds}{\pi^2\tan\pi h} \Upsilon_h^B(\chi,\zeta)\Upsilon_h^B(\chi',\zeta')-
&\sum_{n=1}^\infty \frac{2}{\pi^2}\Bigl(\Upsilon^B_{2n}(\chi,\zeta)\Upsilon^B_{2n}(\chi',\zeta')-\Upsilon^B_{1-2n}(\chi,\zeta)\Upsilon^B_{1-2n}(\chi',
\zeta')\Bigr)\notag \\ &=-(\chi+\zeta)(\zeta+\zeta')\delta(\chi-\chi'). \end{align}
To verify this relation is a simple matter of expanding in powers of $\zeta$ and $\zeta'$ and using the bosonic relation (\ref{combos}).
For example, if we set $\zeta=\zeta'=0$, then $\Upsilon^B_h$ becomes $\Psi_h$, with the familiar  $h\to 1-h$ symmetry.  As a result,
the integral on the left hand side of (\ref{superrel}) vanishes because the integrand is odd under $s\to -s$, and similarly in the sum over discrete
states, the contributions at $h=2n$ and $h=1-2n$ cancel.  On the other hand, the right hand side of eqn.~(\ref{superrel}) trivially vanishes if
$\zeta=\zeta'=0$.  Suppose instead that we set $\zeta'=0$ and look at the term in the equation linear in $\zeta$.  Then we can replace 
$\Upsilon_h^B(\chi,\zeta)\Upsilon_h^B(\chi',\zeta')$ by $(h/\chi)\Psi_h(\chi)\Psi_h(\chi')$.  Here, using the symmetry properties of the integral
and the sum under $s\to 1-s$, we can restrict the integral over $s$ to the half-line $s\geq 0$, and we can consider only the states at $h=2n$
in the sum, if we also replace $h$ by $h - (1-h)=2h-1$.  The desired identity then reduces to the bosonic formula (\ref{combos}).  The term
linear in $\zeta'$ can be treated the same way.  Finally,
to  verify the term in eqn.~(\ref{superrel}) that is proportional to $\zeta\zeta'$, we argue similarly using the identity $h^2-(1-h)^2
= 2h-1$.  

The analog of  the bosonic identity (\ref{eigendecomp}) is the statement  that for any function $F(\chi,\zeta)$, we have 
\begin{align}\label{zeigendecomp}F(\chi,\zeta)=-\int_{-\infty}^\infty &\frac{ds}{\pi^2 \tan\pi h}\Upsilon^B_h(\chi,\zeta) \langle\Upsilon^B_h,F\rangle\\
                        &-\sum_{n=1}^\infty \frac{2}{\pi^2} \left(\Upsilon^B_h(\chi,\zeta)\langle\Upsilon^B_h,F\rangle- \Upsilon^B_{1-2h}(\chi,\zeta)
                        \langle\Upsilon^B_{1-2h},F\rangle\right). \notag\end{align}
The consistency of this with the formulas of section \ref{innerprods} for the inner products can be verified as in the bosonic case.                        

\subsection{The Kernel}\label{kkernel}

\begin{figure}[ht]
 \begin{center}
   \includegraphics[width=2.7in]{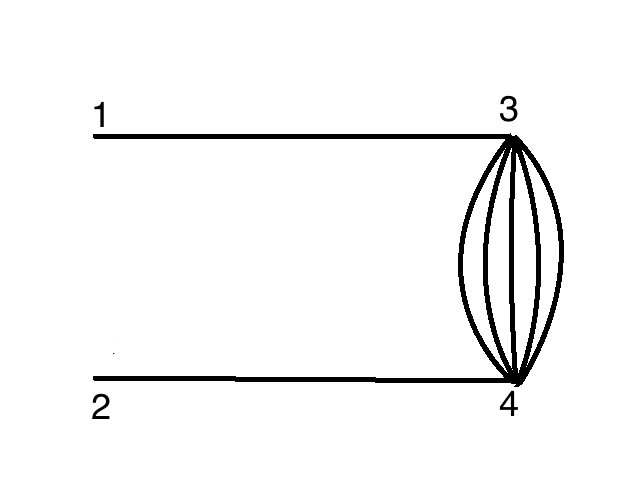}
 \end{center}
\caption{\small A kernel describing propagation  of a two-particle system from $t_3,\theta_3$ and $t_4,\theta_4$ to $t_1,\theta_1$ and $t_2,\theta_2$. 
For the case of a $\h q$-fold interaction in superspace, points 3 and 4 are connected by $\h q-2$ propagators, as here for the case $\h q=7$.  \label{kernel}}
\end{figure}

To follow the procedure of \cite{KitaevTalks,Maldacena:2016hyu} to evaluate the four-point function of the supersymmetric SYK model,
we need to compute the eigenvalue of a certain ladder kernel (fig. \ref{kernel}) that governs propagation in a two-particle channel.\footnote{In  \cite{Maldacena:2016hyu},
a not necessarily conformally-invariant Euclidean kernel $K$ was studied, and its conformally-invariant low energy limit was denoted $K_c$.   In the present
paper, in studying the Euclidean kernel, we are always in the conformal limit, and we omit the subscript $c$ for the kernel $K$ and its eigenvalue $k$.} 
The relevant kernels have already been computed in \cite{Fu:2016vas}, but here we will describe the computation in a supersymmetric language.

 For the case
of $\h q$-fold interactions, with effective coupling $j$, adapting eqn.~(3.44) of \cite{Maldacena:2016hyu}, the kernel is\footnote{It does not make sense
to specify the overall sign of $K$ without also specifying the sign of the integration measure $d\theta_3d\theta_4$ in eqn.~(\ref{theprod}) that will
be used when we define the action of $K$ on a wavefunction.  We choose the measure such that $\int d\theta_3d\theta_4\theta_4\theta_3=1$.}
\be\label{kernis}K(t_1,\theta_1,t_2,\theta_2;t_3,\theta_3,t_4,\theta_4)=(\h q-1)j^2 G(1,3)G(2,4)G(3,4)^{\h q-2},\ee
where the propagator is
\be\label{propa}G(t_i,\theta_i,t_j,\theta_j)=\frac{b_\psi\,\sgn(t_i-t_j)}{|\langle i,j\rangle|^{2\Delta}},~~\Delta=1/2\h q. \ee
According to  eqn.~(2.29) of \cite{Fu:2016vas} (where our $j^2$ is denoted $J$)
\be\label{kertwo}j^2b_\psi^{\h q}=\frac{\tan\frac{\pi}{2\h q}}{2\pi},\ee  
so that\footnote{Since $\h q$ is always odd in the supersymmetric SYK model and there are $\h q-2$ propagators
connecting points 3 and 4 in fig. \ref{kernel}, there is an odd power of $\sgn(t_3-t_4)$ in the numerator of $K$.}
\be\label{kernell} K(t_1,\theta_1,t_2,\theta_2;t_3,\theta_3,t_4,\theta_4)=(\h q-1) \frac{\tan\frac{\pi}{2\h q}}{2\pi}\frac{\sgn(t_1-t_3)\sgn(t_2-t_4)\sgn(t_3-t_4)}
{|\langle 1,3\rangle|^{2\Delta}|\langle 2,4\rangle|^{2\Delta}|\langle 3,4\rangle|^{2\Delta (\h q-2)}},\ee
where we recall that $\Delta=1/2\h q$.
We view this kernel as an operator that maps functions of $t_3,\theta_3$ and $t_4,\theta_4$ to functions of $t_1,\theta_1$ and $t_2,\theta_2$.
By superconformal symmetry, its eigenfunctions are the eigenfunctions of the two-particle Casimir $\C_{12}$ or $\C_{34}$.
For the case that identical fermionic primaries are inserted at 1 and 2 (or at 3 and 4), 
there are two kinds of eigenfunction depending on whether the two operators
fuse to a bosonic primary or a descendant of a fermionic one.  The two types of eigenfunction were already 
described in eqns.~(\ref{nnouble}) and (\ref{ffnouble}), where $t_3,\theta_3$ is arbitrary (the choice of this point does not
affect the eigenvalue of the Casimir).  Taking $t_3\to \infty$ and relabeling particles 1,2 as 3,4 (since we want to think of $K$
as an operator acting on particles 3,4) the eigenfunctions corresponding to a bosonic primary of dimension $h$ are
\begin{equation}\label{bospri}S^B_h(t_3,\theta_3,t_4,\theta_4)=\frac{\sgn(t_3-t_4)}{|\langle 3,4\rangle|^{2\Delta-h}} \ee
while those corresponding to a fermionic primary of dimension $h$ are 
\begin{equation}\label{ferpri}S^F_h(t_3,\theta_3,t_4,\theta_4)=\frac{\theta_3-\theta_4 }{|\langle 3,4\rangle|^{2\Delta-h+1/2}}.\ee
 
To evaluate the eigenvalue $k^B(h)$ with which $K$ acts on $S^B_h$, we evaluate the integral
\be\label{theprod}\int dt_3 dt_4 d\theta_3d\theta_4\,K(t_1,\dots,\theta_4)S_h^B(t_3,\dots,\theta_4).\ee
  The result, by superconformal symmetry, will be a multiple of $S_h^B(t_1,\theta_1,t_2,\theta_2)$.
The coefficient is by definition $k^B(h)$.  Since $S_h^B=1$ at $t_1,\theta_1=1,0$ and $t_2,\theta_2=0,0$, we can compute $k^B(h)$ by simply
setting $t_1,\dots,\theta_2$ to those values and integrating over $t_3,\dots,\theta_4$:
\be\label{kcb} k^B(h)= (\h q-1) \frac{\tan\frac{\pi}{2\h q}}{2\pi}\int_{-\infty}^\infty \!\!\!\!dt_3 dt_4 d\theta_3d\theta_4 \,\,\frac{\sgn(1-t_3)\sgn(-t_4)}{|1-t_3|^{2\Delta}
|t_4|^{2\Delta} |t_3-t_4-\theta_3\theta_4|^{1-2\Delta-h}}.\ee
Integrating over $\theta_3$ and $\theta_4$ gives
\be\label{kcbtwo} k^B(h)=-(\h q-1)(1-2\Delta-h)
 \frac{\tan\frac{\pi}{2\h q}}{2\pi} \int_{-\infty}^\infty \!\!\!\!dt_3 dt_4  \,\,\frac{\sgn(1-t_3)\sgn(-t_4)\sgn(t_3-t_4)}{|1-t_3|^{2\Delta}
|t_4|^{2\Delta} |t_3-t_4|^{2-2\Delta-h}}.\ee
To perform the integrals,\footnote{In fact, eqn.~(\ref{kcbtwo}) coincides apart from a prefactor with eqn.~(3.70) of \cite{Maldacena:2016hyu}, so the integral
can also be performed by the procedure described there.} set $t_3=x$, $t_4=xy$, to get
\be\label{kcbthree}k^B(h)=(\h q-1)(1-2\Delta-h)
 \frac{\tan\frac{\pi}{2\h q}}{2\pi}  I_x(h)I_y(h), \ee
 with
 \begin{align} \label{kcbfour} I_x(h)&= \int_{-\infty}^\infty \!\!\!\!dx \,\,\frac{\sgn(1-x)}{|x|^{1-h}|1-x|^{2\Delta}}\cr
    I_y(h)&=\int_{-\infty}^\infty \!\!\!\!dy \,\,\frac{\sgn(y(1-y))}{|y|^{2\Delta}|1-y|^{2-2\Delta-h}}. \end{align}
  Like the integral in eqn. 
    (\ref{nefl}), $I_x$ is the sum of three beta function integrals.   Evaluating them and using some standard identities, one finds that
    \be\label{evalint}I_x=\frac{1}{\pi}\left(-\sin((1-h)\pi)+\sin 2\Delta\pi +\sin((1+h-2\Delta)\pi)\right)\Gamma(h)\Gamma(1-2\Delta)\Gamma(-h+2\Delta). \ee
    An $SL(2,\R)$ transformation that permutes the three points $0,1,\infty$ can be used to show that $I_y(h)=-I_x(1-h)$.
    Combining these facts and simplifying the result with the help of standard identities, one finds 
    \be\label{kcbresult}k^B(h)=-(\h q-1) \frac{\sin2\pi\Delta-\sin\pi h}{\sin 2\pi\Delta} \frac{\Gamma(-h+2\Delta)\Gamma(h+2\Delta)}{\Gamma^2(2\Delta)}.\ee

 As a check on this formula, we find that $k^B(0)=-(\h q-1)$, as predicted by an argument   described in section 3.2.3 of \cite{Maldacena:2016hyu}.
 (This argument will be explained in section \ref{ladderk}.)  Another check is that $k^B(1)=1$.   This reflects the existence of a supersymmetry-violating
 deformation of the solution of the Schwinger-Dyson equation for the two-point function in the infrared limit.  This deformation was described in \cite{Fu:2016vas} and we will
 return to it in section \ref{enhance}.
  
To compute the eigenvalue $k^F(h)$ of the kernel $K$ acting on $S_h^F$, we consider the integral
\be\label{fermiprod}\int dt_3 dt_4 d\theta_3 d\theta_4 \,\,K(t_1,\dots,\theta_4)S_h^F(t_3,\dots,\theta_4) \ee
and integrate over $t_3,\dots,\theta_4$.  The result will be a multiple -- namely  $k^F(h)$ -- of $S_h^F(t_1,\dots,\theta_2)$.  We cannot evaluate $k^F(h)$ by setting $\theta_1=\theta_2=0$,
because then $S_h^F(t_1,\dots,\theta_2)=0$.  However, we can evaluate $k_h^F$ by setting $t_1=1$, $t_2=\theta_2=0$ in the integral (\ref{fermiprod}).  Since this
sets $S_h^F=\theta_1$, the integral in (\ref{fermiprod}) for these choices of $t_1,t_2$, and $\theta_2$ will equal $\theta_1 k^F$.  
Integrating over $\theta_3$ and $\theta_4$ in eqn.~(\ref{fermiprod}) and extracting the coefficient of $\theta_1$, we arrive at
\be\label{fermiintegral}k^F(h)= (\h q-1) \frac{\tan\frac{\pi}{2\h q}}{2\pi}2\Delta\int_{-\infty}^\infty \!\!\!\!dt_3 dt_4 \,\,\frac{\sgn(-t_4)\sgn(t_3-t_4)}{|1-t_3|^{2\Delta+1}|t_4|^{2\Delta}|t_3-t_4|^{3/2-2\Delta-h}}.\ee
This integral can be evaluated via the same steps as before, with the simple result
\be\label{kcfresult}k^F(h)= k^B(1/2-h). \ee
As an important example, this implies that $k^F(3/2)=k^B(-1)=1$, corresponding to the existence of a superconformal primary that is a fermionic operator with $h=3/2$.
The top component of this multiplet is the bosonic operator of dimension 2 that is related to the chaos exponent.  

The relation $k^B(h)=k^F(1/2-h)$ has a simple explanation by thinking of the kernel $K$ acting on $\Upsilon^B_h$ and $\Upsilon^F_{1/2-h}$. For this purpose we consider $\Upsilon^B_h$
and $\Upsilon^F_{1/2-h}$ as  
functions not of the
invariants $\chi$ and $\zeta$ but of the full set of variables $t_1,\theta_1,\dots,t_4,\theta_4$, and we consider $K$ to act on each function on the 12 variables.
Then  $K$ acts on $\Upsilon^B_h$ with eigenvalue $k^B(h)$, and it acts on $\Upsilon^F_h$ with eigenvalue $k^F(h)$.  (This follows from the shadow representation,
which exhibits $\Upsilon^B_h$ and $\Upsilon^F_h$, in their dependence on the 12 variables, as continuous integrals of the conformal wavefunctions $S^B_h$ and $S^F_h$,
which are eigenfunctions of the kernels.) But $\Upsilon_{h}^B=\Upsilon^F_{1/2-h}$ so
$k^B(h)=k^F(1/2-h)$.    The same argument in the bosonic SYK model, using the relation $\Psi_h=\Psi_{1-h}$, implies that $k(h)=k(1-h)$.

\begin{figure}[ht]
 \begin{center}
   \includegraphics[width=2.7in]{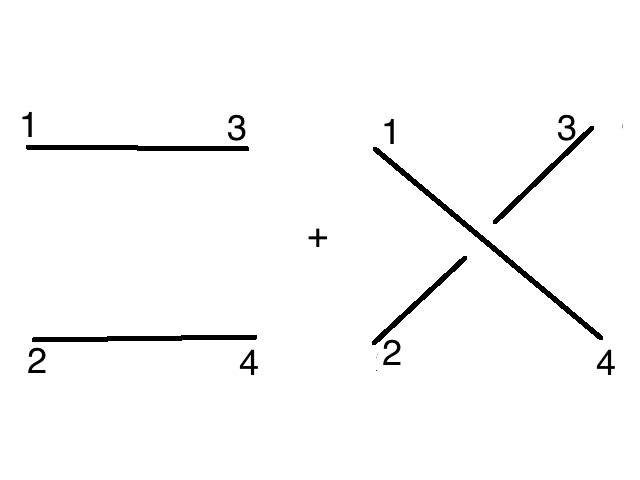}
 \end{center}
\caption{\small The lowest order contribution to a four-point function $\langle 1234\rangle $  comes from these ``zero-rung'' diagrams (a third diagram is missing because as
usual we consider a four-point function with the disconnected contribution $\langle 12\rangle \langle 34\rangle$ subtracted out).  The lines represent exact two-point functions. \label{Zero Rung}}
\end{figure}

Let us now use this to sum the ladder diagrams for the model, following the nonsupersymmetric case \cite{kitaevfirsttalk,Maldacena:2016hyu}. The ``zero-rung'' contribution $\F_0$ to the four-point function $\F$ comes from the diagrams of fig. \ref{Zero Rung}.  Recalling that we normalize the four-point function by
dividing by $G(1,2)G(3,4)$, we have
\be\label{zerorung}\F_0=\frac{-G(1,3)G(2,4)+G(1,4)G(2,3)}{G(1,2)G(3,4)}. \ee
The full normalized  four-point function is
\be\label{fullw}\F=\frac{1}{1-K}\F_0.\ee
To calculate $\F$ by means of an expansion in eigenfunctions of the Casimir, we need to know the inner products
\be\label{ullw}\langle \Upsilon_h^B,\F\rangle = \langle\Upsilon_h^B,\frac{1}{1-K} \F_0\rangle =\frac{1}{1-k^B(h)}\langle \Upsilon_h^B ,\F_0\rangle,\ee
where in the last step we replaced $K$ by its eigenvalue acting on $\Upsilon_h^B$.

In the analysis in \cite{Maldacena:2016hyu} of the ordinary (nonsupersymmetric) SYK model, the inner product $\bigl(\Psi_h,\F_0\bigr)$ was needed for just this reason.  In this model, $\F_0$
is given by the same formula as in eqn.~(\ref{zerorung}), with  now 
\be\label{reff}G(i,j)=\frac{b\, \sgn(t_i-t_j)}{|t_i-t_j|^{2\Delta}},~~~\Delta=1/q\ee
the two-point function of the ordinary SYK model.  
It was observed that the inner product $\left( \Psi_h,\F_0\right)$ is actually a simple (and $h$-independent) multiple of the kernel function $k(h)$, which is defined
as the eigenvalue of the kernel
\be\label{treff} K(t_1,t_2,t_3,t_4) =-J^2(q-1)G(1,3)G(2,4)G(3,4)^{q-2},\ee
acting on the conformal wavefunction 
\be\label{preff}\frac{\sgn(t_3-t_4)}{|t_3-t_4|^{2\Delta-h}}.\ee
This relation can be understood as follows.  As a preliminary simplification, observe that 
the two terms in the numerator of eqn.~(\ref{zerorung}) are
actually exchanged by the discrete symmetry $\chi\to \chi/(\chi-1)$, so if we ignore this discrete symmetry (and integrate over the whole real $\chi$ axis in computing an inner
product involving $\F_0$) we can replace $\F_0$ by $\F_0'=-G(1,3)G(2,4)/G(1,2)G(3,4)$.   Now consider formally the integral
\be\label{pelican}\frac{1}{2\, \mathrm{vol}(SL(2,\R))}\int_{-\infty}^\infty\frac{dt_1dt_2}{(t_1-t_2)^2}\frac{dt_3dt_4}{(t_3-t_4)^2}   dy
\frac{|t_1-t_2|^h|t_3-t_4|^{1-h}}{ |t_1-y|^h|t_2-y|^h|t_3-y|^{1-h}|t_4-y|^{1-h}}  \F_0'. \ee
This is a formal expression: the integral is badly divergent because the integrand  is $SL(2,\R)$-invariant, and of course the volume of $SL(2,\R)$ is also infinite.  As is familiar in the
context of perturbative string theory, one can get a well-defined integral by fixing any three of the five integration variables $t_1,t_2,t_3,t_4,y$ to chosen values $a,b,c$, including also a factor
$|a-b||b-c||c-a|$, and throwing away the prefactor $1/\mathrm{vol}(SL(2,\R)$.    The resulting integral does not depend on the choices that were made.    If we set $t_1=0$, $t_3=1$,
$t_4=\infty$, then $t_2$ can be identified with $\chi$ and the integral over $y$ is the shadow representation of $\Psi_h(\chi)$.  The integral over $\chi$ is then
\be\label{elican}\int_{-\infty}^\infty\frac{d\chi}{\chi^2}\Psi_h(\chi)\F_0' = \bigl(\Psi_h,\F_0\bigr). \ee
An alternative ``gauge-fixing'' is to set $y=\infty$, $t_1=1$, $t_2=0$.   In this case, the integral gives a simple multiple of $k(h)$.  To see this, observe that the factor $1/(t_3-t_4)^2$
in the integrand is $b^{-q}G(3,4)^q$ ($q$ is even in the SYK model so there is no sign factor here).   Also, with this second gauge-fixing, $G(1,2)=b$.  Using these facts and the definitions
of $K$ and $\F_0'$, we find that with this  gauge-fixing, the integral is
\be\label{lican}\frac{1}{2J^2b^q(q-1)}\int_{-\infty}^\infty \!\!\!\!dt_3dt_4 \,\,K(1,0,t_3,t_4)\frac{|t_3-t_4|^{1-h}\sgn(t_3-t_4)}{|t_3-t_4|^{2\Delta}}= \frac{k(1-h)}{2J^2b^q(q-1)}.\ee
Comparing the two calculations and recalling that $k(h)=k(1-h)$, we learn that
\begin{align}\label{twofacts} \bigl(\Psi_h,\F_0\bigr) & = \frac{k(h)}{2J^2b^q(q-1)}=k(h)\frac{\alpha_0}{2},\end{align}
where \be\label{useful}\alpha_0=\frac{1}{J^2b^q(q-1)}=\frac{2\pi q}{(q-1)(q-2)\tan(\pi/q)} \ee
is defined in \cite{Maldacena:2016hyu}.

A similar argument for the supersymmetric SYK model can be modeled on the shadow representation of $\Upsilon_h^F$.  With this in mind, we formally consider the $OSp(1|2)$-invariant
integral
\be\label{tomix}\frac{1}{2}\int_{-\infty}^\infty \frac{dt_1 d\theta_1 dt_2d\theta_2}{t_1-t_2-\theta_1\theta_2} \frac{dt_3 d\theta_3 dt_4d\theta_4}{t_3-t_4-\theta_3\theta_4} dy\,d\theta_y \frac{|\langle 1,2\rangle|^h \sgn(t_1-t_2)P(1,2,y)|\langle 3,4\rangle|^{1/2-h}}{|\langle 1,y\rangle|^h
|\langle 2,y\rangle|^h|\langle 3,y\rangle|^{1/2-h}|\langle 4,y\rangle|^{1/2-h}}\F_0'.\ee
We would like to divide by the divergent volume of $OSp(1|2)$ by fixing some integration variables. Let $t$, $t'$, $t''$ be any three of $t_1,\dots,t_4,y$, and let $\theta,\theta'$ be the fermionic partners of $t,t'$.  The $OSp(1|2)$ symmetry can be fixed by setting $t,t',t''$ to any values
$a,b,c$, setting $\theta=\theta'=0$, and including a factor\footnote{As a quick way to understand this factor, observe that the factor $dt d\theta dt' d\theta' dt''$, which is being removed
from the measure, scales under conformal transformation with weight $1/2$ in the $t$ and $t'$ variables but weight 1 in $t''$.  So $c$ appears twice in $|(a-c)(b-c)|$ but $a$ and $b$ only
once each. The correct sign depends on the ordering within the measure of the $\theta$ variable that are fixed, but for our purposes, it will be $\text{sgn}(a-b)$.} $\pm|(a-c)(b-c)|$. If we choose $t_1=0$, $t_3=1$, $t_4=\infty$, $\theta_3=\theta_4=0$, then the integral over $y,\theta_y$ gives the shadow representation of $\Upsilon^F_h$, and the remaining
integral over $t_2,\theta_1$, and $\theta_2$ computes the inner product $\langle \Upsilon^F_h,\F_0\rangle$.  Alternatively, we can pick $t_1=1$, $t_2=0$, $y=\infty$,
and $\theta_1=\theta_2=0$.  After integrating over $\theta_y$, we are left with
\be\label{omix}\frac{1}{2}\int \frac{dt_3 d\theta_3 dt_4 d\theta_4}{\langle 3,4\rangle} |\langle 3,4\rangle|^{1/2-h} \frac{G(1,3)G(2,4)}{G(1,2)G(3,4)} \ee
Starting with the fact that $1/\langle 3,4\rangle=b_\psi^{-\h q} G(3,4)^{\h q}$, all the previous steps can be repeated to show that this integral is
$\frac{1}{2 j^2 b_\psi^{\h q}(\h q-1)}k^B(1/2-h)$.   Since $k^B(1/2-h)=k^F(h)$, the comparison of these ways to do the integral gives
\be\label{mix}\langle \Upsilon_h^F,\F_0\rangle =\frac{k^F(h) }{2j^2 b_\psi^{\h q}(\h q-1)}=\frac{\h\alpha_0}{2} k^F(h), \hspace{20pt} \h\alpha_0 \equiv \frac{2\pi}{(\h q-1)\tan\frac{\pi}{2\h q}}.\ee
Of course, replacing $h$ by $1/2-h$, the same relation holds between $\langle\Upsilon_h^B,\F_0\rangle$ and $k^B(h)$.

\subsection{The Operator Product Expansion Of The Supersymmetric SYK Model}\label{opesuper}

We can now imitate the derivation in \cite{Maldacena:2016hyu} of the operator product expansion of the bosonic SYK model.   We use the decomposition (\ref{zeigendecomp})
of the four-point function $\F(\chi,\zeta)$ and the formula of eqn.~(\ref{ullw})  for the inner product $\langle \Upsilon_n^B, \F\rangle$:
\begin{align}\label{longfor}\frac{\F(\chi,\zeta)}{\h\alpha_0}=-\int_{-\infty}^\infty &\frac{d s }{2\pi^2\tan \pi h} \Upsilon_h^B(\chi,\zeta)\frac{k^B(h)}{1-k^B(h)}\\-& \sum_{n=1}^\infty \frac{1}{\pi^2}\left(\Upsilon_{2n}^B(\chi,\zeta)\frac{k^B(2n)}{1 - k^B(2n)}-\Upsilon^B_{1-2n}(\chi,\zeta)\frac{k^B(1-2n)}{1-k^B(1-2n)}\right).\notag
 \end{align}
 We further write $\Upsilon_h^B(\chi,\zeta)=(1+h\zeta/\chi)\Psi_h(\chi)=(1+h\zeta/\chi)(\t A(h)F_h+\t B(h)F_{1-h})$, giving
 \begin{align}\label{longerfor} \frac{\F(\chi,\zeta)}{\t\alpha_0}&=-\int_{-\infty}^\infty \frac{ d s }{2\pi^2\tan \pi h} \left(1+\frac{h\zeta}{\chi}\right)\left(\t A(h) F_h(\chi)+\t B(h) F_{1-h}(\chi)\right)
 \frac{k^B(h)}{1-k^B(h)}  \\ &- \sum_{n=1}^\infty \frac{\t A(2n)F_{2n}(\chi)}{\pi^2}\left[\left(1+\frac{2n\zeta}{\chi}\right) \frac{k^B(2n)}{1-k^B(2n)} - \left(1+\frac{(1-2n)\zeta}{\chi} \right) \frac{k^B(1-2n)}{1-k^B(1-2n)}\right].\notag
 \end{align}
 In writing the contribution of the discrete states this way, we have used the fact that $\t B(2n)=0=\t A(1-2n)$, along with $\t B(1-2n)=\t A(2n)$.  As in \cite{Maldacena:2016hyu}, this is a
 formal expression because $k^B(-1)=1 $ ($=k^F(3/2)$) so the contribution of one of the discrete states needs to be analyzed more precisely.
 
 Following \cite{Maldacena:2016hyu}, to derive an operator product expansion of the four-point function, we would like to move the contour in the integral in the direction of
 increasing $\Re\,h$.  The $F_{1-h}(\chi)$ function increases in this this direction, so before we can usefully move the contour, we must first eliminate this term from the integral.   For this, we note that the factor $\t A(h) F_h(\chi)+\t B(h) F_{1-h}(\chi)$
 in (\ref{longerfor}) is symmetric under $h\to 1-h$ and moreover that this transformation exchanges the $\t A$ and $\t B$ terms.  Thus if the rest of the integrand were symmetric
 under $h\to 1-h$, the $\t A$ and $\t B$ terms would contribute equally and we could replace this factor with $2\t A(h) F_h(\chi)$.  The rest of the integrand is not symmetric,
 but precisely because $\t A(h) F_h(\chi)+\t B(h) F_{1-h}(\chi)$ is symmetric under $h\to 1-h$, we can symmetrize the rest of the integrand without changing the integral.
 Thus in (\ref{longerfor}) we can make the replacement
 \be\label{helpful}\frac{\left(1+\frac{h\zeta}{\chi}\right)}{\tan\pi h}
 \frac{k^B(h)}{1-k^B(h)}\longrightarrow  \frac{\left(1+\frac{h\zeta}{\chi}\right)}{2\tan\pi h} \frac{k^B(h)}{1-k^B(h)}- \frac{\left(1+\frac{(1-h)\zeta}{\chi}\right)}{2\tan\pi h} \frac{k^B(1-h)}{1-k^B(1-h)}.\ee
After doing this, the $\t A$ and $\t B$ terms contribute equally so that the integral in (\ref{longerfor}) can be written as
 \be\label{morehelpful}-\int_{-\infty}^\infty \!\!\!\!ds\,\,\frac{ \t A(h) F_h(\chi) }{2\pi^2\tan \pi h}
\left[\left(1+\frac{h\zeta}{\chi}\right) \frac{k^B(h)}{1-k^B(h)} - \left(1+\frac{(1-h)\zeta}{\chi}\right) \frac{k^B(1-h)}{1-k^B(1-h)}\right].\ee
Now we can usefully move the contour in the direction of increasing $\Re\,h$.  The function $\t A(h)/\tan \pi h$ has poles at positive even integers whose
residues just cancel\footnote{In the formulas we have written, this cancellation follows from an apparent coincidence in the normalization coefficients of the continuum and discrete states. In Appendix \ref{normo}, we explain this coincidence.} the discrete sum in eqn.~(\ref{longerfor}).  We are left with contributions only from poles at $k^B(h)=1$ or $k^B(1-h)=1$.  The solutions of this
equation occur for real $h$.  Since $k^B(1-h)=k^F(h-1/2)$, we can write the result as follows:
\begin{align}\label{ope}\frac{\F(\chi,\zeta)}{\h\alpha_0}&=\sum_{k^B(h)=1, \,h>1/2} \mathrm{Res}\left(\frac{ \t A(h) F_h(\chi) }{\pi\tan \pi h}
\left(1+\frac{h\zeta}{\chi}\right) \frac{k^B(h)}{1-k^B(h)}\right)\\ &- \sum_{k^F(h-1/2)=1,\,h>1/2} \mathrm{Res}\left(\frac{ \t A(h) F_h(\chi) }{\pi\tan \pi h}
\left(1+\frac{(1-h)\zeta}{\chi}\right) \frac{k^F(h-1/2)}{1-k^F(h-1/2)}\right) .\nonumber\end{align}

To extract the part of $\F(\chi,\zeta)$ that governs the expectation of a product of  four superconformal primaries (rather than their descendants), we should simply set $\zeta=0$ in this formula.
Then ignoring supersymmetry, the operator product expansion tells us that a term behaving as $\chi^h$ for small $\chi$ is the contribution of an operator of dimension $h$ propagating
in the 12 channel.  This operator must be bosonic, since this channel describes the fusion of two fermions.   Since $F_h(\chi)\sim \chi^h$, we see that a bosonic operator
of dimension $h$ is associated to a solution of either $k^B(h)=1$ or of $k^F(h-1/2)=1$.  The interpretation is clear:  a bosonic operator of dimension $h$ is either a primary or else
the descendant of a fermionic primary of dimension $h-1/2$.

One surprising feature of the formula (\ref{ope}) is that there is no operator contribution at $h = 1$, despite the fact that $k^B(1) = 1$. This is because the factor $\widetilde{A}(h)/\tan\pi h$ has a zero at $h = 1$ that cancels the would-be pole. This is surprising because it means that the supersymmetry-breaking mode described in \cite{Fu:2016vas} does not actually give an operator contribution to the four-point function.

The manipulation we made was slightly formal, because of a divergence in the contribution of the discrete state involving $k^B(1-2n)$ with $n = 1$. The correct procedure is to analyze this particular contribution outside the conformal limit. Presumably this leads to a super-Schwarzian theory, as suggested in \cite{Fu:2016vas}. The four-point function would then be the sum of the (large!) contribution of that sector, plus the contribution written in (\ref{ope}). Note that the sum over residues in (\ref{ope}) should include the double pole at $h = 2$.

\section{The Shadow Representation In Two Dimensions}\label{twod}

\subsection{The Shadow Representation in 2d CFT}\label{boscase}

In two dimensions, a conformal field has left and right dimensions $h, \t h\geq 0$.  The sum of the two is the overall scaling dimension
\be\label{zdelt}\Delta=h+\t h,\ee
and the difference is the spin
\be\label{spin}J=h-\t h.\ee
Here $J$ is always an integer or half-integer (for bosonic or fermionic operators, respectively), and the only general constraint on $\Delta$ is $\Delta\geq |J|$.

We consider the normalized four-point function
\be\label{normfcn}W(z_1,\bar z_1;z_2,\bar z_2;\dots z_4,\bar z_4)=\frac{\langle \Phi_1(z_1,\bar z_1)\Phi_2(z_2,\bar z_2)\Phi_3(z_3,\bar z_3)\Phi_4(z_4,\bar z_4)\rangle}
     {\langle \Phi_1(z_1,\bar z_1)\Phi_2(z_2,\bar z_2)\rangle\langle\Phi_3(z_3,\bar z_3)\Phi_4(z_4,\bar z_4)\rangle},\ee
where $\Phi_1,\dots,\Phi_4$ are conformal primaries of spin 0 and the same dimension $\Delta$.  This choice is motivated by applications to certain 2d bosonic analogs
of the SYK model.   (A more general case can be treated similarly to what follows.)

The shadow representation is obtained with insertion of $\int d^2z \,\V(z,\bar z)\,\t \V(z,\bar z)$, with primary fields $\V$, $\t V$; here $\V$ has
 some dimension $(h,\t h)$ and $\t V$ has the complementary
dimension $(1-h,1-\t h)$.  If not zero, the normalized three-point function $\langle \Phi_1 \Phi_2 \V\rangle/\langle\Phi_1\Phi_2\rangle$ is, for a suitable normalization of $\V$,
\begin{align}\label{ormfn} \frac{\large\langle\Phi_1(z_1,\bz_1)\Phi_2(z_2,\bz_2)\V(z_3,\bz_3)\large\rangle}
{\large\langle\Phi_1(z_1,\bz_1)\Phi_2(z_2,\bz_2)\large\rangle}=\frac{z_{12}^h}{z_{13}^hz_{23}^h}\frac{\bz_{12}^{\th}}{
\bz_{13}^{\t h}\bz_{23}^{\t h}}, \end{align} where $z_{ij}=z_i-z_j$.
Here the operator $\V$ must be bosonic, so its spin $J=h-\th$ is an integer.  The integrality of $J$ ensures that the right hand side of (\ref{ormfn}) is single-valued
if, for example, we interpret $z_{12}^{h}\bz_{12}^{\th}$ as $z_{12}^{J} |z_{12}|^{2\t h}$.  A similar remark applies to many formulas below.

Using (\ref{ormfn}) and the complementary formula for particles 3,4, the 2d analog of the shadow representation (\ref{gruff}), describing a contribution to $W(z_1,\dots,\bz_4)$ due to a primary of dimensions $(h,\th)$, is
\be\label{ruff}\Psi_{h,\th}(z_1,\dots, \bz_4)=\int d^2y\frac{z_{12}^h z_{34}^{1-h}\bz_{12}^{\th} \bz_{34}^{1-\th}}{ z_{y1}^h z_{y2}^hz_{3y}^{1-h}z_{4y}^{1-h}
\bz_{y1}^{\th}\bz_{y2}^{\th }\bz_{3y}^{1-\th}\bz_{4y}^{1-\th}     }.\ee
(We write $z_{iy}=z_i-y = -z_{yi}$. Orderings have been chosen to avoid an inconvenient minus sign in eqn.~(\ref{uff}).)

Of course, $\Psi_{h,\th}$ can only be a function of the cross-ratio $\chi=z_{12}z_{34}/z_{13}z_{24}$ and its complex conjugate $\b\chi$.  Setting $z_1=0$,
$z_3=1$, $z_4=\infty$, and therefore $z_2=\chi$, we get
\be\label{uff}\Psi_{h,\th}(\chi,\b\chi)=\int d^2y \frac{ \chi^h\b\chi^\th   }{y^h(y-\chi)^h(1-y)^{1-h} \b y^\th (\b y -\b\chi)^\th (1-\b y)^{1-\th} }.\ee
This integral converges for $0\leq \mathrm{Re}(h+\th)<2$.    However, the argument given at the end of section \ref{syk} can easily be adapted
to show that $\Psi_{h,\th}$ has a meromorphic continuation in $h$ (with $h-\th$ fixed at an integer value) with its only singularities being simple
poles at certain integer values.

Just as in 1d, if $\Phi_1=\Phi_2$, then 
the four-point function is invariant under  $ z_1\leftrightarrow z_2$, and this leads to
symmetry under $\chi\to \chi/(\chi-1)$.  Explicitly, making the change of variables $y\to y/(y-1)$ in the shadow integral, we find that 
\be\label{zeddo}\Psi_{h,\th}\left(\frac{\chi}{\chi-1},\frac{\b\chi}{\b \chi-1}\right)=(-1)^{h-\th} \Psi_{h,\th}(\chi,\b\chi).\ee
This relation implies that a correlation function that possesses the $z_1\leftrightarrow z_2$ symmetry can only receive contributions from $\Psi_{h,\th}$
with even values of $J=h-\th$.  
Likewise, as in 1d,  the shadow construction is invariant under the exchange of the first two and last two particles, 
together with $(h,\th)\leftrightarrow (1-h,1-\th)$.   This leads to
\be\label{eddo}\Psi_{1-h,1-\th}(\chi,\bar\chi)=\Psi_{h,\th}(\chi,\bar\chi),\ee
which follows explicitly from the change of variables $y\to \chi/y$ in the shadow integral.  

In two dimensions, because of the decomposition of the special conformal group as $SO(2,2)\sim SL_2\times SL_2$, we can
define two Casimirs.  The holomorphic Casimir for two points $z_1,z_2\in\CC$
is defined by the same formula as in the 1d case
\be\label{twoc}C_{12}=-z_{12}^2\frac{\partial^2}{\partial z_1\partial z_2}, \ee but
now with complex coordinates.  Acting on a function that depends only on the cross ratio, $C_{12}$ is again given by the 1d formula:
\be\label{woc}C_{12}=\chi^2(1-\chi)\partial_\chi^2-\chi^2\partial_\chi.\ee
The antiholomorphic Casimir is defined by the complex conjugate formulas,
\be\label{uwoc}\b C_{12}=-\bz_{12}^2\frac{\partial^2}{\partial\b z_1\partial\b z_2}, \ee 
and
\be\label{woco}\b C_{12}=\b\chi^2(1-\b\chi)\partial_{\b\chi}^2-\b\chi^2\partial_{\b\chi}.\ee
The wave function $\Psi_{h,\th}$ is a simultaneous eigenfunction of $C_{12}$ and $\b C_{12}$, with $C_{12}\Psi_{h,\th}=h(h-1)\Psi_{h,\th}$,
$\b C_{12}\Psi_{h,\th}=\th(\th-1)\Psi_{h,\th}$.   This is true because the conformal three-point function (\ref{ormfn}) is an eigenfunction with those
eigenvalues, and integrating over $y$ to construct the shadow representation of $\Psi_{h,\th}$ does not spoil that property.

  Locally, the holomorphic equation $C_{12}\Psi=h(h-1)\Psi$ has the two linearly
independent solutions $F_h$ and $F_{1-h}$, where we continue to use the notation
\be\label{1dblockdef}
F_h(\chi)\equiv \chi^h\,_2F_1(h,h;2h;\chi).
\ee
Likewise, the antiholomorphic equation $\b C_{12}\Psi=\th(\th-1)\Psi$
has  linearly
independent solutions $F_{\th }(\b\chi)$ and $F_{1-\th}(\b\chi)$.  Accordingly, $\Psi_{h,\th}(\chi,\b\chi)$ is a linear combination of
$F_h(\chi)F_{\th}(\b\chi)$ and three more terms with $h$ replaced by $1-h$ and/or $\th$ by $1-\th$.  However, as $\Psi_{h,\th}$ is single-valued
near $\chi=\b\chi=0$, there are actually only two contributions:
\be\label{mido}\Psi_{h,\th}(\chi,\b\chi)=\A(h,\th) F_h(\chi)F_{\t h}(\b\chi)+\B(h,\th) F_{1-h}(\chi)F_{1-\th}(\b\chi). \ee

Since $F_h(\chi)\sim \chi^h$ for small $\chi$, we have \be\label{flox}\Psi_{h,\th}(\chi,\bar\chi)=\A(h,\th)\chi^h\b\chi^\th(1+\dots) +\B(h,\th)\chi^{1-h}\b\chi^{1-\th}(1+\dots),\ee
where the ellipses vanish for small $\chi$.  The $A$ term dominates at small $\chi$ for $\mathrm{Re }(h+\th)<1$.  In this region, assuming also $\mathrm{Re}(h+\th)>0$
so that the shadow integral converges, the small $\chi$ behavior can be determined by simply setting $\chi$ to zero in the denominator in eqn.~(\ref{uff}).  We find
\be\label{luff}\A(h,\th)=\int d^2y  \frac{ 1  }{y^{2h}(1-y)^{1-h} \b y^{2\th} (1-\b y)^{1-\th} } =
\frac{1}{2}\frac{\sin(\pi h)}{\cos(\pi \t{h})}\frac{\Gamma(h)^2}{\Gamma(2h)}\frac{\Gamma(\t{h})^2}{\Gamma(2\t{h})}\ee
See Appendix \ref{eval} for the evaluation of this integral.  
The symmetry under $(h,\th)\leftrightarrow (1-h,1-\th)$ implies that
\be\label{nuff}\B(h,\th)=\A(1-h,1-\th) = -\frac{1}{2}\frac{\sin(\pi h)}{\cos(\pi\th)} \frac{\Gamma(1-h)^2}{\Gamma(2-2h)}\frac{\Gamma(1-\th)^2}
{\Gamma(2-2\th)}.\ee
This implies the remarkably simple
\be\label{nnuff}\A(h,\th)\B(h,\th)=-\frac{\pi^2}{(2h-1)(2\th-1)}.\ee
Note also that
\be\label{lluff}\A(h,\th)=\A(\th,h),~~\B(h,\th)=\B(\th,h).\ee
These last statements depend upon the fact that $h-\th$ is an integer.

\subsection{A Complete Set Of States}\label{complete}

From among the states $\Psi_{h,\th}(\chi,\b\chi)$, we will construct a complete set of states in terms of which the four-point function
can be expanded.  
Before doing this, let us recall the analogous completeness statement \cite{Maldacena:2016hyu} 
for the 1d states $\Psi_h$ defined in eqn.~(\ref{rrefl}). The inner product on functions $\Psi(\chi)$ is, from eqn.~(\ref{invone}),
\be\label{narmo} |\Psi(\chi)|^2=\int_0^2 \frac{d\chi}{\chi^2}\Psi^*(\chi)\Psi(\chi). \ee
As the two-particle Casimir $C_{12}$ is hermitian with respect to this inner product, it must have a complete set
of eigenstates with real values of the eigenvalue $h(h-1)$,
which means that $h$ is either real or is of the form $1/2+is$.  
The eigenfunctions  $\Psi_h$ of $C_{12}$  behave as $\chi^h$ for small $\chi$, so they
  have continuum normalization if $h=1/2+is$ with real $s$.   In searching among the $\Psi_h$ 
for a normalizable state, because of the symmetry $\Psi_h=\Psi_{1-h}$, we can assume $\Re\,h>1/2$.   Given this, since $F_h\sim\chi^h$ for small
$h$, $\Psi_h=\t A(h)F_h+\t B(h)F_{1-h}$ is normalizable if and only if $\t B(h)=0$.  With $\t A(h)$ and $\t B(h)$ as in eqn.~(\ref{zefl}), this
leads to the condition $h=2n$, as in \cite{Maldacena:2016hyu}.

For future reference, it will be helpful to understand the fallacy in the following attempt to find additional normalizable eigenstates of the two-particle
Casimir
in $d=1$.  If $h=n+1/2$ with $n$ a positive integer, then $\t B$ is regular but $\t A$ has a simple pole.   So cannot we form a normalizable
eigenfunction of the Casimir as $S(\chi)=\lim_{h\to n+1/2}(h-(n+1/2))\Psi_h(\chi)$?   Here the factor $h-(n+1/2)$ cancels the pole of $\t A$ 
at $h= n+1/2$ and eliminates the $\t B$ term.  The fallacy here is that although it is true that $S(\chi)$ is normalizable and 
is an eigenfunction of the Casimir, actually $S(\chi)$ is identically 0.  Indeed, we proved at the end of section \ref{syk} 
that $\Psi_h(\chi)$ is meromorphic in $h$ with
poles only at certain integer values of $h$.  In particular, $\Psi_h$ is regular at $h=n+1/2$, implying that $S(\chi)=0$.  Concretely,
 $F_{1-h}$ itself has a pole at $h=n+1/2$.  For non-integer $h$, $F_h$ and $F_{1-h}$ have expansions
\be\label{medx}F_{h}=\chi^{h}\left(1+ u_1(h)\chi+u_2(h)\chi^2 +\dots\right),~~~F_{1-h}=\chi^{1-h}\left(1+v_1(h)\chi+v_2(h)\chi+\dots\right).\ee
A given exponent can appear in both expansions if and only if $h$ is an integer or  half-integer.  At $h=n+1/2$, $n>0$, the coefficients $v_s(h)$,
$s\geq 2n$, have simple poles such that the singular behavior of $F_{1-h}$ is
\be\label{edx}F_{1-h}(\chi){\sim} -\frac{\t A(h)}{\t B(h)} F_{h}(\chi),~~~h\to n+1/2>1, \ee
and thus $\t A F_h+\t BF_{1-h}$ remains regular.

Now we move on to the 2d case.    A repeat of the derivation of eqn.~(\ref{invone}) leads to the inner product
\be\label{meflo}\bigl( F|G\bigr) =\int\frac{d\chi d\b\chi}{|\chi|^4} F^*G. \ee
The holomorphic and antiholomorphic Casimirs commute and are each other's hermitian adjoints with respect to this inner product.
Given this, they will have a complete set of simultaneous eigenfunctions satisfying $C_{12}\Psi=\lambda\Psi$, $\b C_{12}\Psi=\b\lambda\Psi$,
with $\lambda$ and $\b\lambda $ complex conjugate.  In fact, the joint eigenfunctions of $C_{12}$ and $\b C_{12}$ are $\Psi_{h,\th}$
with $\lambda=h(h-1)$, $\b\lambda=\th(\th-1)$.  The condition for $h(h-1)$ and $\th(\th-1)$ to be complex conjugate implies 
that either $h$ and $\th$ are complex conjugate or $\Re(h+\th)=1$.     It must be possible to find a complete set of states from
among the $\Psi_{h,\th}$ with $h,\th$ satisfying those conditions. 

It turns out that the states with $\Re(h+\th)=1$ give a complete set of states.\footnote{This is consistent with a general property of representations of the conformal group: in odd dimensions the completeness relation involves continuum and discrete contributions, while in even dimensions there is only the continuum \cite{Dobrev1977}.}
These states are continuum normalizable, and we will determine their normalization in section \ref{compreln}.
We parametrize them by
\be\label{eflo}h=\frac{1+\ell}{2}+is,~~~ \th =\frac{1-\ell}{2}+is, \ee
with real $s$ and integer $\ell$, and we sometimes denote them as 
 $\Psi_{\ell,s}$.F  These states make up what is known as the principal series for the 2d special conformal group
$SO(2,2)$.  Their  spin is $J=h-\th =\ell$.  

We explain next why there are no discrete states.  In the process, we will also uncover some facts that look special at first sight
but turn out to be important.

\subsection{Absence Of Discrete States}\label{absdisc}

In searching for a normalizable state, because of the symmetry under $(h,\th)\to (1-h,1-\th)$, we can assume that $\Re(h+\th)>1$.  
Then $\Psi_{h,\th}$ can only be normalizable if $\B(h,\th)=0$.  Since 
\be\label{beflo}\B(h,\th)=-\frac{1}{2}\frac{\sin(\pi h)}{\cos(\pi\th)} \frac{\Gamma(1-h)^2}{\Gamma(2-2h)}\frac{\Gamma(1-\th)^2}
{\Gamma(2-2\th)}, \ee
a zero of $\B$ will come from a zero of $\sin(\pi h)$ or from a pole of the denominator.  However, $\sin(\pi h)$ vanishes
only if $h$ is an integer, in which case $\th$ is also an integer and, as $\Re(h+\th)>1$, either $1-h$ or $1-\th$ is nonpositive.
Hence at least one of $\Gamma(1-h)^2$ and $\Gamma(1-\th)^2$ has a double pole and $\B$ does not vanish.  It remains
to consider a zero of $\B$ that comes from a pole of $\Gamma(2-2h)$ and/or of $\Gamma(2-2\th)$ when $h$ and/or $\th$ is
a half-integer greater than 1.  Since $h-\th$ is an integer, when $h$ or $\th$ is a half-integer, so is the other, and therefore
$\cos(\pi\th)$ vanishes.  Accordingly, the only interesting case is that both $h$ and $\th$ are half-integers greater than 1,
leading to a simple zero of $\B$.   However, this does not lead to a normalizable state.  
The contribution of the $\B$ term to $\Psi_{h,\th}$ is $\B(h,\th)F_{1-h}(\chi)F_{1-\th}(\bar\chi)$,
and as we have seen above, when $h$ or $\th$ approaches a half-integer, the functions $F_{1-h}$ and $F_{1-\th}$ have simple
poles, with residues proportional to $F_h$ and $F_{\th}$.  These polar terms survive in the limit that $h$ and $\th$ become half-integral,
and give contributions to the small $\chi$ behavior of $\Psi_{h,\th}$ that are proportional to $\chi^h\b\chi^{1-\th}$ or $\chi^{1-h}\b\chi^\th$.  This behavior
means that $\Psi_{h,\th}$  fails to be square-integrable at these values of $h$, $\th$, though just barely so in the special case $h=\th$.

It may appear from this that $\Psi_{h,\th}$ must have a pole when $h$ and $\th$ are half-integers greater than 1.  If we take the
polar parts of both $F_{1-h}$ and $F_{1-\th}$, we find that $\B  F_{1-h}(\chi) F_{1-\th}(\b\chi)$ does have a polar part proportional
to $F_h(\chi)F_\th(\b\chi)$.  However, eqn.~(\ref{uff}) implies that $\A$ has a pole at the zeroes of $\B$ under discussion, and the resulting
pole in $\A F_h(\chi) F_{\th}(\b\chi)$ cancels the pole just mentioned in $\B  F_{1-h}(\chi) F_{1-\th}(\b\chi)$.  Indeed, as remarked in section \ref{boscase},
the shadow representation can be used to show that $\Psi_{h,\th}$ has no poles except at integer values of $h,\th$.  

Now let us look more closely at $\Psi_{h,\th}$ for the special case that $h$ and $\th$ are both half-integers greater than 1.
As seen above,  the small $\chi$ expansion of $\Psi_{h,\th}$  then has leading terms $\chi^h\b\chi^{1-\th}$ and $\chi^{1-h}\b\chi^\th$.
These are also the leading terms in the expansion of $\Psi_{h,1-\th}$ or equivalently $\Psi_{1-h,\th}$ near $\chi=\bar\chi=0$.
A little thought shows that, for these special values of $h$ and $\th$, $\Psi_{h,\th}$ and $\Psi_{h,1-\th}$ must be multiples of each other.
Indeed, these functions have the same values $h(h-1)$ and $\th(\th-1)$ of the quadratic Casimirs, and the same $\chi\to \chi/(\chi-1)$ symmetry.  
Those properties determine these functions up to a constant multiple.

To determine the constant of proportionality between $\Psi_{h,\th}$ and $\Psi_{h,1-\th}$ when $h,\th$ are half-integers greater than 1, it suffices
to compare the coefficients of $\chi^h\b\chi^{1-\th}$.  For $\Psi_{h,1-\th}$, this is immediate:
\be\label{zelfo} \Psi_{h,1-\th}=\A(h,1-\th)\chi^h\b\chi^{1-\th}+\dots. \ee
On the other hand, a term in $\Psi(h,\th)$ proportional to $\chi^h\b\chi^{1-\th}$ must come from $\B(h,\th)F_{1-h}(\chi)F_{1-\th}(\b\chi)$,
where the leading term of $F_{1-\th}(\b\chi)$ is $\b\chi^{1-\th}$, and according to eqn.  (\ref{edx}), to extract a term proportional to $\chi^h$, 
we can replace  $F_{1-h}(\chi)$ by
$-(\t A(h)/\t B(h))F_h(\chi)\sim -(\t A(h)/\t B(h))\chi^h$.  So comparing coefficients of $\chi^h\b\chi^{1-\th}$, the ratio of the two wavefunctions is
\be\label{elo}\frac{\Psi_{h,\th}}{\Psi_{h,1-\th}} = -\frac{\B(h,\th)}{\A(h,1-\th)} \frac{\t A(h)}{\t B(h)}=1, \ee
where the right hand side was evaluated using eqns.~(\ref{zefl}), (\ref{luff}), and (\ref{nuff}).

Thus when $h,\th$ are half-integers, we have
\be\label{goodrel}\Psi_{h,\th}=\Psi_{h,1-\th}=\Psi_{1-h,\th}=\Psi_{1-h,1-\th}. \ee
In the derivation, we began by assuming that $h,\th>1/2$.  However, the statement (\ref{goodrel}) is trivial (given the general relation $\Psi_{1-h,1-\th}=\Psi_{h,\th}$)
if $h$ or $\th$ equals 1/2, and since eqn.~(\ref{goodrel}) is invariant under $h\leftrightarrow 1-h$ and under $\th \leftrightarrow 1-\th$, it is true for all half-integers
$h,\th$ if it is true for $h,\th>1/2$.

\subsection{The Completeness Relation}\label{compreln}

Because of the symmetry $(h,\th)\to (1-h,1-\th)$, which amounts to $(\ell,s)\to (-\ell,-s)$, in constructing a complete set of states,
we can restrict to $s\geq 0$. In this notation, (\ref{nnuff}) reads
\be\label{duff} \A\B=\frac{\pi^2}{\ell^2+4s^2}. \ee

The inner products (\ref{meflo}) among the states $\Psi_{\ell,s}$ can be computed by following the procedure used in \cite{Maldacena:2016hyu}
in the 1d case.  For a reason that will be explained momentarily, it will suffice to evaluate the integral that controls the inner product
$\bigl( \Psi_{\ell,s},\Psi_{\ell',s'}\bigr) $ in the small $\chi$ region.  In this region, we can approximate $\Psi_{\ell,s}$ as
$\A(\ell,s)\chi^h\b\chi^{\th}+\B(\ell,s)\chi^{1-h}\b\chi^{1-\th}$.  We write $\chi=e^{\rho+i\varphi}$, $-\infty<\rho$, $0\leq \varphi\leq 2\pi$,
so 
\be\label{zapprox}\Psi_{\ell,s}\sim e^\rho\left(\A(\ell,s)e^{2i\rho s+i\ell\varphi} +\B(\ell,s) e^{-2i\rho s -i\ell\varphi}\right).\ee 
The small $\chi$ region produces a delta function contribution to the inner product:\footnote{Because we restrict to $s,s'>0$, the terms proportional to $\b\A \B$ or $\b\B\A$ are oscillatory for $\rho\to -\infty$ and do not produce delta functions.  They
have therefore been dropped in the following. Also,
the upper cutoff at $\rho=0$ is arbitrary and does not affect the delta function terms.}
\begin{align}\label{murot} \bigl( \Psi_{\ell,s},\Psi_{\ell',s'}\bigr) & \sim \int_{-\infty}^0 \!\!\!\!d\rho \int_0^{2\pi} \!\!\!\!d\phi\left(\b\A(\ell,s)\A(\ell',s') e^{2i\rho(s'-s)+i\varphi(\ell'-\ell)}
+\b\B(\ell,s)\B(\ell',s')e^{-2i\rho(s'-s)-i\varphi(\ell'-\ell)}\right)\notag \\ & \sim \pi^2\delta_{\ell,\ell'}\delta(s-s')\left(\bar \A(\ell,s)\A(\ell,s)+\bar\B(\ell,s)\B(\ell,s)\right)
    . \end{align}
    
In general, $\bar \A(h,\th)=\A(\b h,\b{\th})$.  For real $\ell, s$, this gives $\bar\A(h,\th)=\A(1-\th,1-h)=\B(\th,h)=\B(h,\th)$, where we used the fact that eqn.~(\ref{eflo}) implies
$\b{\th}=1-h$, $\b h=1-\th$, and we used
 eqn.~(\ref{lluff})
in the last step.   Using also (\ref{duff}), the coefficient of the delta function in the inner product becomes $4\pi^4/(\ell^2+4s^2)$.  From this
point of view, it appears that the inner product could have a smooth contribution in addition to the delta function, but just as in the 1d case,
conformal invariance ensures that this is not the case.  (The states $\Psi_{\ell,s}$ and $\Psi_{\ell',s'}$ for $(\ell,s)\not=(\ell',s')$ have different
values of  $C_{12}$ and $\b C_{12}$ and therefore are orthogonal.)  Thus the inner product is
\be\label{inpr}\bigl( \Psi_{\ell,s},\Psi_{\ell',s'}\bigr) = \frac{2\pi^4}{\ell^2+4s^2}\delta_{\ell,\ell'}\delta(s-s').\ee 
The corresponding completeness relation is
\be\label{2dcomplete}
\sum_{\ell = -\infty}^\infty\int_0^\infty \frac{ds}{2\pi}\frac{\ell^2+4s^2}{\pi^3}\Psi_{h,\bar h}(\chi,\bar\chi)\bar\Psi_{h,\t{h}}(\chi,\bar\chi') = |\chi|^4\delta^{(2)}(\chi-\chi').
\ee

In this derivation, we started with the hermitian inner product (\ref{meflo}), with complex conjugation of one argument, but we would have arrived at the same
formulas if we instead started with a bilinear form
\be\label{meflox}\bigl( F|G\bigr){}' =\int\frac{d\chi d\b\chi}{|\chi|^4} FG\ee
defined without any complex conjugation.
The reason for this is that the wavefunctions $\Psi_{h,\th}$ with $h,\th$ as in eqn.~(\ref{eflo}) are actually real.  In general, $\b\Psi_{h,\th}=\Psi_{\b\th,\b h}$.
But  eqn.~(\ref{eflo}) gives $\b\th=1-h$, $\b h=1-\th$, so that $\b\Psi_{h,\th}=\Psi_{1-h,1-\th}=\Psi_{h,\th}$.

\subsection{The Operator Product Expansion}\label{expan}

Let us suppose that we are given a 2d bosonic model in which, as in the SYK model, 
the  four-point function $\F(z_1,\dots,\bar z_4)$  (defined now as in eqn.~(\ref{normalfn})) is given by a ladder sum,
\be\label{ladsum}\F=\frac{1}{1-K}\F_0, \ee
with $K$ a conformally-invariant ladder kernel and $\F_0$ a lowest order contribution.  Conformal invariance implies that the functions $\Psi_{\ell,s}$
are eigenfunctions of $K$, say with eigenvalue $k(\ell,s)$.  
Since $\Psi_{\ell,s}=\Psi_{-\ell,-s}$, inevitably $k(\ell,s)=k(-\ell,-s)$ or
\be\label{addsum}k(1-h,1-\th)=k(h,\th). \ee
Suppose further that by an argument similar to what is explained at the
end of section \ref{kkernel}, the inner product $\left( \Psi_{\ell,s},\F_0\right)$ is $\lambda k(\ell,s)$ for some constant $\lambda$.   Examples
of models with these properties will be described in section \ref{bosemod}.  Then using eqn.~(\ref{2dcomplete}), 
\be\label{adsum}\F(z_1,\dots,\bar z_4)=\frac{\lambda}{\pi^3} \sum_{\ell=-\infty}^\infty \int_0^\infty \frac{ds}{2\pi} (\ell^2+4s^2)\,\Psi_{s,\ell}(\chi,\bar\chi) \frac{k(\ell,s)}{1-k(\ell,s)}.\ee
In this formula, $\ell$ is an arbitrary integer in general, but in the case of a correlation function symmetric under the exchange $z_1\leftrightarrow z_2$,
$\ell$ is an even integer, for a reason explained in the discussion of eqn.~(\ref{zeddo}).

We now attempt to proceed as in \cite{Maldacena:2016hyu}.  We write $\Psi_{s,l}=\A(s,\ell) F_{s,\ell} +\B(s,\ell)F_{-s,-\ell}$.   Inserting this
in eqn.~(\ref{adsum}), we can drop the $\B$ term if we also extend the $s$ integral over the whole real line:\footnote{Here we momentarily overlook
a detail that is explained in the last paragraph of this section.}
\be\label{nadsum}\F=\frac{\lambda}{\pi^3} \sum_{\ell=-\infty}^\infty \int_{-\infty}^\infty \frac{ds}{2\pi} (\ell^2+4s^2)\A(s,\ell)F_{s,\ell}(\chi,\bar\chi) \frac{k(\ell,s)}{1-k(\ell,s)}.\ee
To construct an operator product expansion, we now deform the contour in the direction of decreasing $\mathrm{Im}\,s$, or
equivalently increasing $\Re\,h$, $\Re\,\th$.  In doing this,
we encounter poles associated to solutions of the equation $1-k(\ell,s)=0$.   Such a pole gives a contribution to $\F$ that behaves for small $\chi$ as $F_{h,\th}\sim \chi^h\b\chi^\th$,
corresponding to an operator of dimension $(h,\th)$. 
In a unitary theory, these poles occur only for imaginary $s$,
corresponding to real $h,\th$. 
 If the integrand in eqn.~(\ref{nadsum}) had no additional poles in the lower half
$s$-plane (and also a reasonable behavior at infinity), then the representation of $\F$ as a sum over solutions of $1-k(\ell,s)=0$ would correspond
to the operator product expansion or OPE: it gives an expansion of $\F$ as a sum of terms with increasing powers of $h,\th$.

\begin{figure}[ht]
\begin{center}
\includegraphics[width=.4\textwidth]{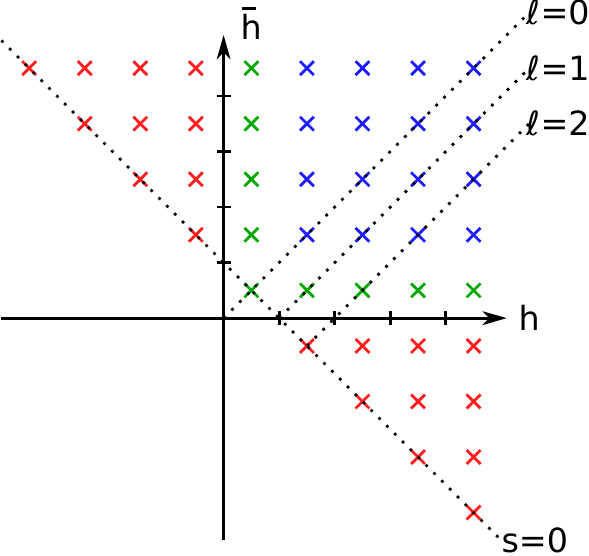}
\caption{This diagram shows the arrangement of spurious poles that we meet in deforming the contour of integration in the direction of increasing
$\Re\,h,\th$.  The red poles (with $h$ or $\th$ negative) are poles of the conformal block $F_h(\chi)F_\th(\b\chi)$, and the blue and green
ones (with $h,\th$ positive) are poles of $\A(h,\th)$.  With the exception of the poles on the line $\ell=0$, which have to be treated more carefully, red and blue poles
are related in pairs by $h\leftrightarrow 1-h$ or by $\th\leftrightarrow 1-\th$, but not both.  A cancellation occurs between these pairs.  Green poles on the lines $h=1/2$ and $\th=1/2$ do not contribute because the measure factor vanishes.}\label{cancellingPolesFig}
\end{center}
\end{figure}

In actuality, there are additional poles, which are related to poles that were discussed in  section \ref{complete}.
  The function $\A(h,\th)$ (eqn.~(\ref{luff})) has poles,
coming from zeroes of $\cos(\pi\th)$, when $h$ and $\th$ are both positive half-integers.  (If either of them is a negative half-integer,
then the vanishing of $\cos(\pi\th)$ is canceled by a pole of $\Gamma(2h)$ or $\Gamma(2\th)$.)  
In addition, we have to consider the poles of the function $F_{s,\ell}(\chi,\bar\chi)=F_h(\chi) F_\th(\b\chi)$.   As we learned in section \ref{complete},
this function has poles if $h$ or $\th$ is a negative half-integer.  Note that the condition $\Re(h+\th)\geq 1$ means that either $h$ or $\th$
may be a negative half-integer, but not both.  

The cancellation of these spurious poles is a little elaborate. It has been previously discussed in \cite{Dobrev:1975ru,Cornalba:2007fs,Costa:2012cb}.  First, poles on the lines $h=1/2$ or $\th=1/2$ (shown in green in fig. \ref{cancellingPolesFig})
do not contribute because $\ell^2+4s^2=-(2h-1)(2\th-1)$ vanishes along these lines.  If we omit these lines and also the diagonal $h=\th$ and the
antidiagonal $h+\th=1$, we observe the following: the remaining poles are paired under either $h\leftrightarrow 1-h$ or $\th\leftrightarrow 1-\th$, but not both.

The residues coming from pairs of poles that are related by $h\leftrightarrow 1-h$ or $\th\leftrightarrow 1-\th$ cancel in pairs.  This results from the following
facts.  First of all, in  2d models in which the four-point function can be computed as a ladder sum, one finds in addition to the general symmetry (\ref{addsum}) that 
\be\label{oddrel}k(h,\th)=k(1-h,\th)=k(h,1-\th),~~~h,\th\in \frac{1}{2}+\Z.\ee
(Note that because of the constraint $h-\th\in\Z$, a relation $k(h,\th)=k(1-h,\th)$ is only meaningful if $h$ and $\th$ are integers or half-integers.)
This is a consequence of eqn.~(\ref{goodrel}), in the same way that $k(h,\th)=k(1-h,1-\th)$ follows from $\Psi_{h,\th}=\Psi_{1-h,1-\th}$.
  Given this, the condition that pairs of poles related by $h\leftrightarrow 1-h$
make no net contribution is that
\be\label{nopole} \sin\pi h \frac{\Gamma(h)^2}{\Gamma(2h)} F_h - \sin\pi(1-h)\frac{\Gamma(1-h)^2}{\Gamma(2-2h)} F_{1-h} \ee
vanishes when $h$ is a positive half-integer. 
A relative minus sign between the two terms results from the fact that the factor $\ell^2+4s^2=-(2h-1)(2\th-1)$ in eqn.~(\ref{nadsum}) is odd under $h\to 1-h$.
By contrast, in section \ref{complete}, we discussed the importance of the fact that the 1d function $\Psi_h(\chi)$ has no pole at half-integral $h$.  This 
amounts to the absence of a pole in 
\be\label{opole}\frac{\tan\pi h}{\tan(\pi h/2)}\frac{\Gamma(h)^2}{\Gamma(2h)}F_h +\frac{\tan\pi(1-h)}{\tan(\pi(1-h)/2)}\frac{\Gamma(1-h)^2}{\Gamma(2-2h)}F_{1-h}\ee
when $h$ is a positive half-integer.   This condition is equivalent to the vanishing of (\ref{nopole}).

It remains to explain how to treat the poles on the lines $h+\th=1$ and $h=\th$.    The poles on the line $h+\th=1$ are on the contour for the original
$s$ integral in eqn.~(\ref{adsum}), so they need special treatment.  The function $\Psi_{h,\th}$ has no poles on the integration contour, so the original
integral eqn.~(\ref{adsum}) is well-defined.   However, when we write $\Psi_{h,\th}$ as the sum of an $\A$ term and a $\B$ term, each term separately has a pole,
and as a result the manipulation leading to eqn.~(\ref{nadsum}) is not quite valid.   We can justify such a manipulation if we first go back to eqn.~(\ref{adsum})
and deform the contour to go just above or just below the poles that the $\A$ term and $\B$ term have on the real axis.  However, in passing from eqn.~(\ref{adsum})
to eqn.~(\ref{nadsum}), we used the symmetry of the integrand under $(h,\th)\to (1-h,1-\th)$.  To justify the manipulation that leads to (\ref{nadsum}), the integration
contour has to have this symmetry.  To achieve this, we interpret the integral in eqn.~(\ref{nadsum}) as one-half of the sum of two integrals, one on a contour that goes
just above the poles on the real $s$ axis, and one on a contour that goes just below them.  The result of this is that the poles on the real $s$ axis, that is
the line $\Re(h+\th)=1$, contribute with weight 1/2 in the residue sum.  By $h\to 1-h$ or $\th \to 1-\th$, such a pole can be mapped to the line $h=\th$.
The mechanism  described in the last paragraph leads to a cancellation between two poles symmetrically placed
on the line $\Re(h+\th)=1$ and a third one on the line $h=\th$.

\section{Two-Dimensional Superconformal Field Theory}\label{twoscft}

In this section, we generalize the considerations of section \ref{twod} to $\N=1$ superconformal
field theory in two dimensions. We formulate such a theory on the supersymmetric analog of Euclidean space, with holomorphic
coordinates $z,\theta$ and antiholomorphic ones $\b z,\b\theta$.

We will consider a normalized four-point function
\be\label{normsuper}W(z_1,\dots,\b\theta_4 )=\frac{\langle 
\Phi_1(z_1,\theta_1,\bar z_1,\b\theta_1)\Phi_2(z_2,\theta_2,\bar z_2,\b\theta_2)\Phi_3(z_3,\theta_3,
\bar z_3,\b\theta_3)\Phi_4(z_4,\theta_4,\bar z_4,\b\theta_4)\rangle}
     {\langle \Phi_1(z_1,\theta_1,\bar z_1,\b\theta_1)\Phi_2(z_2,\theta_2\bar z_2,\b z_2)\rangle\langle\Phi_3(z_3,\theta_3,\bar z_3,\b\theta_3)\Phi_4(z_4,\theta_4,\bar z_4,\b\theta_4)\rangle}\ee
where $\Phi_1,\dots,\Phi_4$ are superconformal primaries of spin 0 and dimension $\Delta$.  This choice is motivated by  2d $\N=1$
supersymmetric analogs of the SYK model that will be described in section \ref{supermod}.

Some basic facts can be essentially borrowed from the 1d analysis of section \ref{supercasimir}.  A conformally covariant function of superspace
coordinates is $\la i,j\ra=z_i-z_j-\theta_i\theta_j$.  The superconformal cross ratio $\chi$ and the nilpotent invariant $\zeta$ can
be defined by the formulas (\ref{zelf}).  However, because
we are in two dimensions, these invariants all have complex conjugates, which are defined using the same formulas but with 
$\bar{\la i,j\ra}=\b z_i-\b z_j-\b\theta_i\b\theta_j$ replacing $\la i,j\ra$.    For particles 1,2, there is a superconformal Casimir
$\C_{12}$ that is defined by eqn.~(\ref{zelo}) (now with real coordinates $t_i$ replaced by complex coordinates $z_i$) or
as an operator acting on functions of $\chi$ and $\zeta$ by eqn.~(\ref{Casimir-system}).  There is also a corresponding antiholomorphic Casimir
$\b\C_{12}$, defined by complex conjugate formulas.  $\C_{12}$ and $\b\C_{12}$ commute, and they are hermitian adjoints
with respect to the natural measure
\be\label{nuzem}\langle A|B\rangle =\int\frac{d^2\chi d^2\zeta}{|\chi+\zeta|^2}A(\chi,\zeta,\b\chi,\b\zeta) 
B(\chi,\zeta,\b\chi,\b\zeta),\ee 
which can be found by adapting the derivation of eqn.~(\ref{uzem}) to the 2d case.\footnote{As in the 1d
case, this inner product is not positive-definite whether or not we include complex conjugation as part of its definition.  Since complex
conjugation does not seem to help, we have defined a bilinear product without any complex conjugation. The measure $d^2\zeta$ is defined by $\int d^2\zeta\, \zeta\b\zeta=1$,
along with $\int d^2\zeta\, p(\zeta,\bar\zeta)=0$ if $p(\zeta,\bar\zeta)$ is a linear function.}

As in the other examples that we have studied, simultaneous eigenfunctions of the Casimirs can be found using the shadow
representation. The shadow representation of the four-point function is obtained by formally inserting $\int d^2y d^2\theta \,\V(y,\theta,\b y,\b\theta)
\t \V(y,\theta,\b y, \b\theta)$, where $\V$ is a superconformal primary of dimension $(h,\t h)$, and $\t\V$ has the complementary dimension,
which in the $\N=1$ superconformal case is $(1/2-h,1/2-\t h)$.  However, we run into a subtlety whose 1d analog was described in 
section \ref{susysyk}.  The operator $\V$ might be either bosonic or fermionic from the holomorphic point of view, and likewise
it might be either bosonic or fermionic from the antiholomorphic point of view.  Since the measure $d^2y d^2\theta=dy d \theta d\b y d \b \theta$
is fermionic both in holomorphic coordinates $y, \theta$ and  antiholomorphic coordinates $\b y,\b \theta$, the operator $\t\V$ will
have opposite properties from $\V$, both holomorphically and antiholomorphically.  

In  the 1d case, we found in section \ref{susysyk} two shadow integrals, leading to functions $\Upsilon_h^B$ and $\Upsilon_h^F$.
In $d=2$, there are four possibilities, which we will denote as $\Upsilon_{h,\th}^{BB}$, $\Upsilon_{h,\th}^{FB}$, $\Upsilon_{h,\th}^{BF}$,
and $\Upsilon_{h,\th}^{FF}$, where the first and second superscripts show respectively whether $\V$ is bosonic or fermionic from
a holomorphic or antiholomorphic point of view.   For example, if $\V$ is supposed to be of type $BB$ (that is, it is bosonic
both holomorphically and antiholomorphically) then the normalized three-point function 
\be\label{normal}\frac{\la \Phi_1(z_1,\theta_1,\b z_1,\b\theta_1)\Phi_2(z_2,\theta_2,\b z_2,\b \theta_2)\V(z,\theta_3,\b z_3,\b\theta_3)\ra}{\la  \Phi_1(z_1,\theta_1,\b z_1,\b\theta_1)\Phi_2(z_2,\theta_2,\b z_2,\b \theta_2)\ra}\ee  is given by the same formula as in eqn.~(\ref{ormfn}),
except that $z_i-z_j$ has to be replaced by $\la ij\ra=z_i-z_j-\theta_i\theta_j$:
\be\label{ormal}\frac{\la 12\ra^h}{\la 13\ra^h\la 23\ra^h}\frac{\b{\la 12\ra}^\th}{\b{\la 13\ra}\,^\th\,\b{\la 23\ra}\,^\th}.\ee
In general, $J=h-\th$ can be any integer, but if $\Phi_1$ and $\Phi_2$ are identical, then this formula is consistent with bose symmetry
only if $J$ is even.
If $\V$ is supposed to be of type $FB$, we include in the three-point function a holomorphic factor 
\be\label{onormal}P(1,2,3) = \frac{\theta_1(z_2-z_3)+\theta_2(z_3-z_1)+\theta_3(z_1-z_2)-2\theta_1\theta_2\theta_3}{
\bigl((z_1-z_2)(z_2-z_3)(z_3-z_1)\bigr)^{1/2}}, \ee
familiar from eqn.~(\ref{pdef}).  In this case,  the descendant  $\partial_\theta\V$ is a boson with a nonzero correlator $\la \Phi_1\Phi_2\partial_\theta\V\ra$.  It has  dimension $(h+1/2,\th)$,
so $h+1/2-\th$ must be an integer.  If $\Phi_1$ and $\Phi_2$ are identical, then $h+1/2-\th$ must be even. 
If $\V$ is of type $BF$, then we include in the three-point function a corresponding antiholomorphic factor $\b{P(1,2,3)}$ (and the constraints
on $h$ and $\th$ are as in the $FB$ case, with $h$ and $\th$ exchanged).  For $\V$ of
type $FF$, we include $P(1,2,3)\b{ P(1,2,3)}$ (and $h-\th$ is an integer, or an even one if $\Phi_1$ and $\Phi_2$ are identical).  

The shadow integrals can be constructed by replacing $z_{ij}$ in the 2d bosonic formula (\ref{ruff}) with $\la i,j\ra$, changing the complementary dimension to $1/2-h,1/2-\th$ and including appropriate factors of $P$ and $\b P$.
For example,
\be\label{forex}\Upsilon_{h,\th}^{BB}=\int d^2y d^2\theta_y \frac{\la 1,2\ra^h\la 3,4\ra^{1/2-h}\b{\la 1,2\ra}\,^\th\, \b{\la 3,4\ra}\,^{1/2-\th}\,P(3,4,y)\,\b{P(3,4,y)}}{\la 1,y\ra^h\la 2,y\ra^h\la 3,y\ra^{1/2-h}
\la 4,y\ra^{1/2-h}\b{\la 1,y\ra}\,^\th\, \b{\la 2,y\ra}\,^\th \,\b{\la 3,y\ra}\,^{1/2-\th}\,\b{\la 4,y\ra}\,^{1/2-\th}}. \ee
To get the other functions $\Upsilon_{h,\th}^{FB}$, etc., one replaces $P(3,4,y)$ with $P(1,2,y)$ and/or one replaces $\b{P(3,4,y)}$ with $\b{P(1,2,y)}$.

As always, the shadow representation has a symmetry that exchanges particles 12 with 34 and exchanges the roles of $\V$ and $\t \V$.  In the present context, this implies
\be\label{norex}\Upsilon^{FF}_{h,\th}=\Upsilon^{BB}_{1/2-h,1/2-\th}, ~~~~\Upsilon^{BF}_{h,\th}=\Upsilon^{FB}_{1/2-h,1/2-\th}.\ee
This is the analog of the 1d formula $\Upsilon^F_h=\Upsilon^B_{1/2-h}$.  

As in the 1d superconformal case, the functions $\Upsilon^{BB}_{h,\th}$, etc.,  can be conveniently expressed in terms of the corresponding bosonic functions $\Psi_{h,\th}$
(eqn.~(\ref{uff})).  For example, to analyze
$\Upsilon^{BB}_{h,\th}$, we pick coordinates $z_1=0,$ $z_3=1$, $ z_4=\infty$, and $\theta_3=\theta_4=\b\theta_3=\b\theta_4=0$.   Imitating the derivation of eqn.~(\ref{shadowf}), we find
\be\label{ladow}\Upsilon^{BB}_{h,\th}(\chi,\zeta,\b\chi,\b\zeta)=\left(1+\frac{h\zeta}{\chi}\right)\left(1+\frac{\th \b\zeta}{\b\chi}\right)\Psi_{h,\th}
(\chi,\b\chi). \ee  To analyze $\Upsilon^{FB}_{h,\th}$, it is convenient to make the same choice of the bosonic coordinates as before, but to choose the fermionic coordinates to satisfy\footnote{Calculus with
 anticommuting coordinates is a formal construction, and there is no reason for the coordinate choice for the fermionic coordinates
 to respect any reality condition.}    $\theta_1=\theta_2=\b\theta_3
 =\b\theta_4=0$.   Then a similar calculation gives 
 \be\label{lladow}\Upsilon^{FB}_{h,\th}(\chi,\zeta,\b\chi,\b\zeta)=\left(1+\frac{(\tfrac{1}{2}-h)\zeta}{\chi}\right)\left(1+\frac{\th \b\zeta}{\b\chi}\right)\Psi_{h+1/2,\th}
(\chi,\b\chi). \ee 
The remaining functions $\Upsilon_{h,\th}^{BF}$ and $\Upsilon^{FF}_{h,\th}$ are determined in terms of these by eqn.~(\ref{norex}).

Comparing eqns.~(\ref{ladow}) and (\ref{lladow}), and making use of eqn.~(\ref{goodrel}), we deduce that for half-integer $h,\th$, 
\be\label{superreln}\Upsilon^{FB}_{1/2-h,\th}=\Upsilon^{BB}_{h,\th}. \ee
This can be combined with eqn.~(\ref{norex}) to generate additional identities.

 All of these functions are eigenfunctions of the holomorphic and antiholomorphic Casimirs $\C_{12}$ and $\b\C_{12}$ with
 eigenvalues $h(h-1/2)$ and $\th(\th-1/2)$.   This follows from the considerations
 in section \ref{Solving}, together with the fact that the bosonic function $\Psi_{h,\th}$ is an eigenfunction of the bosonic Casimirs
 $C_{12}$ and $\b C_{12}$ with eigenvalues $h(h-1)$ and $\th(\th-1)$.  Alternatively, it follows from the fact that the building blocks of the supersymmetric shadow representation
 are three-point functions that themselves are eigenfunctions of the  supersymmetric Casimirs.
 
 In the supersymmetric case, because the measure (\ref{nuzem}) is not positive-definite, we cannot invoke general theorems
 to guarantee that every function of the invariants $\chi,\zeta,\b\chi,\b\zeta$ can be expanded in terms of a complete set of eigenfunctions
 of the Casimirs.  However, as in the 1d case,  the relation between the functions $\Upsilon^{BB}_{h,\th}$, etc., and their bosonic counterparts
 $\Psi_{h,\th}$ means that general theorems are not needed.  Indeed, a complete set of states consists of the functions
 $\Upsilon^{BB}_{h,\th}$ and $\Upsilon^{FB}_{h-1/2,\th}$ with $h,\th$ expressed as in eqn.~(\ref{eflo}) in terms of  $\ell$ and
  $s$. Here $\ell$ is an integer; $s$ is real and not restricted to be positive.\footnote{Alternatively, we could constrain $s$ to be positive and include also
  $\Upsilon^{BF}$ and $\Upsilon^{FF}$.}  A straightforward calculation using (\ref{ladow}), (\ref{nuzem})
 and (\ref{meflo}) shows that
 \be\label{yurtz}\bigl\la \Upsilon^{BB}_{h,\th},\Upsilon^{BB}_{h',\th}\bigr\ra = (h+h'-1)(\th+\th'-1) \bigl(\Psi_{h,\th},\Psi_{h',\th'}\bigr). \ee
 Using also the bosonic inner product (\ref{inpr}), we arrive at\footnote{Eqn.~(\ref{inpr}) is written for $s,s'\geq 0$.  If we drop this restriction, the
 right hand side of eqn.~(\ref{inpr}) has an additional term proportional to $\delta_{\ell,-\ell'}\delta(s+s')$.  There is no such term in eqn.~(\ref{urtz}), because
 on the support of this delta function, $h+h'-1=\th+\th'-1=0$.}
 \be\label{urtz} \bigl\la \Upsilon^{BB}_{s,\ell},\Upsilon^{BB}_{s',\ell'}\bigr\ra = -2\pi^4\delta_{\ell,\ell'}\delta(s-s').  \ee
 Similarly
  \be\label{nurtz} \bigl\la \Upsilon^{FB}_{s,\ell},\Upsilon^{FB}_{s',\ell'}\bigr\ra = 2\pi^4\delta_{\ell,\ell'}\delta(s-s') \ee
  and
  \be\label{vurtz} \bigl\la \Upsilon^{BB}_{s,\ell},\Upsilon^{FB}_{s',\ell'}\bigr\ra = 0.  \ee 
  
  The completeness relation is 
  \begin{align}\label{comprel}\frac{1}{\pi^3} \sum_{\ell\in\Z}\int_{-\infty}^\infty\frac{ds}{2\pi}\Big(-\Upsilon^{BB}_{h,\th}(\chi,\zeta)\Upsilon^{BB}_{h,\th}(\chi',\zeta')
  +\Upsilon^{FB}_{h-1/2,\th}&(\chi,\zeta)\Upsilon^{FB}_{h-1/2,\th}(\chi',\zeta')\Big)\\ & =|\chi+\zeta|^2\delta^2(\zeta-\zeta')\delta^2(\chi-\chi'), \notag\end{align}
 where $\delta^2(\zeta-\zeta')=|\zeta+\zeta'|^2$.   (We have written $\Upsilon^{FB}$ with arguments $h-1/2,\th$ rather than $h,\th$ so that
 in eqn.~(\ref{comprel}), we can simply take $\ell$ to be an integer.)   This relation is a simple consequence of the bosonic completeness relation (\ref{2dcomplete}),
expanding in powers of $\zeta$ and $\zeta'$ and following steps similar to those that are involved in verifying eqn.~(\ref{superrel}) in the 1d case.  
For example, if we just set $\zeta=\zeta'=0$, the left and right hand sides of eqn.~(\ref{comprel}) both trivially vanish.  At the opposite extreme, to verify the
component of eqn.~(\ref{comprel}) that is proportional to $\zeta\b \zeta\zeta'\b\zeta{}'$, the identity that we have to use is that $-h^2\th^2+(1-h)^2\th^2$, after being
symmetrized under $\ell,s\to -\ell,-s$, is equal to $-\frac{1}{2}(2h-1)(2\th-1)$.
 
 Now we can explain the superanalog of the considerations of section \ref{expan}, in which we described the OPE for a bosonic 2d analog of the
 SYK model.  We assume again that the  four-point function $\F$ of some model (with the SYK-like normalization of eqn.~(\ref{normalfn})) is a ladder sum,
 \be\label{ladtwo}\F=\frac{1}{1-K}\F_0, \ee
 where $\F_0$ is some lowest order contribution and $K$ is a ladder kernel.   Superconformal invariance insures that  $\Upsilon^{BB}_{h,\th}$ is
 an  eigenfunction of $K$, with some eigenvalue $k^{BB}(h,\th)$,  and similarly for the other functions:
 \be\label{madtwo}K\Upsilon^{BB}_{h,\th}=k^{BB}(h,\th) \Upsilon^{BB}_{h,\th},~~~~K\Upsilon^{FB}_{h,\th}=k^{FB}(h,\th)\Upsilon^{FB}_{h,\th},~~~~\mathrm{etc.}
 \ee
Eqn.~(\ref{norex}) implies that
 \be\label{wadtwo} k^{BB}(\tfrac{1}{2}-h,\tfrac{1}{2}-\th)=k^{FF}(h,\th),~~~k^{FB}(\tfrac{1}{2}-h,\tfrac{1}{2}-\th)=k^{BF}(h,\th). \ee
 We assume further that, for reasons similar to what is explained at the end of section \ref{kkernel}, $\la \Upsilon_{h,\th}^{BB},\F_0\ra=\lambda k^{BB}(h,\th)$, $\la\Upsilon_{h,\th}^{FB},\F_0\ra=\lambda k^{FB}(h,\th)$, for some constant $\lambda$.  Given these assumptions, we have
 an analog of eqn.~(\ref{adsum}):
 \be\label{padsum}\F=\frac{\lambda}{\pi^3} \sum_{\ell=-\infty}^\infty \int_{-\infty}^\infty \frac{ds}{2\pi} \left(-\Upsilon^{BB}_{h,\th}(\chi,\zeta,\bar\chi,\bar\zeta) \frac{k^{BB}(h,\th)}{1-k^{BB}(h,\th)}+\Upsilon^{FB}_{h-1/2,\th}(\chi,\zeta,\bar\chi,\bar\zeta) \frac{k^{FB}(h-\tfrac{1}{2},\th)}{1-k^{FB}(h-\tfrac{1}{2},\th)} \right).\ee
 
Next we write
\begin{align}\label{nexr}\Upsilon^{BB}_{h,\th} =& \left(1+\frac{h\zeta}{\chi}\right)\left(1+\frac{\th\b\zeta}{\b\chi}\right)\Psi_{h,\th}(\chi,\b\chi)
\cr=& \left(1+\frac{h\zeta}{\chi}\right)\left(1+\frac{\th\b\zeta}{\b\chi}\right)\left(\A(h,\th) F_h(\chi)F_{\th}(\b\chi) +\B(h,\th)F_{1-h}(\chi)F_{1-\th}(\b\chi)\right).\end{align}
Similarly
 \be\label{wexr}\Upsilon^{FB}_{h-1/2,\th}=\left(1+\frac{(1-h)\zeta}{\chi}\right) \left(1+\frac{\th\b\zeta}{\b\chi}\right)\left(\A(h,\th)F_{h}(\chi)F_\th(\b\chi) +
 \B(h,\th)F_{1-h}x(\chi)F_{1-\th}(\b\chi)\right). \ee
Inserting this in eqn.~(\ref{padsum}), we can drop the $\B$ terms if we symmetrize what multiplies the $\A$ terms under  $(h,\th)\to (1-h,1-\th)$.   Since
$k^{BB}(1-h,1-\th)=k^{FF}(h-\tfrac{1}{2},\th-\tfrac{1}{2})$, and $k^{FB}(1-h-\tfrac{1}{2},1-\th)=k^{BF}(h,\th-\tfrac{1}{2})$, the upshot is that 
  \begin{align}\label{radsum}\F=\frac{\lambda}{\pi^3} \sum_{\ell=-\infty}^\infty \int_{-\infty}^\infty \frac{ds}{2\pi}  \A(h,\th)&F_h(\chi)F_\th(\b\chi)
  \left\{-\left(1+\frac{h\zeta}{\chi}\right)\left(1+\frac{\th\b\zeta}{\b\chi}\right)\frac{k^{BB}(h,\th)}{1-k^{BB}(h,\th)}\right.\cr &+\left(1+\frac{(1-h)\zeta}{\chi}\right) \left(1+\frac{\th\b\zeta}{\b\chi}\right)\frac{k^{FB}(h-\tfrac{1}{2},\th)}{1-k^{FB}(h-\tfrac{1}{2},\th)}  \cr & 
 -\left(1+\frac{(1-h)\zeta}{\chi}\right) \left(1+\frac{(1-\th)\b\zeta}{\b\chi}\right)\frac{k^{FF}(h-\tfrac{1}{2},\th-\tfrac{1}{2})}{1-k^{FF}(h-\tfrac{1}{2},\th-\tfrac{1}{2})}\cr &+\left.
 \left(1+\frac{h\zeta}{\chi}\right) \left(1+\frac{(1-\th)\b\zeta}{\b\chi}\right)\frac{k^{BF}(h,\th-\tfrac{1}{2})}{1-k^{BF}(h,\th-\tfrac{1}{2})}   \right\}.\end{align}
 
 To demonstrate an operator product expansion, we now shift the contour in the $s$ integral in the direction of increasing $\mathrm{Im}\,h,\th$.  As usual, in this process, operators
 contributing to the OPE are associated to  poles
 that arise when one of the kernel functions $k^{BB}$, etc., is equal to 1.  The function $\A(h,\th)F_h(\chi)F_\th(\b\chi)$ also has poles, but the residues of
 those poles cancel by the mechanism
 described  in section \ref{expan} if  the quantity in braces in eqn.~(\ref{radsum}) is antisymmetric under $h\leftrightarrow 1-h$ (or equivalently
 $\th \leftrightarrow 1-\th$) when $h$ and $\th$ are half-integers. This is true because when $h$ and $\th$ are half-integers,
 \be\label{reln} k^{FB}(h-1/2,\th)=k^{BB}(1-h,\th),~~k^{FF}(h-1/2,\th -1/2)=k^{BF}(1-h,\th-1/2). \ee  This relation between eigenvalues of the kernel is
 actually a consequence of the relation (\ref{superreln}) between the eigenfunctions, just as eqn.~(\ref{norex}) implies (\ref{wadtwo}).
  (The two formulas in 
 eqn.~(\ref{reln}) are equivalent modulo eqn.~(\ref{wadtwo}).)      See section \ref{superkernel}, where the kernel functions are computed in some explicit
 models, with results consistent with eqn.~(\ref{reln}).

 As in the 1d case, the shifts in the arguments of the kernel functions by $h\to h-1/2$ or $\th\to \th-1/2$ in eqn.~(\ref{radsum}) have a simple interpretation.
 To get a contribution to the OPE that behaves as $\chi^h\b\chi^\th$ for small $\chi$, we need an operator of dimension $(h,\th)$.  This operator,
 however, may be a primary of type $BB$, or a descendant of one of the other types.  For example, an $FF$ primary of dimension $(h-1/2,\th -1/2)$ has
 a $BB$ descendant of dimension $(h,\th)$, accounting for the shift in eqn.~(\ref{radsum}).

 \section{Bosonic Models In Two Dimensions}\label{bosemod}
 
 In this section, we will consider two-dimensional theories of scalar fields $\phi_1,\dots,\phi_N$ with a potential $V(\phi_1,\dots,\phi_N)$.  Picking $V$ to
 be a homogeneous polynomial of degree $q$, we will look for a large $N$ fixed point described by a ladder sum, similar to that of the SYK model.   The considerations
 of section \ref{twod} are applicable to these models.
 
Various  instabilities might frustrate this program.  Obviously, if $V(\phi_1,\dots,\phi_N)$ is not positive-definite,
 we should not expect the model to be stable.  Even if $V$ is positive-definite, something more subtle may go wrong. In an interacting theory of scalar
 fields, a relevant mass deformation $\frac{1}{2}m^2\phi^2$ is always possible.  A fixed point, if there is one, is found by adjusting the coefficient $m^2$ (and possibly the coefficients of other relevant operators).  However,
 as a function of $m^2$, the theory may undergo a first-order phase transition and so there may be no fixed point. 
 
 In section \ref{firsttry}, we first consider a simple model in which $V$ is not positive-definite. In section \ref{secondtry} we consider an alternative with positive-definite $V$. In both cases we run into difficulties that will be resolved in the supersymmetric model studied in the next section.
 
 \subsection{A Naive Model}\label{firsttry}
 
 \subsubsection{Schwinger-Dyson Equations}
 
 We consider a bosonic model with $N$ scalar fields and a random potential of degree $q$.  The Euclidean action is\footnote{Here $\x=(\x_1,\x_2)$ is a Euclidean two-vector
 that we will  write in terms of a complex coordinate $x= \x_1+i\x_2$.   The measure for integration in two dimensions will be  $d^2\x=d\x_1 d\x_2$. Similarly in
 momentum space $\p=(\p_1,\p_2)$ is a Euclidean two-vector, and we define $p=\p_1+i\p_2$, so that $\p\cdot \x =\frac{1}{2}(p\b x+\b p x)$.}
\be\label{boseunstable}
I = \int_{\R^2} d^2 \x\left[\frac{1}{2}(\partial \phi^i)^2 + \sum_{i_1\cdots i_q}J_{i_1\cdots i_q}\phi^{i_1}\cdots\phi^{i_q}\right].
\ee
A tensor version of this model was considered for $q = 4$ in \cite{KT}. The couplings $J_{i_1\dots i_q}$ are  independent Gaussian variables with mean zero and variance $\langle J_{i_1\cdots i_q}^2\rangle = \frac{J^2}{q N^{q-1}}$.
In the large $N$ limit, the Schwinger-Dyson equations for the propagator $G$ and self-energy $\Sigma$ read, after averaging over disorder,
\be\label{sdequations}
\Sigma = J^2 G^{q-1}, \hspace{20pt} G = \frac{1}{-\partial^2 - \Sigma}.
\ee
(Diagrammatically, $\Sigma=J^2G^{q-1}$ is represented by  fig. \ref{argument} below, in which the lines represent exact propagators.)  
These equations have to be treated with care; in the first equation, $G^{q-1}$ is defined by pointwise multiplication in position space, while in the
second equation, $1/(\partial^2+\Sigma)$ is the inverse of $\partial^2+\Sigma$ as a matrix.  After diagonalizing the matrix $\partial^2+\Sigma$ by going to momentum
space, it can be inverted pointwise, $G(\p) = 1/(\p^2-\Sigma(\p))$.

Generically, the action (\ref{boseunstable}) will not be bounded from below, and we are not guaranteed to find a reasonable solution to the Schwinger-Dyson equations. In fact, it seems that there is no reasonable exact solution. For example, for $q = 2$, one finds the momentum space answer
\be
G(\p) = \frac{\p^2 - \sqrt{\p^4-4J^2}}{2J^2} \hspace{20pt} (q=2),
\ee
which is complex for $\p^2<2J$. For larger $q$, we have attempted to solve these equations by numerical iteration, and we found that the iteration either did not converge (in $d = 1$) or converged to a position-space correlator that was negative in places (in $d = 2$), violating reflection positivity. However, we will proceed formally and, as in the SYK model, drop the $\partial^2$ term in (\ref{sdequations}). Then there is a solution
\be\label{solution}
G = \frac{b}{|x|^{2\Delta}}, \hspace{20pt} \Delta = \frac{2}{q}, \hspace{20pt} b^qJ^2 = \frac{(1 - \Delta)^2}{\pi^2}.
\ee
In deriving this equation, the integral (\ref{lotag}) is helpful.

\subsubsection{The Ladder Kernel}\label{ladderk}
The usual ladder kernel, constructed from Feynman diagrams as in fig. \ref{kernel} of section \ref{kkernel}, is in this model
\be
K(x_1,x_2;x_3,x_4) = \frac{(q-1)(1-\Delta)^2}{\pi^2}\frac{1}{|x_{13}|^{2\Delta}|x_{24}|^{2\Delta}|x_{34}|^{4-4\Delta}}.
\ee
As in the SYK model, conformal invariance implies that the kernel has eigenfunctions given by a  conformal  three-point function $\la\Phi(x_3)\Phi(x_4)V(y)\ra$, where $V(y)$ may have any dimension $h,\th$ and the eigenvalue of $K$ will not depend on $y$.  Taking $y\to\infty$, we
can consider the eigenfunction 
\be\label{tref}S_{h,\th}=\frac {x_{34}^h\bar x_{34}^\th}{|x_{34}|^{2\Delta}}.\ee
The eigenvalue of $K$ acting on this eigenfunction can be computed as described in 1d in section \ref{kkernel}, with the result
\begin{align}\label{zolt}
k(h,\t{h}) &= \frac{(q-1)(1-\Delta)^2}{\pi^2}\int d^2\x d^2\x' \frac{1}{|1-x|^{2\Delta}|x'|^{2\Delta}|x-x'|^{4-4\Delta}}\frac{(x-x')^h(\bar{x}-\bar{x}')^{\t{h}}}{|x-x'|^{2\Delta}}.
\end{align}
This integral can be evaluated by the change of variables $x'=xy$ followed by use of eqn. \ref{KLT}:
\begin{align}\label{bosker}k(h,\t{h})&=\frac{(q-1)(1-\Delta)^2}{\pi^2}\left(\int\frac{x^h\bar{x}^{\t{h}}d^2\x}{|x|^{2}|1-x|^{2\Delta}}\right)\left(\int \frac{(1-y)^h(1-\bar{y})^{\t{h}}d^2\y}{|y|^{2\Delta}|1-y|^{4-2\Delta}}\right)\notag\\
&=\Delta(2-\Delta)\frac{\Gamma(2-\Delta)^2}{\Gamma(1+\Delta)^2}\frac{\Gamma(-1+h+\Delta)}{\Gamma(1+h-\Delta)}\frac{\Gamma(-\t{h}+\Delta)}{\Gamma(2-\t{h}-\Delta)}.
\end{align}

  \begin{figure}[ht]
 \begin{center}
   \includegraphics[width=3.2in]{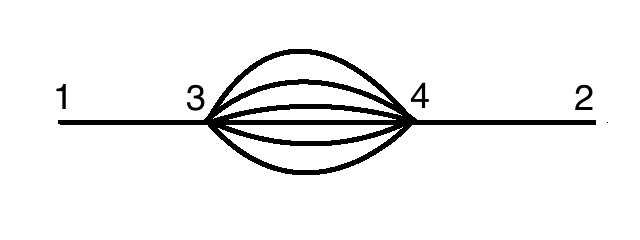}
 \end{center}
\caption{\small Gluing in one more propagator connecting vertices 3 and 4 in fig. \ref{kernel}, we arrive at this diagram, with $q-1$ propagators connecting
points 3 and 4.  It is essentially the
same diagram that one encounters in the Schwinger-Dyson equation that determines the two-point function of the model for large $N$.  \label{argument}}
\end{figure}  
    
This formula agrees with that of \cite{KT} for the case of $h = \th$ and $\Delta = 1/2$. As a check on this formula, we find that $k(0,0)=-( q-1)$.  This holds because of an argument   described in section 3.2.3 of \cite{Maldacena:2016hyu}
 that we will describe briefly. 
 For $h=\th=0$, the wavefunction $S_{h,\th}$ is, up to a factor of $b$, the same as the propagator $G$ of eqn.~(\ref{solution}), so
 what we want to prove is that $K G=-(q-1)G$.   To calculate $KG$, we have to glue in one more propagator connecting the vertices labeled 3 and 4 in fig. \ref{kernel}.  The
 resulting Feynman diagram is shown in fig. \ref{argument}.  Apart from a factor of $q-1$ that arises because one of the $q-1$ propagators between points 3 and 4
 (the one that was just glued in) is distinguished, 
 this diagram appears in constructing the Schwinger-Dyson equation (\ref{sdequations}); if the external lines are amputated, this diagram  
 represents  the self-energy $\Sigma=J^2 G^{q-1}$.  Because of the factor of $q-1$ and the fact that in fig. \ref{argument}, the external lines are not
 amputated, the diagram actually represents $(q-1)G\Sigma G=-(q-1)G$, confirming that $KG=-(q-1)G$.  The same reasoning applies 
 in all SYK analog models considered in this paper.

We also see that $k(2,0) = 1=k(0,2)$, indicating the  existence in the spectrum of a holomorphic stress tensor, as one would expect
in a  2d conformal field theory. As in the original SYK model, existence of this operator
 is a consequence of the reparametrization-invariance of the low energy Schwinger-Dyson equations.
We return to this point in section \ref{stress}.

In addition to the stress tensor, there are other states corresponding to other solutions to the equation $k = 1$. For generic values of $\Delta$, 
the situation is as follows: for each value of the spin $J = h - \t{h}$ we have infinitely many solutions, corresponding to different energies $E = h + \t{h}$. 
However, for the special case of $q = 4$ so that $\Delta = \frac{1}{2}$, the kernel has the very simple expression
\be \label{halfk}
k = \frac{3}{-1 + 2(h + \t{h}) - 4 h\t{h}}, \hspace{20pt} (q = 4).
\ee
For this case there is a single state corresponding to each spin $J\ge 2$, with dimension
\be\label{soldelta}
E = 1 +\sqrt{J^2 - 3}.
\ee
This formula, however, implies  a drawback that was pointed out by I. Klebanov, see \cite{GKT}.   If we set $J=0$, eqn.~(\ref{soldelta}) gives a solution with complex
dimension $E=1\pm \sqrt{-3}$, contradicting unitarity of the theory.  This solution is on the integration contour $\Re\,(h+\th)=1$ of the integral (\ref{nadsum})
for the four-point function, and cannot be ignored.   

To decide if a similar phenomenon happens for $q>4$, we can proceed as follows.  Eqn.~(\ref{bosker}) shows that 
$k(h,\th)$ is positive on the line $h=\th=1/2+is$.  It also vanishes at infinity,
so $k(1/2+is,1/2+is)$ will have to pass through 1  for some value of $s$ if $k(1/2,1/2)>1$.   From eqn.~(\ref{halfk}), $k(1/2,1/2)>1$
 for $q=4$ and therefore the same is true  for $q$ near 4.   But for sufficiently large 
$q$, the factor of $\Delta=1/2q$ in eqn.~(\ref{bosker}) ensures that $k(1/2,1/2)<1$.  Numerically, we find that for $q\geq 65$, there is no solution to $k=1$ on the
line $h=\th=1/2+is$.   At a critical value of $q$ close to 65, a complex conjugate pair of solutions of $k=1$ meet at $h=\th=1$ and then become real.  It is conceivable
that the model is sensible for $q\geq 65$, though this would sound rather strange, especially since we have made no effort to make the potential positive-definite. 


Something similar  happens in the AdS/CFT correspondence if in AdS space,
one assumes the existence of a spin zero field with a mass squared so negative as to violate the Breitenlohner-Freedman bound.  The dual CFT then
formally has an operator of complex dimension.

We should also ask about the convergence of the integral (\ref{bosker}) by which we computed $k(h,\th)$.  The ladder diagrams that we are trying
to sum are convergent for  $q<4$, but for $q\geq 4$, if we evaluate them using the IR propagator (\ref{solution}), they are UV divergent.  (A variant of this
remark was made by K. Bulycheva.)  Related to this, on the principal series $\Re\,(h+\th)=1$ (which are the values we need in order to use the kernel
in eqn.~(\ref{nadsum}) for the four-point function), the integral (\ref{zolt}) is divergent at $x=x'$ if $q\geq 4$.  
It has effectively been defined by analytic continuation (in $q$ or in  $h,\th$) to arrive at the above formulas.  In the 1d SYK model, 
the ladder integrals (computed with propagators of the IR theory) similarly look
 divergent by power counting, but there actually is no problem because the divergence would multiply an operator that vanishes
by fermi statistics.  In the bosonic model considered here, there is no such rescue.  For $q>4$, the formula  (\ref{bosker}) for the kernel may be physically justifiable in a model with counterterms appropriately tuned to cancel the
divergences.  For $q=4$, however, 
 the expression for the ladder diagrams that we get from the kernel does not make sense, because $k$ has a double pole at $h = \t{h} = 1/2$, which is on the original integration contour. This does not  lead to a divergence in the zero-rung ladder, because of the factors of $(2h-1)(2\t{h}-1)$ from the norm of the wave 
 functions. But it does lead to a divergence in the one-rung and higher terms, where we have two or more factors of $k$.  The integral (\ref{nadsum}) for the four-point function
 still makes sense, because $k/(1-k)\to -1$ for $k\to \infty$, but it is not clear how this should be interpreted.

\subsubsection{Normalization Of Four-Point Function And Central Charge}\label{normF}
We would also like to work out a more complete expression for the four-point function, including the normalization coefficient $\lambda$ in (\ref{nadsum}). To determine this coefficient, we follow the steps described near the end of section \ref{kkernel}. In this case, we do not have any minus signs coming from fermi statistics, so the zero-rung ladder is given simply by
\be\label{f0bose}
\mathcal{F}_0(\chi,\overline{\chi}) = \frac{G(1,3)G(2,4) + G(1,4)G(2,3)}{G(1,2)G(3,4)} = |\chi|^{2\Delta} + \left|\frac{\chi}{1-\chi}\right|^{2\Delta}.
\ee
The inner product of the first term with a basis function is
\begin{align}
\left(\Psi_{h,\t h},|\chi|^{2\Delta}\right) &= \int \frac{d^2\chi d^2y}{|\chi|^4}\frac{|\chi|^{2\Delta}\chi^h\bar{\chi}^{\t h}}{y^h\bar{y}^{\t h}(y-\chi)^h(\bar{y} - \bar{\chi})^{\t h}(1-y)^{1-h}(1-\bar{y})^{1-\t h}}\\
&=\left(\int d^2x \frac{x^{\Delta+h-2}\bar{x}^{\Delta+\t h-2}}{(1-x)^h(1-\bar{x})^{\t h}}\right)\left(\int d^2 y \frac{y^{\Delta-h-1}\bar{y}^{\Delta-\t h-1}}{(1-y)^{1-h}(1-\bar{y})^{1-\t h}}\right).\label{actualans}
\end{align}
In the second line, we defined $x = \chi/y$. These integrals can be evaluated in terms of gamma functions (see Appendix \ref{eval}). For integer $h-\t h$, the answer is a multiple of the kernel (\ref{bosker});
this fact  can be understood along the lines explained in section \ref{kkernel}. It is straightforward to show that the inner product with the second term in (\ref{f0bose}) is given by $(-1)^\ell$ times the inner product with the first term. Putting in the actual answer for the integral (\ref{actualans}) and combining terms, we get
\be\label{actualans2}
\left( \Psi_{h,\widetilde{h}},\mathcal{F}_0\right)= k(h,\widetilde{h})\frac{\pi^2\Delta}{(2{-}\Delta)(1{-}\Delta)^2}\left(1 + (-1)^\ell\right).
\ee
We can now formally write a formula for the four-point function including the numerical coefficient in (\ref{nadsum}) as\footnote{Since this is a formal expression, we will suppress a subtlety regarding the integration contour that is analyzed for the supersymmetric model in section \ref{innerpr}.}
\be\label{withcoeff}
\F=\frac{2\Delta}{\pi(2{-}\Delta)(1{-}\Delta)^2} \sum_{\ell=\text{ even}}\int_{-\infty}^\infty \frac{ds}{2\pi} (\ell^2+4s^2)\A(s,\ell)F_{s,\ell}(\chi,\bar\chi) \frac{k(\ell,s)}{1-k(\ell,s)}.
\ee
As before, the variables $h,\widetilde{h}$ and $\ell,s$ are being used interchangeably, with the understanding that in all cases
\be
h = \frac{1+\ell}{2}+is, \hspace{20pt} \widetilde{h} = \frac{1-\ell}{2}+is.
\ee

One point of writing out this expression in detail is that we can check that the residues are positive for the operator contributions from poles of $\frac{k}{1-k}$ that arise when we shift the $s$ contour. We have not checked this sytematically, but in all cases that we looked at this was the case. Sometimes it is little bit nontrivial, relying on sign changes in the $A(s,\ell)$ factor. So in this sense, at least, most operators of the theory make contributions consistent with unitarity.

We can also compute the central charge $c$, by evaluating the contribution of the stress tensor. To do this, one finds the residue of the pole at $h = 2,\widetilde{h} = 0$, and compares to the expected contribution of the stress tensor. See Appendix \ref{app:c} for details. This leads to the formula
\be
c = (1-\Delta)^3N = \left(1 - \frac{2}{q}\right)^3 N.
\ee
We will compute and analyze a similar formula in the better defined supersymmetric model in section \ref{fpf} below.

 \subsection{Another Model}\label{secondtry}
Searching for a bosonic model that might be stable, it is natural to assume that the potential is the square of a random function.  For example,
we consider the interaction
\be\label{morestable} 
\sum_{a=1}^M\Bigg(\sum_{i_1,\cdots, i_{q/2}=1}^NC_{i_1i_2\cdots i_{q/2}}^a\phi_{i_1}\cdots \phi_{i_{q/2}}\Bigg)^2 \sim \frac{1}{4}(B^a)^2 + i C_{i_1\cdots i_{q/2}}^a\phi_{i_1}
\cdots \phi_{i_{q/2}} B^a,
\ee
where integrating out $B$ on the right hand side reproduces the left hand side.   As in the previous model, there are $N$ scalar fields $\phi_i$,  and now the index $a$ takes $M$
values.  We will take $N,M$ large with fixed ratio $r=M/N$.
The $C$'s are independent Gaussian random variables, with zero mean and variance of an individual element given by $\langle (C_{i_1\cdots i_{\frac{q}{2}}}^a)^2\rangle = J^2/N^{\frac{q}{2}}$.

\subsubsection{Schwinger-Dyson Equations}

 Let us begin by analyzing the disorder-averaged
  two-point functions $G_\phi=\la \phi\phi\ra$, $G_B=\la B B\ra$ for this model in $d$ dimensions. We can get the Schwinger-Dyson (SD) equations either by studying the Feynman diagrams or by finding the disorder-averaged effective action.  The Feynman diagram route is quickest, and (from the diagram of fig. \ref{argument},
  but now with an internal or external $B$ line) we find the equations
\begin{align}\label{sdreal}\notag
G_\phi = \frac{1}{-\partial^2 - \Sigma_\phi}, \hspace{20pt} G_B = \frac{1}{\frac{1}{2} - \Sigma_B}\\
\Sigma_\phi = -\frac{qr}{2}J^2G_\phi^{\frac{q}{2}-1}G_B, \hspace{20pt} \Sigma_B = - J^2 G_{\phi}^\frac{q}{2}.
\end{align}
Since the model now has a stable potential, we would expect these equations to have reflection positive solutions. And indeed, one can check that they do, by solving the equations exactly for $q = 2$ and numerically for larger values of $q$.

Let us proceed naively and drop the free terms $\frac{1}{2}$ and $-\partial^2$ from the above equations. Then they become soluble via an ansatz:
\be\label{betterboseansatz}
G_\phi = \frac{b_\phi}{|\x|^{2\Delta}}, \hspace{20pt} G_{B} = \frac{b_B}{|\x|^{2\Delta_B}},
\ee
where $\x$ is a Euclidean $d$-vector. To determine $\Sigma_\phi$ and $\Sigma_B$, one Fourier transforms these expressions 
using eqn.~(\ref{fourierd}) of Appendix \ref{KLT}.
In momentum space, one has
simply $\Sigma_\phi=-1/G_\phi$, $\Sigma_B=-1/G_B$.  Then $\Sigma_\phi$ and $\Sigma_B$ can be Fourier transformed back to coordinate space. The Schwinger-Dyson equations (\ref{sdreal})  imply
\begin{align}\label{sdanswer}
 \Delta_B = d - \frac{q}{2}\Delta, \hspace{20pt} 
\frac{\c_d(d - \Delta)\c_d(\Delta)}{\c_d(q\Delta/2)\c_d(d-q\Delta/2)} = \frac{2}{qr},\hspace{20pt}
J^2 b_B b_\phi^{\frac{q}{2}} = \frac{1}{\c_d(q\Delta/2)\c_d(d - q\Delta/2)}.
\end{align}
Here $\c_d$ is a ratio of gamma functions defined in (\ref{fourierd}). These equations do not determine the solutions completely, because they do not fix $b_\phi$ and $b_B$ independently, only the product $b_\phi^\frac{q}{2}b_B$. Fortunately, it will turn out that this product is the only thing that is important for determining the four-point function. The equations simplify somewhat in $d = 2$, where we find the solution
\be\label{simp1}
\Delta = \frac{8 - 2 q^2 r + 2\sqrt{2qr}(q-2)}{8-rq^3}, \hspace{20pt} J^2 b_B b_\phi^{\frac{q}{2}} = -\frac{(q\Delta-2)^2}{4\pi^2}.
\ee
The last formula is consistent with $b_\phi>0$, $b_B<0$.  This is compatible with unitarity, since $\phi$ is hermitian and $B\sim i \phi^{q/2}$ is antihermitian.

  \begin{figure}[t]
 \begin{center}
   \includegraphics[width=3.2in]{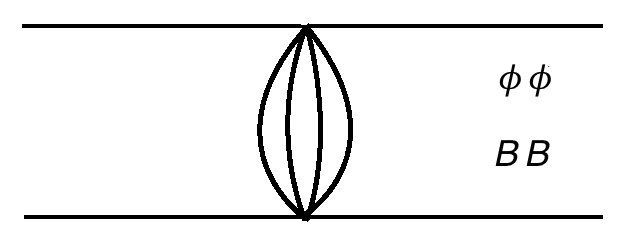}
 \end{center}
\caption{\small Propagation of $\phi\phi$ or $BB$ in a ladder diagram. \label{TwoLadder}}
\end{figure}  

\subsubsection{The Ladder Kernel And Four-Point Function}
Let us continue naively and compute the ladder kernel, assuming for the moment that the IR ansatz is valid. Since we have two types of field in the model, $\phi$ and $B$, the ladder kernel is more complicated than in the previous bosonic model. The most interesting channel is the one in which $\phi\phi$ or $BB$ propagate along the ladder (fig. \ref{TwoLadder}). For a given set of particle coordinates, the kernel acting in this space is a $2\times 2$ matrix. Multiplying the appropriate propagators together and collecting factors, we find 
\be
{\bf K}(x_1,x_2;x_3,x_4) = -\frac{\Gamma(d-\Delta)\Gamma(\Delta)}{\pi^d\Gamma(\frac{d}{2}-\Delta)\Gamma(\Delta-\frac{d}{2})}\left( \begin{array}{cc}
\frac{q/2-1}{|x_{13}|^{2\Delta}|x_{24}|^{2\Delta}|x_{34}|^{2(d-2\Delta)}}& \frac{b_\phi/(b_B\sqrt{r})}{|x_{13}|^{2\Delta}|x_{24}|^{2\Delta}|x_{34}|^{(\frac{q}{2}-1)2\Delta}} \\
\frac{b_B/(b_\phi\sqrt{r})}{|x_{13}|^{2\Delta_B}|x_{24}|^{2\Delta_B}|x_{34}|^{(\frac{q}{2}-1)2\Delta}}& 0  \end{array} \right).
\ee
The top left entry represents $\phi\phi\rightarrow \phi\phi$, the top right represents $BB\rightarrow \phi\phi$, etc. There is a minus sign from two factors of $i$ in the interaction (\ref{morestable}), but the overall sign of the kernel is actually positive, because the ratio of gamma functions is negative. This particular ratio of gamma functions comes from using (\ref{sdanswer}) to simplify the prefactor $\frac{rq}{2}J^2 b_\phi^{\frac{q}{2}}b_B$ that multiplies the kernel.

To diagonalize ${\mathbf K}$, we are supposed to act with it on conformal vectors $(v_{\phi\phi},v_{BB})$ with the appropriate external weights. At this point we specialize to $d = 2$, although the eigenvalues of the kernel can be worked out in general dimensions using (\ref{symmtrac}). In $d = 2$ the conformal vectors are
\be
v_{\phi\phi} = \frac{x_{34}^h\bar{x}_{34}^{\t{h}}}{(x_{30}x_{40})^h (\bar{x}_{30}\bar{x}_{40})^{\t{h}}}\frac{b_\phi}{|x_{34}|^{2\Delta}}, \hspace{20pt} v_{BB} = \frac{x_{34}^h\bar{x}_{34}^{\t{h}}}{(x_{30}x_{40})^h (\bar{x}_{30}\bar{x}_{40})^{\t{h}}}\frac{b_B}{|x_{34}|^{2\Delta_B}}.
\ee
Each matrix element of the kernel simplifies to the type of expression we evaluated before in the naive bosonic model. To write the answer, let us define the function
\begin{align}
I(\Delta,h,\t{h}) &
=\pi^2\frac{\Gamma(1-\Delta)^2}{\Gamma(\Delta)^2}\frac{\Gamma(h+\Delta-1)}{\Gamma(h-\Delta+1)}\frac{\Gamma(-\t{h}+\Delta)}{\Gamma(2-\t{h}-\Delta)}.
\end{align}
Then one finds that the kernel ${\bf K}$ acts on the functions $(v_{\phi\phi},v_{BB})$ as matrix multiplication by
\be \label{longkernel}
{\bf k}(h,\t{h}) = \frac{\pi^2}{(1-\Delta)^2}\left( \begin{array}{cc}
(\frac{q}{2}-1)I(\Delta,h,\t{h}) & \frac{1}{\sqrt{r}}I(\Delta,h,\t{h}) \\
\frac{1}{\sqrt{r}}I(\Delta_B,h,\t{h}) & 0  \end{array} \right).
\ee
Let us briefly summarize the meaning of this formula. The model (\ref{morestable}) has two parameters, $q$ and $r$. We find $\Delta,\Delta_B$ from (\ref{simp1}) and (\ref{sdanswer}). Then we plug these values into eqn.~(\ref{longkernel})  and find the eigenvalues of this two-by-two matrix to determine the eigenvalues of the conformal ladder kernel ${\bf K}$.

There are some simple checks of (\ref{longkernel}). The upper left entry at $h=\th=0$ is $-(\frac{q}{2}-1)$, as one would predict by adapting an argument explained in section \ref{ladderk}. Also, one of the eigenvalues is equal to one for $h = 2,\t{h} = 0$, corresponding to the existence of a stress tensor.
After repeating the steps of Appendix \ref{app:c}, one can evaluate the contribution of the stress tensor and find the central charge
\be
c = (1-\Delta)^2\left(1-\frac{2}{q}\right)N.
\ee
Here $\Delta$ is determined by $q,r$ using (\ref{simp1}). We will not dwell on this formula because a more urgent question is whether or not it was reasonable to proceed with the IR ansatz we made above.

\subsubsection{Does The Model Actually Flow To The IR Ansatz?}
Ultimately, we do not have a satisfactory answer to the question of whether the scaling form (\ref{betterboseansatz}) is actually realized in this theory. One basic condition is that the ansatz should be self-consistent, in the sense that the self-energies should dominate over the free terms for small momentum. This means that $\Sigma_\phi$ should vanish more slowly than $|p|^2$ for small $p$, and $\Sigma_B$ should grow for small $p$, leading to
\be\label{condforflow}
d-2\Delta  < 2, \hspace{20pt} q\Delta - d < 0.
\ee
These inequalities give us some basic constraints. By combining them, we find that $d < \frac{2q}{q-2}$. Nontrivial models have $q\geq 4$, which implies $d<4$. Because $q$ is even, $d=3$ is possible only if we take $q=4$. For $d = 2$, we could have solutions for any value of $q$, provided $r > \frac{2}{q}$. For $d = 1$, we could have solutions for any value of $q$, provided $r > 1$.

However, this is not the end of the story. An important point is that within the IR ansatz, the self-energy $\Sigma_\phi$ is UV divergent. Concretely, using (\ref{sdanswer}), one finds that the self energy $\Sigma_\phi$ is proportional to $1/|x|^{2(d-\Delta)}$, and using the second inequality in (\ref{condforflow}), we find that for $q>2$ the integral over $x$ to give the fourier transform $\Sigma_\phi(\p)$ will diverge at $x = 0$. This divergence has been implicitly subtracted above, where we define the integrals by analytic continuation in $\Delta$ or $d$. Now, if we imagine using the exact propagators, there is no divergence because the short-distance behavior crosses over to the free theory, where $\Sigma_\phi$ behaves like a power of a logarithm for small $|x|$. However, if we optimistically imagine that the solution is close to the scaling answer for $|x|\ll J^{-1}$, then the divergence gets cut off at a scale $J^{-1}$, giving a mass of order $J$ to $\phi$. To have a hope of reaching the scaling solution, we would have to try to cancel this by tuning a negative mass-squared counterterm.

\begin{figure}
\begin{center}
\includegraphics[width=0.4\textwidth]{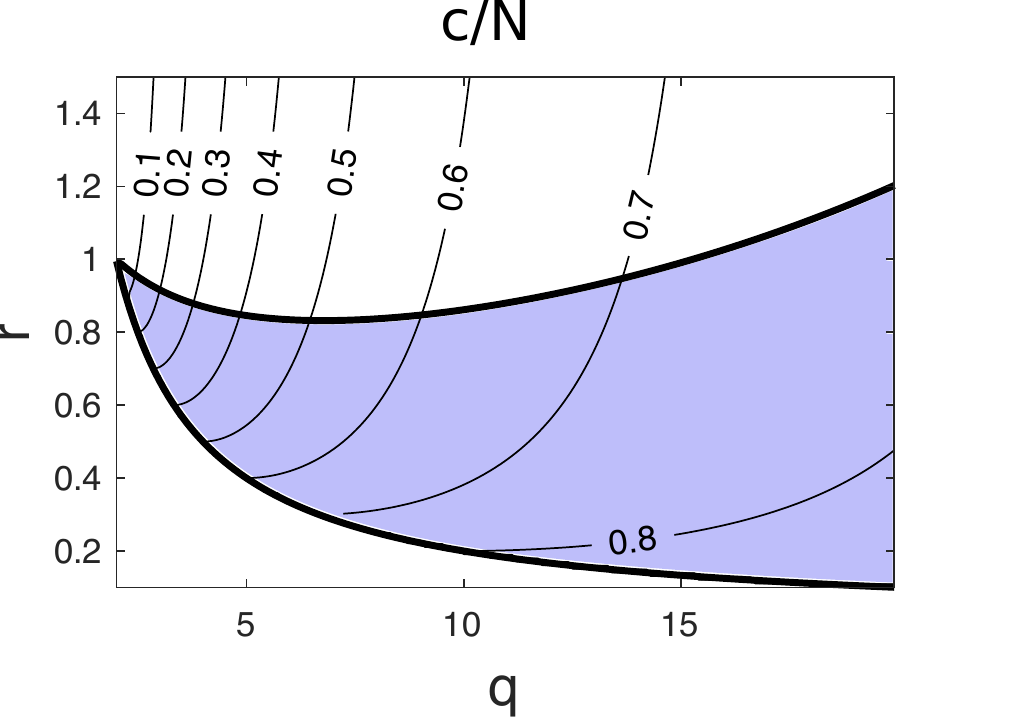}
\includegraphics[width=0.4\textwidth]{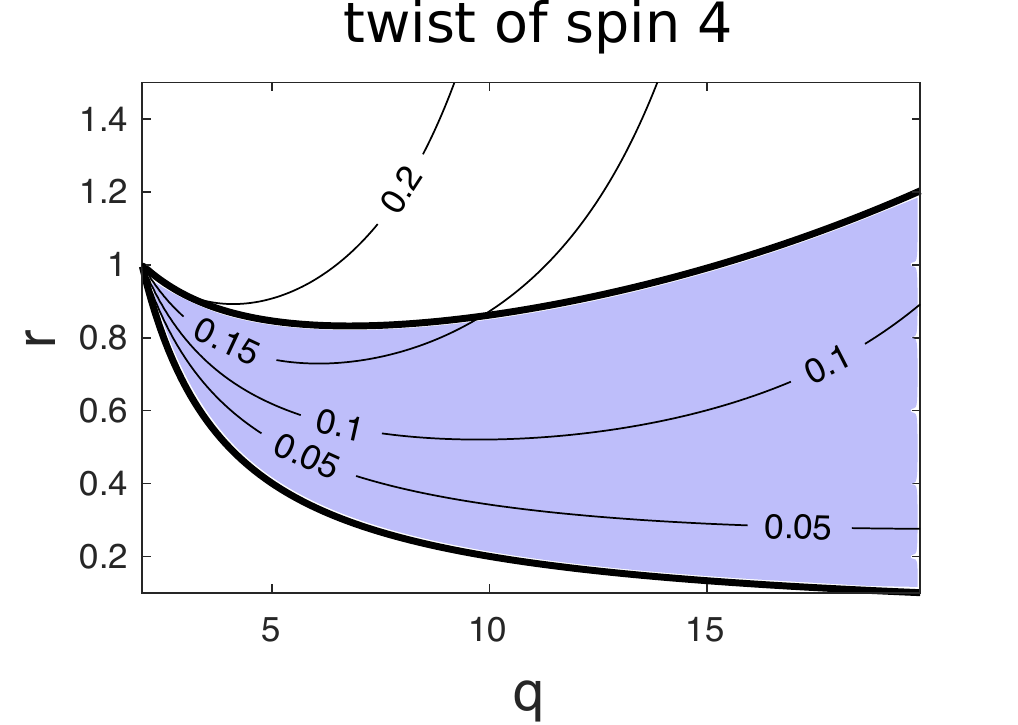}
\caption{\small{The $(q,r)$ parameter space of the $d = 2$ model (\ref{morestable}). We draw contours of the central charge per degree of freedom $c/N$ and the twist $2\widetilde{h}$ of the lightest spin-four operator. At the very least, the region outside the shaded region is disallowed: the lower boundary is the constraint (\ref{condforflow}) and the upper boundary is the constraint that the eigenvalues of ${\bf k}(1/2,1/2)$ are less than one. It is not clear if the model (\ref{morestable}) can be made to flow to desired
IR fixed point  even within the shaded region.}}\label{allowed}
\end{center}
\end{figure}
There is no guarantee that if we do this, we will actually reach the critical solution. Of couse, one necessary condition is that the putative IR theory should be well-defined. This requires that the eigenvalues of ${\bf k}$ be less than one for the principal series representations. In practice, this reduces to checking that the maximum eigenvalue of ${\bf k}(1/2,1/2)$ is less than one. This is not true for all values of $q,r$, but in figure \ref{allowed}, we plot a sizeable region of $q,r$ parameter space such that this holds, and also such that the inequalities (\ref{condforflow}) hold. The complement of this region is ruled out, but the region inside is unclear.

To determine the fate of the shaded region, the correct procedure would be to numerically solve the exact Schwinger-Dyson equations and try to find the critical solution by tuning a mass counterterm (and possibly other coefficients). In numerical experiments, we have not succeeded in this. One possible conclusion is that there is a first order transition and the model does not flow to the critical solution, even within the shaded region of parameter space in figure \ref{allowed}. However, we are not confident in this conclusion, and we hope the question will be studied more systematically.

In a special case (with $r=1$ and $C_{i_1\cdots i_{q/2}}^a$ completely symmetric in all its indices), the potential energy that we have assumed is the bosonic potential of a supersymmetric model.  The difficulties that we have encountered give a motivation to study the supersymmetric model in section \ref{supermod}.
 
 \subsection{The Stress Tensor}\label{stress}
 
 Finally, we will explain why these models satisfy $k(2,0)=1$, corresponding to the existence of a holomorphic stress tensor.  The mechanism is the same as in the 1d
 SYK model.
 
 First let us explain why any infinitesimal variation of the two-point function that preserves the low energy Schwinger-Dyson equations corresponds to an eigenfunction of the kernel
 with eigenvalue 1.  We write the Schwinger-Dyson equations in the IR limit
 as $\Sigma\star G=-1$, $\Sigma=J^2 G^{q-1}$, where $\star$ represents matrix multiplication, and in the absence of a star, pointwise
 multiplication in position space is intended.  We thus have $\Sigma=-G^{-1}$ (as a matrix) and therefore under an arbitrary first order deformation
 of $G$, we get 
 \be\label{miff}\delta\Sigma=-\delta(G^{-1})=G^{-1}\star \delta G\star G^{-1}.\ee
A given deformation of $G$ preserves the condition $\Sigma=J^2 G^{q-1}$ if and only if $\delta\Sigma=(q-1)J^2G^{q-2}\delta G$, or, using (\ref{miff}),
\be\label{iff}\delta G= (q-1)J^2 G\star G^{q-2}\delta G\star G . \ee
But this is equivalent to the statement that $\delta G$ is an eigenfunction of $K$ with eigenvalue 1.

Now we construct deformations of $G$  that preserve the IR Schwinger-Dyson equations.  Those equations have a large symmetry group that the
full UV equations lack.  In going to the IR limit, we dropped the ordinary kinetic energy and kept only the potential energy.  The potential energy was
a homogeneous polynomial in $\phi$ of degree $q$, schematically
\be\label{potential} V=\int d^2x \,\phi^q. \ee
 More precise details of the two models are not relevant at the moment.  Such a potential has the symmetry
\be\label{oddsym}\delta \phi= a(x,\b x)\partial_x \phi +\frac{1}{q}\partial_x a(x,\b x)\,\phi,\ee
for any function $a(x,\b x)$, not necessarily holomorphic or antiholomorphic.  These symmetries, including also the corresponding symmetry with $\partial_x $ replaced by
$\partial_{\b x}$, simply generate the full diffeomorphism group in two dimensions.  The corresponding deformation of $G$ is
\be\label{defg}\delta G(x,y)=\left(a(x,\b x)\partial_x +a(y,\b y)\partial_y+\frac{1}{q}\left(\partial_x a(x,\b x)+\partial_y a(y,\b y)\right)\right)G(x,y). \ee
So this is, for any $a(x,\b x)$, an eigenfunction of the kernel with eigenvalue 1.  

We set $a(x,\b  x)=-q/(x-z)$, for some $z$, and compute
\begin{align}\label{efg}\delta G(x,y) &= G(x,y)\left(\frac{1}{(x-z)^2}+\frac{1}{(y-z)^2}+\frac{2}{x-y}\left(\frac{1}{x-z}-\frac{1}{y-z}\right)\right)\cr&=G(x,y)\frac{(x-y)^2}{(x-z)^2(y-z)^2}.
\end{align}
But with $G(x,y)\sim 1/|x-y|^{2\Delta}$ as in eqn.~(\ref{solution}), this is the conformal three-point function $\langle\Phi(x)\Phi(y)\V(z)\rangle$
where $\V(z)$ has dimension $(2,0)$.  Thus we have learned that $k(2,0)=1$.

 \section{Supersymmetric Models In Two Dimensions}\label{supermod}
 
 \subsection{Random Superpotential}\label{rs}

We ran into difficulty with both of the bosonic models considered in section \ref{bosemod}. In looking for simple but stable SYK-like models, it is natural to consider a supersymmetric theory.  We will consider a two-dimensional model with $\N=1$ supersymmetry, which can be formulated in a superspace with two bosonic
 and two fermionic coordinates.  We write the bosonic coordinates as a real two-vector $\x=(\x_1,\x_2)$ or a complex variable $x=\x_1+i\x_2$.  For the fermionic
 coordinates, we will use a complex basis of coordinates $\theta, \,\b\theta$ of positive or negative chirality, with  the measure $\int d^2\theta\,\b\theta\theta=1$.

A scalar superfield in this space takes the form
 \be\label{scalsup}\uphi(\x,\theta,\b\theta)=\phi(x)+i\theta\psi(x)+i\b\theta\, \b\psi(x)+\theta\b\theta F(x). \ee
 As usual,  $\phi$ is a real scalar field, $\psi$ and $\b\psi$ are chiral fermions of opposite chirality, and $F$ is an auxiliary field.   
 The supersymmetry generators are
\be\label{supergen}Q=\partial_\theta-\theta\partial_x, ~~\bar Q=\partial_{\b\theta}-\b\theta\partial_{\b x}. \ee  
 Supersymmetric models are constructed with the help of the superspace derivatives
 \be\label{supder}D_\theta=\partial_\theta+\theta\partial_x,~~~ D_{\b\theta}=\partial_{\b\theta}+\b\theta\partial_{\b x}, \ee
which anticommute with $Q$ and $\b Q$.   We sometimes write $\X$ for the whole set of coordinates $\x, \theta,\b\theta$.

 We introduce $N$ scalar superfields $\uphi_i$, $i=1,\dots , N$, with a standard kinetic energy and a superpotential that will be a homogeneous polynomial
 of some degree $\h q$. The action is\footnote{The $i$ in the second term ensures that (as in eqn.~(\ref{morestable}))
 after integrating out auxiliary fields the bosonic potential is positive-definite.   This is related by supersymmetry to the fact that the Yukawa couplings are hermitian in Lorentz
 signature. Our normalization of the kinetic energy differs from the standard one by a factor of 4.}
  \be\label{superaction}I=\int \!\!d^2\x\, d^2\theta \left(\frac{1}{2}\sum_i D_{\b\theta}\uphi_i D_\theta\uphi_i +i\!\!\!\sum_{i_1i_2\cdots i_{\h q}}C_{i_1i_2\cdots i_{\h q}}\uphi_{i_1}
 \uphi_{i_2}\cdots \uphi_{i_{\h q}}  \right) .\ee
 Here the $C_{i_1i_2\cdots i_{\h q}}$ are independent Gaussian variables with  mean zero and variance $\la C_{i_1i_2\cdots i_q}^2\rangle =J^2/(\h q N^{\h q-1})$.

 For two particles with holomorphic coordinates $x_1,\theta_1$ and $x_2,\theta_2$,  the supersymmetric invariants are $\la 1,2\ra=x_{12}-\theta_1\theta_2$
 and $\theta_1-\theta_2$ (and their complex conjugates).  
 Accordingly a supersymmetric two-point function $\la \uphi_i(x_1,\theta_1)\uphi_i(x_2,\theta_2)\ra$ can be a function only of
 $|\la 1,2\ra|^2$ and $|\theta_1-\theta_2|^2$.  There is an important difference between the two terms.  If $\h q$ is odd, the model (\ref{superaction}) has a discrete
 $R$-symmetry with $\theta\to -\theta$, $\bar\theta\to +\bar\theta$, $\uphi_i\to -\uphi_i$.  When
 $\h q$ is even, there may still be a discrete $R$-symmetry if there is some
 linear transformation of the $\uphi_i$ under which the coupling tensor $C_{i_1\cdots i_{\h q}}$ is odd.  Typically there is no such exact symmetry, but for large $N$
 there is almost always an approximate one because the probability distribution for $C_{i_1\cdots i_{\h q}}$ is invariant under $C\to -C$.  Thus, the large $N$ theory
 always has the discrete $R$-symmmetry.   Since $|\theta_1-\theta_2|^2$ is odd under this symmetry (and its square is zero) while $\la 1,2\ra$ is even, 
 the large $N$ disorder-averaged 
 two-point correlator is a function only of $|\la 1,2\ra|^2$.  For even $\h q$, we should expect to see effects of $R$-symmetry violation in higher orders in $1/N$.
 
 \subsection{Solving For The Two-Point Function}\label{solvetwo}
 
 In the large $N$ limit, the two-point function is determined by the Schwinger-Dyson equations, which can be written
 \be\label{sdeqnone} -D{}_{\b\theta_1}D_{\theta_1} G(\X_1,\X_3)-\int d\X_2 \,\Sigma(\X_1,\X_2)G(\X_2,\X_3)=\delta^2(x_{13})\delta^2(\theta_1-\theta_3)\,,
 \ee
 and 
 \be\label{sdeqntwo}\Sigma(\X_1,\X_2)= -J^2G(\X_1,\X_2)^{\h q-1}.\ee
 The minus sign in the second equation comes from the explicit factors of $i$ in the superpotential.   We have set $\delta^2(\theta)=\b\theta\theta$ and the
 measure is $d \X=d^2\x \,d^2\theta$.
 
 Because $G(\X_1,\X_2)$ is a function only of $\la 1,2\ra=x_1-x_2-\theta_1\theta_2$ and its complex conjugate, a short calculation shows that
 \be\label{shortone}D{}_{\b\theta_1}D_{\theta_1} G(\X_1,\X_2)=\delta^2(\theta_1-\theta_2)\frac{\partial^2 G(\X_1,\X_2)}{\partial x_1\partial \b x_1}. \ee
 Now setting $\x=\x_1-\x_2$, we Fourier transform $G(\X_1,\X_2)$ with respect to $\x$.  Since $G(\X_1,\X_2)$ is a function only of $x-\theta_1\theta_2$
 and $\b x-\b\theta_1\b\theta_2$, the Fourier transform can be written
 \be\label{ftrans}\h G(\p, \theta_1\theta_2,\b\theta_1\b\theta_2)=\int d^2\x \,e^{i\p\cdot \x}G(x-\theta_1\theta_2,\b x-\b\theta_1\b\theta_2). \ee
 Shifting the integration variable and remembering that $\p\cdot\x=\frac{1}{2}\left(p\b x+\b p x\right)$, we get 
 \begin{align}\label{gtrans}\h G(\p,\theta_1\theta_2,\b\theta_1\b\theta_2)&=\exp\left(\frac{i}{2}\left(p\b\theta_1\b\theta_2+\b p \theta_1\theta_2\right)\right)\int d^2\x
 e^{i\p\cdot \x} G(x,\b x)\cr &= \exp\left(\frac{i}{2}\left(p\b\theta_1\b\theta_2+\b p \theta_1\theta_2\right)\right)G(\p), \end{align}
 with $G(\p)$ a function of $\p $ only. 
 Inverting this, 
 \be\label{inve}G(\X_1,\X_2)=\int\frac{d^2\p}{(2\pi)^2} e^{-i\p\cdot (\x_1-\x_2)} \exp\left(\frac{i}{2}\left(p\b\theta_1\b\theta_2+\b p \theta_1\theta_2\right)\right)G(\p).\ee
 Likewise, $\Sigma(\X_1,\X_2)$ (which also is a function only of $\la 1,2\ra$ because of the $R$-symmetry) can be written
  \be\label{inveq}\Sigma(\X_1,\X_2)=\int\frac{d^2\p}{(2\pi)^2} e^{-i\p\cdot (\x_1-\x_2)} \exp\left(\frac{i}{2}\left(p\b\theta_1\b\theta_2+\b p \theta_1\theta_2\right)\right)\Sigma(\p).\ee
  Inserting these formulas in the second term in eqn.~(\ref{sdeqnone}), one can conveniently integrate over $\X_2$:
  \be\label{nve}\int d\X_2 \,\Sigma(\X_1,\X_2)G(\X_2,\X_3)=\delta^2(\theta_1-\theta_3)\int\frac{d^2\p}{(2\pi)^2}e^{-i\p\cdot(\x_1-\x_3)}\frac{p^2}{4}\Sigma(\p)G(\p).\ee
  
  In momentum space, because $\p\cdot\x=\frac{1}{2}\left(p\b x+\b p x\right)$, the $\partial_{x_1}\partial_{\b x_1}$ in eqn.~(\ref{shortone}) is equivalent to $-\p^2/4$.  
  Combining this fact with eqn.~(\ref{nve}), we see that
  in momentum space,  eqn.~(\ref{sdeqnone}) becomes simply 
  \be\label{simpleonep} G(\p)=\frac{4 }{\p^2(1-\Sigma(\p))}. \ee
  To get the IR solution, we drop the 1 in the denominator.   Also, at this point it suffices to set the fermionic coordinates to zero in eqn.~(\ref{sdeqntwo}) and to view
  $\Sigma$ and $G$ just as functions of $\x=\x_{12}$.  So the equations we have to solve are
  \be\label{longones} G(\p)=-\frac{4}{\p^2\Sigma(\p)},~~~ \Sigma(\x)=-J^2 G(\x)^{\hat{q}-1}. \ee
  Note that what we have called $\Sigma(\p)$ and $G(\p)$ are in fact the Fourier transforms of $\Sigma(\x)$ and $G(\x)$.
  With the help of eqn.~(\ref{lotag}), one can verify the following solution:
  \be\label{versol}G(\X_1,\X_2)=\frac{b}{|x_1-x_2-\theta_1\theta_2|^{2\Delta}},~~~\Delta=\frac{1}{\h q}, ~~~b^{\h q}J^2=\frac{1}{\pi^2}. \ee
  In writing the solution, we have restored the $\theta$'s. 

  Unlike in the bosonic models considered previously, the self-energy is not UV divergent here. Explicitly, if we plug in the conformal answer $G(\x)\propto |x|^{-2/\hat{q}}$, then $\Sigma(\x)\propto |x|^{-2+2/\hat{q}}$, and we do not find a divergence in computing $\Sigma(\p)$. This means that we do not need to tune a counterterm in order to reach the conformal solution. Instead, we expect that the solution of these equations automatically approaches (\ref{versol}) in the IR. This seems to be consistent with numerics; see figure \ref{superq3plot}.
\begin{figure}
\begin{center}
\includegraphics[width=0.4\textwidth]{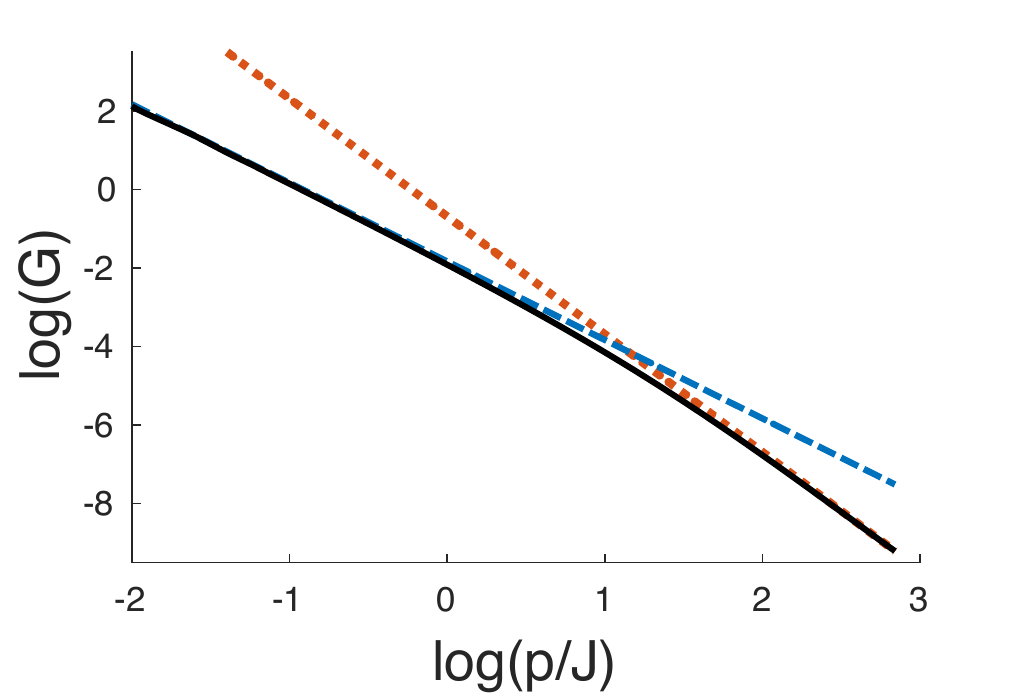}
\caption{\small{The numerical solution of the exact supersymmetric SD equations (black) interpolates between the free behavior (red, dotted) and the IR behavior (blue, dashed). This plot is for $\hat{q}=3$.}}\label{superq3plot}
\end{center}
\end{figure}
  
 \subsection{The Kernel}\label{superkernel}
Now let us evaluate the eigenvalues of the kernel that generates ladder diagrams.  Similarly to eqn.~(\ref{kernis}), the kernel is 
\be
K = -\frac{(\q-1)}{\pi^2}\frac{1}{|x_{13} - \theta_1\theta_3|^{2\Delta}|x_{24} - \theta_2\theta_4|^{2\Delta}|x_{34} - \theta_3\theta_4|^{2-4\Delta}},
\ee
where the minus sign arises from two factors of $i$ coming from the two interaction vertices. As discussed in section \ref{twoscft}, a primary may be either
bosonic or fermionic from either a holomorphic or antiholomorphic point of view.
 First we consider a purely bosonic primary, of type $BB$. The eigenfunction is as in eqn.~(\ref{tref}), with the  obvious substitution $x_{34}\to x_{34}-\theta_3\theta_4$.
 The eigenvalue of the kernel is 
\begin{align}\notag
k^{BB}(h,\t h) &= -\frac{(\q-1)}{\pi^2}\int d^2x d^2x' d^2\theta d^2\theta' \frac{(x-x' - \theta\theta')^h(\bar{x} - \bar{x}' - \bar{\theta}\bar{\theta}')^{\t h}}{|1-x|^{2\Delta}|x'|^{2\Delta}|x-x' - \theta\theta'|^{2-2\Delta}}\\ \notag
&=\frac{(\q-1)}{\pi^2}\int d^2x d^2x' \frac{(x-x')^h(\bar{x} - \bar{x}')^{\t h}}{|1-x|^{2\Delta}|x'|^{2\Delta}|x-x' |^{4-2\Delta}}(h+\Delta-1)(\t h+\Delta-1)\\ 
&=\frac{\Gamma(1-\Delta)^2}{\Gamma(1+\Delta)\Gamma(\Delta-1)}\frac{\Gamma(h+\Delta)\Gamma(-\t h+\Delta)}{\Gamma(1+h-\Delta)\Gamma(1-\t h-\Delta)}.\label{kBB}
\end{align}
The minus sign disappeared in the second line because of an additional minus sign from $\int d^2\theta d^2\theta' \theta \theta' \bar{\theta}\,\bar{\theta}' = -1$. 
As a check, we find $k^{BB}(0,0)=-(\h q-1)$.  (The explanation of this value that was reviewed in section \ref{ladderk} is applicable here.)   We also note
that for $\h q>2$, $k^{BB}(\tfrac{1}{2},\tfrac{1}{2})<1$.  This avoids some of the difficulties of bosonic models studied in section \ref{bosemod}.

Now let us consider  the kernel acting on a primary of type $FB$.  In this case, the appropriate eigenfunction is
\be
\frac{(x_{34} - \theta_3\theta_4)^h(\bar{x}_{34} - \bar{\theta}_3\bar{\theta}_4)^{\t h}}{|x_{34} - \theta_3\theta_4|^{2\Delta}}\frac{(\theta_3-\theta_4)}{(x_{34})^{\frac{1}{2}}},
\ee
and the eigenvalue is
\begin{align}
k^{FB}(h,\t h) &= \frac{(\q-1)}{\pi^2}\Delta(\t h+\Delta-1)\int  \frac{(x-x')^{h-\frac{1}{2}}(\bar{x}-\bar{x}')^{\t h-1}d^2 x d^2x'}{(1-x)|1-x|^{2\Delta}|x'|^{2\Delta}|x-x'|^{2-2\Delta}}\\
&= -\frac{\Gamma(1-\Delta)^2}{\Gamma(1+\Delta)\Gamma(\Delta-1)}\frac{\Gamma(\frac{1}{2}-h + \Delta)\Gamma(\t h+\Delta)}{\Gamma(\frac{3}{2}-h-\Delta)\Gamma(1 + \t h-\Delta)}.\label{kFB}
\end{align}
In particular, $k^{FB}(3/2,0)=1$.  This result is expected in a model that flows in the IR to a superconformal field theory.  An $\N=1$ superconformal field theory
has a supercurrent multiplet ${\mathcal S}(x,\theta)=S+\theta T$, where $S$ and $T$ are the holomorphic supercurrent and stress tensor; $S$ is a primary
with dimension $(3/2,0)$.    For this multiplet
to exist, we need $k^{FB}(3/2,0)=1$. 


Finally, let us consider the $FF$ channel, with eigenfunction
\be
\frac{(x_{34} - \theta_3\theta_4)^h(\bar{x}_{34} - \bar{\theta}_3\bar{\theta}_4)^{\t h}}{|x_{34} - \theta_3\theta_4|^{2\Delta}}\frac{|\theta_3-\theta_4|^2}{|x_{34}|}
\ee
and eigenvalue 
\be\label{oddfn}
k^{FF}(h,\t h) = \frac{\Gamma(1-\Delta)^2}{\Gamma(1+\Delta)\Gamma(\Delta-1)}\frac{\Gamma(\frac{1}{2}-h+\Delta)\Gamma(-\frac{1}{2}+\t h+\Delta)}{\Gamma(\frac{1}{2}+\t h-\Delta)\Gamma(\frac{3}{2}-h-\Delta)}.
\ee

We will now discuss some relationships between the different kernel eigenvalues. Several relations follow from identifies relating the wave functions $\Upsilon$, and have been discussed above. In particular, one can verify (\ref{wadtwo}) and (\ref{reln}). There are also some more trivial relationships between the kernels that follow from exchanging holomorphic and antiholomorphic variables, which is a symmetry
of the theory under study. This exchanges $k^{FB}$ and $k^{BF}$, while also exchanging $h$ and $\th$. So
\be\label{wexpres}k^{FB}(h,\t h)=k^{BF}(\t h, h). \ee
Combining (\ref{wadtwo}) and (\ref{wexpres}), we have
\be\label{xprest}k^{FB}(\tfrac{1}{2}-h,\tfrac{1}{2}-\t h)=k^{FB}(\t h, h). \ee
Symmetry under exchanging holomorphic and antiholomorphic variables also leads to\footnote{The following holds if $h-\th$ is an integer,
as assumed in all of our derivations.  If, for example, $k^{FF}$ is written as in eqn.~(\ref{oddfn}), then in this form it is defined for arbitrary $h,\th$,
but eqn.~(\ref{yxpres}) does not hold.}
\be\label{yxpres}k^{BB}(h,\t h)=k^{BB}(\th , h),~~~k^{FF}(h,\t h)=k^{FF}(\t h , h). \ee

Finally, there is an interesting relationship between the kernel eigenfunctions  that does not follow from an identity of the $\Upsilon$ wave functions. For integer spin, we have
\be\label{relationKernels}
k^{BB}(h,\th) = -k^{FB}(h+\tfrac{1}{2},\th)= -k^{BF}(h,\th+\tfrac{1}{2})=k^{FF}(h+\tfrac{1}{2},\th+\tfrac{1}{2}).
\ee
These relations appear to be a special property of the SYK model under study. As we will see, they go part way towards putting the operators in $\mathcal{N} = 2$ multiplets. 

For the next section, it will be helpful to know the range of $h,\th$ for which the integrals that define the various kernels converge. Finding these ranges directly from the integrals is slightly subtle. For example, consider the spin-zero sector ($h = \th$) of the $k^{BB}$ kernel. There are two possible divergences, a UV divergence at $x = x'$ and an IR divergence where both $x$ coordinates go to infinity. The possible UV divergence is best analyzed before doing the integral over the $\theta$ coordinates: we find that the integral is convergent if $\text{Re}(h+\th) > -2\Delta$. The possible IR divergence is best analyzed after the $\theta$ integrals, and we find convergence if $\text{Re}(h+\th) < 2\Delta$. For general spin, and for the other kernels studied below, we can more quickly find the region of convergence from the locations of the poles in the gamma functions in the numerator. It is convenient to give the answer for shifted dimensions so that we study $k^{BB}(h,\th), k^{FB}(h{-}\tfrac{1}{2},\th), k^{BF}(h,\th{-}\tfrac{1}{2}),k^{FF}(h{-}\tfrac{1}{2},\th{-}\tfrac{1}{2})$. Then one finds that for nonzero $J$, the region of convergence always includes the principal series $h = \frac{1+J}{2}+is,\th = \frac{1-J}{2}+is$ with real $s$. However, there are important exceptions for $J = 0$, where the $BB$ kernel converges for $\frac{1}{2}-\Delta<\text{Im}(s)<\frac{1}{2}+\Delta$ and the $FF$ kernel converges for $-\frac{1}{2}-\Delta<\text{Im}(s)<-\frac{1}{2}+\Delta$.


 \subsection{Inner Products And The Integration Contour}\label{innerpr}
 
 As a step in understanding the four-point function, we need to compute
 the inner products between the zero-rung ladder and the wave functions $\Upsilon^{BB}$ and $\Upsilon^{FB}$.   We will perform a direct computation. The simple
 relation to the kernel functions that results can be understood along lines explained near the end of section \ref{kkernel}.

 The zero-rung ladder is given by
\be\label{zerorungdef}
\mathcal{F}_0 = \frac{G(1,3)G(2,4)+G(1,4)G(2,3)}{G(1,2)G(3,4)} = |\chi+\zeta|^{2\Delta} + \left|\frac{\chi+\zeta}{\chi-1}\right|^{2\Delta}.
\ee
We would like to compute the inner product, as defined in (\ref{nuzem}), of this contribution with the $\Upsilon$ basis functions. This can be reduced to a similar inner product in the bosonic theory as follows. First, we use (\ref{ladow}) and (\ref{lladow}) to write the $\Upsilon$ functions in terms of $\Psi$. Then we ``integrate'' over $\zeta,\bar{\zeta}$ by taking the part of the integrand proportional to $\zeta\bar{\zeta}$. The resulting integrals over $\chi,\bar{\chi}$ are proportional to bosonic inner products. Explicitly,
\begin{align}
\langle \Upsilon^{BB}_{h,\t h},\mathcal{F}_0\rangle &= (1-h-\Delta)(1-\t h-\Delta)\left(\Psi_{h,\t h},|\chi|^{2\Delta} + |\frac{\chi}{\chi-1}|^{2\Delta}\right)\label{extrafactors1}\\
\langle \Upsilon^{FB}_{h-\frac{1}{2},\t h},\mathcal{F}_0\rangle &= (h-\Delta)(1-\t h-\Delta)\left(\Psi_{h,\t h},|\chi|^{2\Delta}+ |\frac{\chi}{\chi-1}|^{2\Delta}\right).\label{extrafactors2}
\end{align}
We evaluated these bosonic inner products in (\ref{actualans}) and (\ref{actualans2}). There it was convenient to relate the answer to the bosonic kernel function $k(h,\widetilde{h})$. Here it is more convenient to relate the answer to the supersymmetric kernels. After including the extra prefactors in (\ref{extrafactors1}) and (\ref{extrafactors2}), we find that the full inner products can be written (for integer $h-\t h$) as
\begin{align} \label{poffo}
\langle \Upsilon^{BB}_{h,\t h},\mathcal{F}_0\rangle &= \frac{\pi^2}{\h{q}-1}k^{BB}(h,\t h)\left(1 + (-1)^\ell\right)\\ \label{boffo}
\langle \Upsilon^{FB}_{h-\frac{1}{2},\t h},\mathcal{F}_0\rangle &= \frac{\pi^2}{\h{q}-1}k^{FB}(h-\tfrac{1}{2},\t h)\left(1 + (-1)^\ell\right).
\end{align}

These inner products determine the constant $\lambda$ appearing in (\ref{padsum}), and (\ref{radsum}), which apparently completes our derivation of the four-point function. In fact, there is an important subtlety related to the convergence of the integrals that define the kernel eigenvalues. To discuss this, let us write a candidate expression for the zero rung ladder that follows from these inner products and the manipulations that led to (\ref{radsum}):
\begin{align}\label{F0candidate}
\mathcal{F}_0\Big|_{\zeta = \b\zeta = 0} &\stackrel{?}{=} \frac{2}{\pi(\q-1)}\sum_{\ell = \text{even}} \int_{-\infty}^\infty \frac{ds}{2\pi}A(h,\th) F_h(\chi)F_{\th}(\bar{\chi})\\&\hspace{20pt}\left[k^{FB}(h{-}\tfrac{1}{2},\th)+k^{BF}(h,\th{-}\tfrac{1}{2})- k^{BB}(h,\th)-k^{FF}(h{-}\tfrac{1}{2},\th{-}\tfrac{1}{2})\right].\notag
\end{align}
In fact, this expression is not correct. One way to see this is to note that from (\ref{zerorungdef}) $\mathcal{F}_0$ contains an operator of dimension $2\Delta<1$. On the other hand, when we shift the contour of integration in (\ref{F0candidate}) to get an OPE expansion, we will only pick up poles to the right of the principal series, with dimension $h+\th > 1$. The fix for this problem is as follows: we leave the $k^{FB}$ and $k^{BF}$ terms alone, but for the $k^{BB}$ term in the spin zero sector, we shift the defining contour from real
$s$ to $s = i/2 + \R$. Similarly, for the $k^{FF}$ term in the spin zero sector, we shift the defining contour to $s = -i/2 + \R$. It can then be checked that (\ref{F0candidate}) gives the correct answer for $\mathcal{F}_0$, by evaluating the sum over residues from the poles in the various kernels.\footnote{D.~Simmons-Duffin pointed out to us that the zero rung-ladder is essentially the four-point function in ``generalized free field theory'' or ``mean field theory'' and this sum over poles in the kernels can be understood as the OPE expansion for this theory, which was previously discussed in \cite{Heemskerk:2009pn,Fitzpatrick:2011dm}. Note that the sum of the supersymmetric kernels in (\ref{F0candidate}) is proportional to the kernel in the bosonic theory.}

This contour prescription appears ad hoc, but it can be explained as follows. We can get a formula for  the zero rung ladder by acting with the kernel on a delta function. To get a formula involving an integral over conformal wave functions, we represent the delta function as in (\ref{comprel}), as an integral over the principal series. If we try to act with the kernel inside the integral over $s$, we will find a divergence in the integrals that define $k^{BB}$ in the spin zero sector. This is because the principal series does not satisfy $-2\Delta < \text{Re}(h+\th) <2\Delta$, which we identified in the previous section as the criterion for convergence of the integral that defines $k^{BB}$ for $J = 0$. However, before we act with the kernel, we can deform the contour in the $BB$ channel so that $s = i/2+\R$ for the spin zero states. Then the action of the kernel will be well defined. Similarly, one finds that the contour for $k^{FF}$ has to be shifted in the opposite direction for the integrals to converge: $s = -i/2 + \R$ is a  simple choice that works for all values of $\Delta$.

\subsection{The Four-Point Function}\label{fpf}

The results of section \ref{innerpr} give the remaining information we need to justify the derivation of eqn.~(\ref{radsum}) (where relations of the form of eqns.~(\ref{poffo}), (\ref{boffo}) were
assumed).  Moreover, we can now identify the constant $\lambda$ in that formula as $\pi^2/(\h q-1)$.  Thus we can now write the four-point function more explicitly.
For brevity we do this only at $\zeta=\bar\zeta=0$ (also writing out the $\A$ function explicitly):  
\begin{align}\label{fourpt}
\mathcal{F}\Big|_{\zeta = \bar{\zeta} = 0} &= \frac{1}{\pi(\hat{q}{-}1)}\sum_{\ell = \text{even}} \int_{-\infty}^\infty \frac{ds}{2\pi}\frac{\sin(\pi h)}{\cos(\pi \th)}\frac{\Gamma(h)^2}{\Gamma(2h)}\frac{\Gamma(\th)^2}{\Gamma(2\th)} F_h(\chi)F_{\th}(\bar{\chi})\\&\left[\frac{k^{FB}(h{-}\frac{1}{2},\th)}{1-k^{FB}(h{-}\frac{1}{2},\th)}+\frac{k^{BF}(h,\th{-}\frac{1}{2})}{1-k^{BF}(h,\th{-}\frac{1}{2})}- \frac{k^{BB}(h,\th)}{1-k^{BB}(h,\th)}-\frac{k^{FF}(h{-}\frac{1}{2},\th{-}\frac{1}{2})}{1-k^{FF}(h{-}\frac{1}{2},\th{-}\frac{1}{2})}\right].\notag
\end{align}
For the $\ell = 0$ terms, the contour integral over $s$ has to be defined as in the previous section: for the $BB$ terms we integrate over $s = i/2+\R$, for the $FB$ and $BF$ terms we integrate over $s = \R$, and for the $FF$ terms we integrate over $s = -i/2 + \R$. For nonzero values of $\ell$ we can integrate all terms over $s = \R$. The function $F_h$ is the $SL(2,\R)$ conformal block, as defined in (\ref{1dblockdef}). As before, we are using $h,\widetilde{h}$ interchangeably with $\ell,s$, with the understanding that
\be
h = \frac{1+\ell}{2}+is,\hspace{20pt}\widetilde{h} = \frac{1-\ell}{2}+is.
\ee
An important feature of this expression and the kernels derived above is that for even $\ell$, the kernels are never equal to one when evaluated on the relevant $s$ integration contour. By plotting the functions, one finds that this is true as long as $\hat{q}>2$.

To derive the operator product expansion, we shift the $s$ contour as described in section \ref{opesuper}, but keeping in mind the slightly shifted defining contour for the $BB$ and $FF$ spin zero sectors. It appears that solutions of $k=1$ (for any of the kernel functions) in the half-plane in which we deform the contour occur only for real $h,\th$ and lead to positive residues.
One finds that the prefactor $\A$ is positive (this depends on the fact that $\ell$ is even)  and that the derivative of $k$ is negative for $FF,\,BB$ and positive for $FB,\,BF$ at
the point with $k=1$. Since the OPE coefficient is minus the residue, and because of the explicit signs in the above, this leads to a positive coefficient in the OPE. 

A  possibly surprising fact  is that $k^{FB}(\frac{1}{2},1) = 1$, suggesting a bosonic operator with $h = 1,\t h =1$, which would represent a conformal (but not supersymmetric) modulus of the theory. In fact, the $\A$ coefficient vanishes at this point, and the derivative of the kernel is nonzero, so the residue vanishes and we do not have an operator there. (We saw a similar thing in the 1d supersymmetric model.)

We can use the above formulas to determine the central charge by computing the contribution of the stress tensor to the four-point function.  The stress tensor is the $(h,\t h) = (2,0)$ descendant of
an $FB$ primary at $(h,\th)=(3/2,0)$.  Near $h = 2$, we have
\be
k^{FB}\left(h-\frac{1}{2},h-2\right) = 1 + \frac{1-2\Delta}{\Delta(1-\Delta)}(h-2)+...
\ee
Taking the residue of the pole in $k^{FB}/(1-k^{FB})$, and collecting the other factors, we get the contribution \be\label{elbow}
\mathcal{F}\supset \frac{\Delta^2}{3-6\Delta}F_2(\chi).
\ee
The expected contribution of the stress tensor would be\footnote{See Appendix \ref{app:c}.}
 $\frac{2h_1h_2N}{c}F_2(\chi)$, where in our case $h_1 = h_2 = \frac{\Delta}{2}$. Matching the two expressions, we find
\be
c = \frac{3}{2}\left(1-2\Delta\right)N = \frac{3}{2}\left(1-\frac{2}{\hat{q}}\right)N.
\ee
The answer passes some simple checks. First, as $\hat{q}\rightarrow \infty$, we get the free-field answer, suggesting that the large $\h q$ theory is weakly coupled. This seems consistent with the fact that anomalous dimensions and the chaos exponent become small at large $\h q$, as we discuss below. Next, the answer is an increasing 
function of $\hat{q}$. This is in keeping with the $c$ theorem, because with a suitable perturbation of the superpotential, we can flow from a larger value of 
$\hat{q}$ in the UV to a smaller value in the IR, causing $c$ to decrease. Finally, the central charge vanishes at $\hat{q} = 2$.  This is in keeping 
with the fact that we do not get a nontrivial CFT at $\h q=2$.

Besides the stress tensor, there are of course many other operators that contribute to the four-point function. In particular, one finds an infinite family of operators of each spin in each of the $BB,FB,BF,FF$ channels. The twists of the bosonic operators that appear can be written as $E-J = 2n + 2\Delta+ \epsilon(n,J)$ where $\epsilon(n,J)$ is at most of order one, and becomes small for large $n$ or large $J$. One can check that $\epsilon(n,J)$ approaches zero from above in the $BB,FF$ channels, and from below in the $BF,FB$ channels. 

We can roughly think of the operators appearing in the OPE as being the following
\be
\uphi_i (\partial\b\partial)^n \partial^m \uphi_i, \hspace{20pt} \uphi_i (\partial\b\partial)^n \partial^m Q\uphi_i, \hspace{20pt} \uphi_i (\partial\b\partial)^n \partial^m \b Q\uphi_i, \hspace{20pt} \uphi_i (\partial\b\partial)^n \partial^m Q \b Q\uphi_i,
\ee
where the four cases are respectively for $BB,FB,BF,FF$, and $Q$ is the differential operator defined in (\ref{supergen}). These expressions are schematic for two reasons. First, even in a (generalized) free field theory, we would have to take a particular linear combination of terms with the derivatives and $Q$s acting on the two $\uphi$ fields in order to form a conformal primary. In order for such a primary to be present among the various combinations of such terms, we have to take $m$ even in the $BB,FF$ cases, and odd in the $FB,BF$ cases. Second, our theory is interacting, so these simple expressions are not true primaries. The point is simply that the actual primaries can be labeled by these free operators, and the dimensions agree for large values of $n$ or $m$. In this labeling scheme, the stress tensor appears within the multiplet of the $n,m = 0,1$ contribution in the $FB$ channel.

In some ways, one would like to think about these ``single-sum'' operators as being analogous to single-trace operators of a large $N$ matrix field theory. However, the number of operators and the behavior of their anomalous dimensions is more reminiscent of the behavior of double-twist families that were discussed in \cite{Alday:2007mf,Fitzpatrick:2012yx,Komargodski:2012ek}; see also \cite{Simmons-Duffin:2016wlq} for data in the 3d Ising model.

In figure \ref{spin4Fig} we plot some information about the anomalous dimensions, including the twist of the lightest spin-four operator, as a function of $\Delta$. Note that the twist becomes small for small $\Delta$, consistent with the idea that the large $\h q$ theory is weakly interacting. We also plot a few of the light operators in different channels for $\Delta = \frac{1}{3}$. 
\begin{figure}[t]
\begin{center}
\includegraphics[width=.328\textwidth]{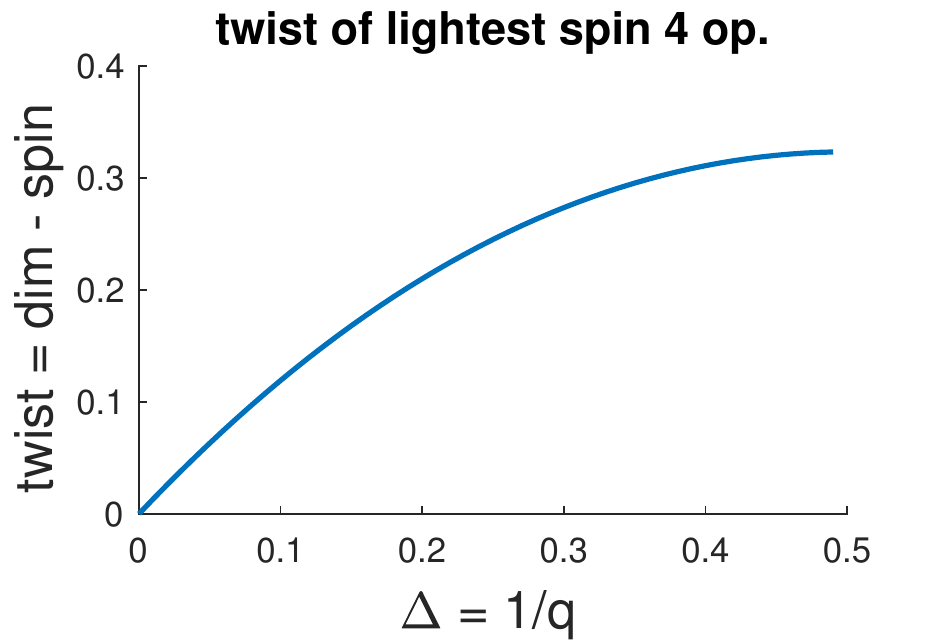}
\includegraphics[width=.328\textwidth]{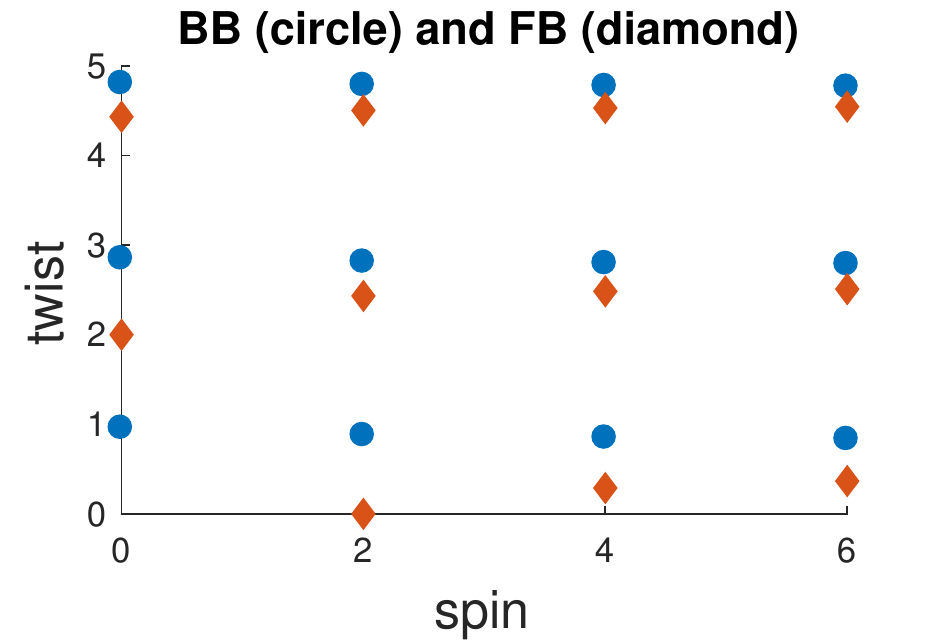}
\includegraphics[width=.328\textwidth]{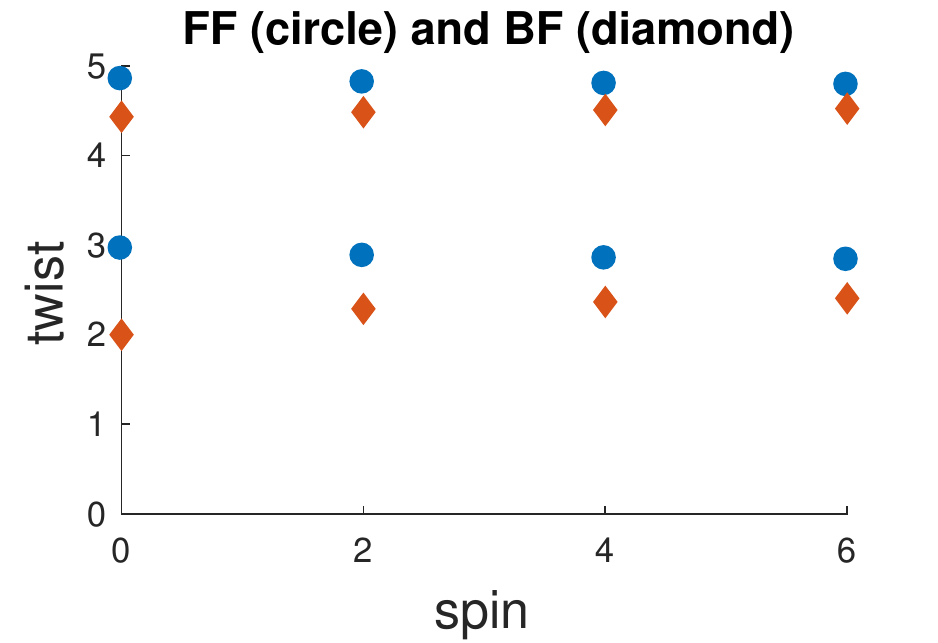}
\caption{At left we plot the twist  of the lowest-twist spin four operator, as a function of $\Delta = \frac{1}{\hat{q}}$. This operator is in the $FB$ channel. In the middle and right, we show the twists of several low-lying bosonic operators in different channels for the $\h q = 3$ theory. For the $FF$ channel, these are the top component, and for the $BB$ channel these are the bottom.}\label{spin4Fig}
\end{center}
\end{figure}

There is a somewhat surprising pattern in the spectrum that can be derived from the fact that the kernels satisfy $k^{FF}(h,\t h) = k^{BB}(h-\frac{1}{2},\t h-\frac{1}{2})$ for integer spin. What this means is that $FF$ and $BB$ operators come in pairs, such that the $FF$ primary has dimension one larger than the corresponding $BB$ primary. There is a similar pattern within the $FB$ and $BF$ channels, where the dimensions of the primaries are the same, but the spin differs by one.\footnote{In the four-point function $\F$,
restricted to primary fields (that is, with $\zeta=\b\zeta=0$), 
the only bosonic operator within a given  $FF$ multiplet that contributes is the top component, so this operator has dimension two more than the bottom component of the associated $BB$ operator. Similarly, in a pair of  $FB$ and $BF$ multiplets, the operators contributing to this four-point function have spins that
differ by two.} We are not sure of the significance of this, but it is interesting to note the following. A generic $\mathcal{N} = 2$ multiplet contains four generic $\mathcal{N} = 1$ multiplets, one of each type $BB,BF,FB,FF$. The relations between the dimensions that we just described imply that each $BB/FF$ pair or $FB/BF$ pair separately forms half of an $\mathcal{N} = 2$ multiplet.

 \subsection{Enhancement to $\N=2$?}\label{enhance}
 
We will next discuss  a subtlety in the spectrum.
 
 We start with the supersymmetric SYK model in one dimension.  In the IR limit, one discards the ordinary kinetic energy, leaving the interaction
 \be\label{nouble} \int dt\,d \theta \,J_{i_1i_2\dots i_{\h q}}\uppsi_{i_1}(t,\theta)\dots \uppsi_{i_{\h q}}(t,\theta). \ee
This interaction is invariant under arbitrary reparametrizations of the coordinates $t,\theta$, accompanied by suitable scaling of $\uppsi$.
The group of reparametrizations of one bosonic variable $t$ and one fermionic variable $\theta$ is actually, somewhat
surprisingly, the $\N=2$
superconformal algebra in one dimension.\footnote{This is related to the fact that the $\N=2$ algebra can act on a chiral superspace with
one bosonic and one fermionic dimension.}  Therefore, the IR limit of the supersymmetric SYK model has a spontaneously broken $\N=2$
supersymmetry.  From an $\N=1$ point of view, the $\N=2$ algebra is generated by a fermionic primary of dimension $3/2$ and a bosonic
primary of dimension 1.  As explained in \cite{Fu:2016vas}, both types of modes are present as solutions to the Schwinger-Dyson equations, since $k^F(3/2) = k^B(1) = 1$. However, the implications for the four-point function of these modes are very different. The fermionic primary of dimension
$3/2$ is associated to a formally infinite contribution to the four-point function, which must be treated outside the conformal limit. By contrast, the would-be bosonic state of dimension $1$ does not actually contribute to the four-point function at all, as we saw in section \ref{opesuper}.

Similarly, in two dimensions, in the IR limit the only part of the action that is relevant is the superpotential coupling:
\be\label{trouble}\int d^2\x d^2\theta\left(i\sum_{i_1i_2\cdots i_{\h q}}C_{i_1i_2\cdots i_{\h q}}\uphi_{i_1}
 \uphi_{i_2}\cdots \uphi_{i_{\h q}} \right).\ee
 Again, this is invariant under arbitrary volume-preserving diffeomorphisms of the whole set of bosonic and fermionic coordinates, with suitable rescaling of $\uphi$.
 Naively, this implies that the model has an $\N=2$ superconformal symmetry in the IR, spontaneously broken down to a finite-dimensional
 supergroup that is an invariance of the saddle point solution for the two-point function.  This would correspond to the existence of
 holomorphic primaries of dimension $(3/2,0)$ and $(1,0)$ and antiholomorphic primaries of dimension $(0,3/2)$ and $(0,1)$.
 
 In fact, the primaries of dimension $(3/2,0)$ and $(0,3/2)$ are present, associated with $k^{FB}(3/2,0)=k^{BF}(0,3/2)=1$.
 But in the model as we have defined it so far, there are no primaries of dimension $(1,0)$ or $(0,1)$, since actually $k^{BB}(1,0)$ and $k^{BB}(0,1)$
 equal $-1$, not $+1$.  We will now explain what is wrong technically with the argument that claims to prove that such primaries should exist.
 We begin by explaining the situation at $h=1$ in the 1-dimensional model, where the formal argument does give the right answer, and then we move on to two dimensions.
 
 The interaction (\ref{nouble}) has the symmetry $\delta\theta=\alpha(t)\theta$, for any function $\alpha(t)$.
 Writing $\uppsi(t,\theta)=\psi(t)+\theta b(t)$, the symmetry, in terms of components, is
 \be\label{dolf} \delta\psi(t)=\frac{\alpha(t) \psi(t)}{\h q},~~~\delta b=\frac{(1-\h q)\alpha(t)}{\h q} b. \ee
  The corresponding deformation of the two-point function $G(t_1,\theta_1, t_2,\theta_2)$ is
 \be\label{olf}\delta G=\frac{G}{\h q}\left(\left(\alpha(t_1)+\alpha(t_2)\right)\left(1-\frac{\theta_1\theta_2}{t_1-t_2}\right)\right). \ee
 To decide if this deformation corresponds to a state of $h=1$, we will simply use the explicit formula of eqn.~(\ref{zelo}) for the Casimir.   
 We  recall that $\C_{12}$ as defined
 in that formula is the Casimir acting on a two-particle wavefunction that has been normalized by dividing by $G$.  So the candidate wavefunction that
 we have to investigate is
 \be\label{nolf}\Lambda(t_1,\theta_1,t_2,\theta_2)=\left(\alpha(t_1)+\alpha(t_2)\right)\left(1-\frac{\theta_1\theta_2}{t_1-t_2}\right). \ee
 A short calculation reveals that, regardless of the choice of $\alpha(t)$, this function obeys
 \be\label{woff}\C_{12}\Lambda=\frac{1}{2}\Lambda.\ee
 The eigenvalue $1/2$ is $h(h-1/2)$ with $h=1$, confirming that these deformations represent states of $h=1$.
 
 Now let us see what goes wrong with the corresponding argument in two dimensions.   The interaction (\ref{trouble}) now has a symmetry $\theta\to \alpha(x,\b x)\theta$,
 acting on component fields in a manner similar to (\ref{dolf}).  The corresponding deformation of the two-point function is just as in (\ref{olf}):
  \be\label{olof}\delta G=\frac{G}{\h q}\left(\bigl(\alpha(x_1,\bar x_1)+\alpha(x_2,\bar x_2)\bigr)\left(1-\frac{\theta_1\theta_2}{x_1-x_2}\right)\right). \ee
 We want to know if this deformation corresponds to a state of $(h,\th)=(1,0)$.  For this, we have to ask if the corresponding wavefunction
 $\Lambda=\left(\alpha(x_1,\b x_1)+\alpha(x_2,\b x_2)\right)(1-\theta_1\theta_2/(x_1-x_2))$ satisfies the eigenvalue equations
 \be\label{eigenq}\C_{12}\Lambda=\frac{1}{2}\Lambda, ~~~ \b\C_{12}\Lambda=0.  \ee
 There is actually no trouble with the first equation.  The same calculation that demonstrated eqn.~(\ref{woff}) in the 1d case carries over here to show
 that $\C_{12}\Lambda=\frac{1}{2}\Lambda$.  The problem is in the antiholomorphic equation.  The antiholomorphic Casimir $\b\C_{12}$ is simply given by
 the same formula as in eqn.~(\ref{zelo}), with the obvious substitutions $t,\theta\to \b x,\b\theta$.  Because of the term 
 $(\b x_1-\b x_2)\b\theta_1\b\theta_2\frac{ \partial^2}{\partial \b x_1\partial\b x_2}$
 in $\b\C_{12}$,  we get $\b\C_{12}\Lambda\sim |\theta_1|^2|\theta_2|^2\delta^2(x_1-x_2)$.  
 
 Thus contrary to naive expectations, even though the deformation $\Lambda$ preserves the Schwinger-Dyson equations, it does not represent a state of the expected
 dimensions.  
 In hindsight, this is a relief, since if we did find a holomorphic primary of dimension $(1,0)$, this would extend the $\N=1$ superconformal algebra to $\N=2$
 and force all operators to be in multiplets of $\N=2$ supersymmetry.
 This is not  the case, though the states do form partial $\N=2$ multiplets as explained in section \ref{fpf}.

\subsection{Model With $U(1)$ Symmetry}\label{modone}

Consider in two dimensions a CFT or SCFT with a continuous symmetry and thus a conserved current $J$.  Conformal invariance implies that the holomorphic
and antiholomorphic parts of $J$ are separately conserved.

Thus, if we start in the UV with a $U(1)$ symmetry generated by a conserved 
current, one might expect to find $U(1)\times U(1)$ in the IR, with the two factors associated 
with the $(1,0)$ and $(0,1)$ parts of the current.   In an $\N=1$  superconformal field theory, the $(1,0)$ current would be a descendant of a $(1/2,0)$ primary
in the $FB$ channel, and similarly for the $(0,1)$ current.    Thus in particular, in a 2d supersymmetric model with $U(1)$ symmetry that  has the ladder
structure of the SYK model, we expect to find an $FB$ primary of dimension $(1/2,0)$.

It is not difficult to construct a candidate model.  We take $\h q=2r$ for some $r\geq 2$.  We introduce $N/2$ complex scalar superfields $\uphi_i$
and their complex conjugates $\b\uphi_i$, and we consider a superpotential
$i C_{i_1\cdots i_r \b j_1\cdots \b j_r}\uphi_{i_1}\cdots \uphi_{i_r}\b\uphi_{j_1}\cdots \b\uphi_{j_r}$, with the coefficients $C_{\cdots}$ being as usual
independent Gaussian variables.  In this model, the normalized four-point function  $\F$ 
is given by the same ladder diagrams as before and can be studied by the same methods.  There is just one important difference.  In contrast to the
previous cases that we have studied, this four-point function is not symmetric or antisymmetric under $\X_1\leftrightarrow \X_2$.  It can be symmetrized
or antisymmetrized under $\X_1\leftrightarrow \X_2$, if we also symmetrize or antisymmetrize under $\uphi\leftrightarrow \b\uphi$.  Let us call the
two channels symmetric and antisymmetric.  The four-point function in the symmetric channel will receive contributions only with even $\ell$, as in
eqn.~(\ref{fourpt}), while the four-point function in the antisymmetric channel is governed by a formula of the same form but with contributions only for odd $\ell$.

We write $k^{FB}_S$ and $k^{FB}_A$ for the $FB$ kernels of this model in the symmetric and antisymmetric channels.   The relation of these functions to
$k^{FB}$ of the model without $U(1)$ symmetry, as defined in eqn.~(\ref{oddfn}), is simply\footnote{The derivation of eqn.~(\ref{oddfn}) was valid for all
$\ell$, even or odd.  In the derivation of eqn.~(\ref{fourpt}), the projection on even $\ell$ came from the computation of $\la \Psi^{h,\th},\F_0\ra$ in section
\ref{innerpr}. In the antisymmetric channel, the same computation gives a projection onto odd $\ell$.}
\be\label{relker}k^{FB}_S(h,\th)=k^{FB}(h,\th),~~~~k^{FB}_A(h,\th)=\frac{1}{\h q-1}k^{FB}(h,\th). \ee
These formulas may be understood as folows.\footnote{See section 3.1 of \cite{KT} for discussion of a 1d model in which the same formulas are relevant.}
If we symmetrize under $\uphi\leftrightarrow \bar\uphi$, we are effectively back to the ladder diagrams of the model without a $U(1)$ symmetry, so
$k^{FB}_S=k^{FB}$.  However, antisymmetrizing gives a result that is smaller by a factor of $\h q-1$, for a reason that is explained in fig. \ref{symmetry}.

\begin{figure}[ht]
\begin{center}
\includegraphics[width=.9\textwidth]{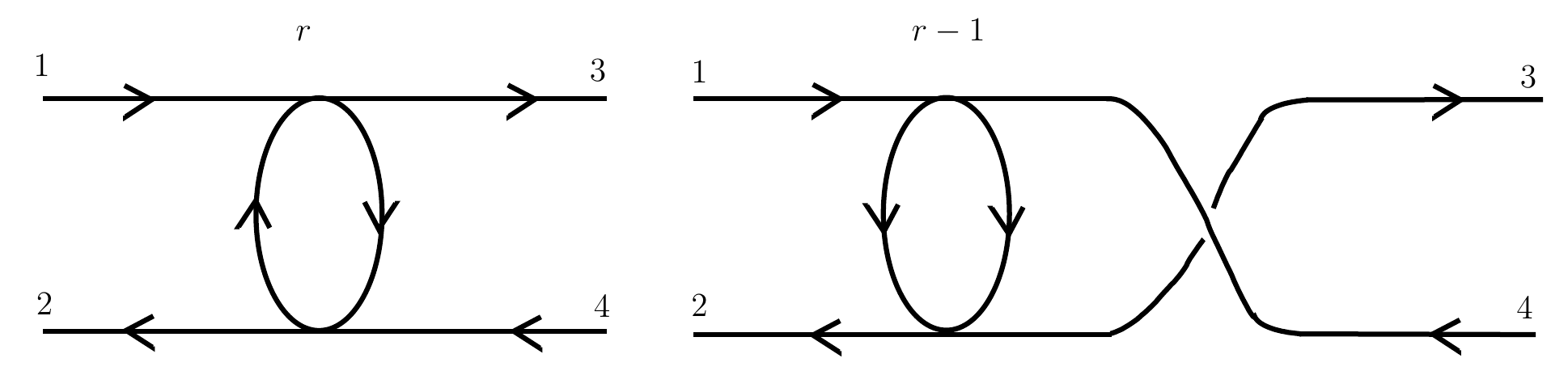}
\caption{Ladder diagrams in a theory with $U(1)$ symmetry and $\upphi^r\b\upphi{}^r$ couplings (sketched here for $r=2$), 
either (a) without charge exchange or (b) with charge exchange.   The two diagrams
 have relative combinatoric factors $r$ and $r-1$, respectively, from counting Wick contractions.  
 In the symmetric channel the resulting contribution is a factor of $r+(r-1)=\h q-1$,
but in the antisymmetric channel, one gets instead $r-(r-1)=1$.  So the kernel is smaller in the antisymmetric channel by a factor of $\h q-1$. }\label{symmetry}
\end{center}
\end{figure}

A look back to eqn.~(\ref{oddfn}) reveals that $k^{FB}(1/2,0)=\h q-1$, so $k^{FB}_A(1/2,0)=1$.  The $(1,0)$ part of the $U(1)$ current is indeed expected
to appear in the antisymmetric channel, since the $U(1)$ current is odd under $\uphi\leftrightarrow\b\uphi$.  So at first sight it seems that we have found the 
expected mode.  However, there is a snag.  The derivative of the function $k_A^{FB}(h,\th)$ vanishes at $(h,\th)=(1/2,0)$, so that the function
$1/(1-k_A^{FB}(h,\th))$ has a double pole at $(h,\th)=(1/2,0)$.  Moreover, because $k^{FB}$ appears in eqn.~(\ref{fourpt}) with a shift by $-1/2$, 
the double pole at $(h,\th)=(1/2,0)$ is actually on the original integration contour for the four-point function.  Thus the IR formula (\ref{fourpt})
is actually divergent.

We tentatively offer the following interpretation (suggested in part by I. Klebanov).  The operators $\sum_i \b\uphi{}_i\uphi_i$,
$\sum_i \b\uphi{}_i D\uphi_i$, etc., that appear in the OPE for the ladder sum are rough analogs of the usual single-trace operators of the AdS/CFT
correspondence.  As in section \ref{fpf}, we call  them single-sum operators.   At the level of single-sum operators, the model under discussion is conformally-invariant in the
large $N$ limit,
and its operator spectrum is given by the solutions of $k(h,\th)=1$, where $k$ can be any of the kernel functions.  In particular, $k^{FB}_A(1/2,0)=1$,
and there is a corresponding single-sum primary field $\Sigma=\sigma+\theta J+\dots $ of dimension $(1/2,0)$.  Its complex conjugate is
$\b\Sigma=\b\sigma+\b\theta\,\b J+\dots$, a primary of dimension $(0,1/2)$.  Here $J$ and $\b J$ are holomorphic and antiholomorphic currents.
The model also has, in the large $N$ limit, a double-sum primary $\b\Sigma\Sigma$ of dimension $(1/2,1/2)$.  Since it is a double-sum operator,
it does not correspond to a solution of $k=1$ for any of the kernel functions.  Because it has dimension $(1/2,1/2)$, its integral is a marginal deformation:
\be\label{marg}\Delta I=\int d^2\x d^2\theta \,\b\Sigma\Sigma=-\int d^2\x \left(\b J J +\dots\right).  \ee
The idea now is that at the double-sum level, the model is not conformally-invariant at large $N$, but rather has a renormalization group flow of
the coefficient of the double-sum interaction (\ref{marg}).  If so, the four-point function $\F(\X_1,\dots,\X_4)$ will have no IR limit
for large $N$; rather, one expects a divergence in $\F(\X_1,\dots,\X_4)$ that will be just proportional to the effects of an insertion of $\Delta I$.
This is in accord with the divergence in $\F(\X_1,\dots,\X_4)$ that results from the double pole in $k^{FB}_A$.

\def\veps{\varepsilon}
\section{Retarded Kernel And Chaos Exponent}\label{ret}

\subsection{Background}\label{background}

In general, to give an operator interpretation to Euclidean correlation functions, one selects one direction in space that is viewed as imaginary
``time.'' We parametrize imaginary time by $\tau$.  Then one introduces a Hamiltonian $H$, such that the transfer matrix $e^{-\lambda H}$ propagates fields by a distance $\lambda$ in the  $\tau$
direction.   A correlation function
\be\label{corrfn}\bigl\la \phi(\vec x_1,\tau_1) \phi(\vec x_2,\tau_2)\cdots \phi(\vec x_n,\tau_n)\bigr\ra,\ee
where $\vec x$ are the Euclidean coordinates orthogonal to $\tau$, can then in  a standard way be interpreted in terms of a matrix element of
a product of operators.
In this interpretation, the operators are automatically $\tau$-ordered.  For example, assuming that we are on $\R^n$
and denoting the ground state as $|\Omega\ra$,  if $\tau_1<\tau_2<\dots<\tau_n$  then we interpret
the correlator (\ref{corrfn}) as \be\label{orrfn}F(\tau_1,\tau_2,\cdots,\tau_n)=\la\Omega|\phi(\vec x_n,\tau_n) \phi(\vec x_{n-1},\tau_{n-1})\cdots \phi(\vec x_1,\tau_1))|\Omega\ra.\ee
 The important
point is that the operators in eqn.~(\ref{orrfn}) are $\tau$-ordered: from right to left, they appear in order of increasing $\tau$. 

The correlator (\ref{corrfn}) can be analytically continued from real values of the $\tau$'s to complex values, say
\be\label{norrfn}\tau_k=\phi_k+i t_k,\ee
with $\phi_k$ and $t_k$ real.  As long as  $H$ is hermitian and nonnegative,
and as long as we only vary the $\tau_k$ in a way that preserves
the ordering of their real parts $\phi_k$, the correlators (\ref{orrfn}) extend to holomorphic functions of $\tau_k$.  For example, for $n=2$,
the correlator can be written as \be\label{tolf}F(\tau_1,\tau_2)=\bigl\la\Omega|\phi(\vec x_2,0)\exp(-(\tau_2-\tau_1)H)\phi(\vec x_1,0)|\Omega\bigr\ra. \ee  Here the factor
$\exp(-(\tau_1-\tau_1)H)=\exp(-(\phi_2-\phi_1)H)\exp(-i(t_2-t_1)H)$ provides convergence as long as $\phi_2-\phi_1>0$.  (The unitary operator 
$\exp(-i(t_2-t_1)H)$ does not spoil this convergence.)  The same argument
applies for any $n$.  

Now suppose that we want to compute a real time correlation function \be\la\Omega|\phi(\vec x_n,t_n)\phi(\vec x_{n-1},t_{n-1})\cdots \phi(\vec x_1,t_1)|\Omega\ra.\ee
Note that this is {\it not}  necessarily a time-ordered product; we want the vacuum expectation value of the operators
taken in the indicated order,  {\it without} assuming that $t_n>t_{n-1}>\dots>t_1$.   Thus in general we want an out of time order (OTO)
correlation function.

Naively, we can get a real time correlator by starting from  the Euclidean  correlator and setting  $\tau_k=it_k$.  But if we do take the $\tau_k$
to be purely imaginary, we will lose the convergence that was just mentioned, and we will lose the information needed to order the operators
in the desired fashion.   Instead we should give the $\tau_k$ small real parts $\veps_k$, 
\be\label{comptime}\tau_k=\veps_k+it_k,\ee
chosen so that $\veps_n>\veps_{n-1}>\cdots
>\veps_1$.  The imaginary time differences between the operator insertions will ensure that the Euclidean correlation function can be analytically continued to arbitrary real values of the $t_k$, and, with the
given inequalities among the $\veps_k$, the operators will be ordered in the desired fashion. In the limit $\veps_k\to 0$, one will get the desired
OTO Lorentz signature correlator.   Feynman's recipe to construct time-ordered products
in Lorentz signature is equivalent to
the special case of this with $\veps_k=\veps t_k$ for all $k$, with $\veps>0$.  This condition ensures that ordering the operators
with increasing $\veps_k$ is the same as ordering them with increasing $t_k$.  

All of these considerations apply equally well if the $\tau$ direction is compactified on a circle. Then the ground state expectation value is 
replaced by a thermal trace in the standard way.  The procedure described above can be used to compute OTO real time correlators at nonzero temperature.

It has been argued \cite{larkin,Almheiri:2013hfa,Shenker:2013pqa,Kitaev:2014t1,Roberts:2014ifa} that the behavior of OTO thermal correlators provides a criterion for quantum chaos in large $N$ systems. At temperature $1/\beta$, one can consider a correlator 
such as \be \label{vwvw}F(t)=\bigl \la V(0)W(\beta/4+it)V(\beta/2)W(3\beta/4+it)\bigr\ra\ee (here $V,W$ are generic hermitian operators, we suppress the spatial coordinates,
and $\la~~~\ra$ denotes a thermal trace).   
The precise imaginary time values $\beta/4$, etc., of  the operators are not important here.  What is important is how they are ordered in
imaginary time -- which differs from  how they are ordered in real time. As a criterion for chaos, we require that $F(t)$ should become small for large values of $t$, regardless of the choice of $V,W$. The manner in which it becomes small can be used as a more fine-grained measure of the strength of chaos. In large $N$ systems one expects an exponentially growing deviation from a nonzero initial value, schematically
\be
F(t) = 1 - \frac{1}{N}e^{\lambda_L t} + \dots
\ee
where the dots resum to make the late-time behavior small. The parameter $\lambda_L$ is sometimes called the ``chaos exponent,'' and is bounded by $\lambda_L \le \frac{2\pi}{\beta}$ \cite{Maldacena:2015waa}. In this paper we consider the four-point function at order $1/N$, so we expect to find exponential growth in the OTO correlator, but we will not see the higher order effects that eventually make $F$ small.

A related
quantity that is believed to contain roughly the same information is the double commutator \be\bigl\la \bigl[V(0),W(it)] [V(\beta/2),W(\beta/2+it)\bigr]
\bigr\ra,\ee where again the precise value $\beta/2$ is not important.
Here in view of the preceeding discussion, a matrix element of the commutator $[V(0),W(it)]=V(0)W(it)-W(it)V(0)$  can be evaluated by an
(analytically continued) Euclidean correlator with insertion of $(V(\veps)-V(-\veps))W(it)$, $\veps>0$.  The infinitesimal displacement of the operators
in Euclidean time gives the desired ordering.  Likewise the second commutator $[V(\beta/2),W(\beta/2+it)]$ can be derived from 
$(V(\beta/2+\veps)-V(\beta/2-\veps))W(\beta/2+it)$.   So the double commutator corresponds to a correlator
\be\label{dbc}\bigl\la \bigl(V(\veps)-V(-\veps)\bigr)\;W(it) \;\bigl(V(\beta/2+\veps)-V(\beta/2-\veps)\bigr)\;W(\beta/2+it)\bigr\ra \ee
that can be constructed by analytic continuation from a Euclidean correlator.

The chaos region of the SYK model was originally described in \cite{kitaevfirsttalk} using a retarded kernel and an integral equation in real time.
This method was further described in \cite{Maldacena:2016hyu}, and in that paper another approach was introduced based on direct analytic continuation
of the Euclidean correlation functions. This second method in principle gives more information. In the rest of this section, we review and explain 
the retarded kernel approach, and apply
this method to two-dimensional bosonic and supersymmetric models.   The alternative approach via direct analytic continuation  will
be  the topic of section \ref{sec:anacon}.

\subsection{Ladder Diagrams And Double Commutators}\label{ladder}  

\begin{figure}[ht]
\begin{center}
\includegraphics[width=.7\textwidth]{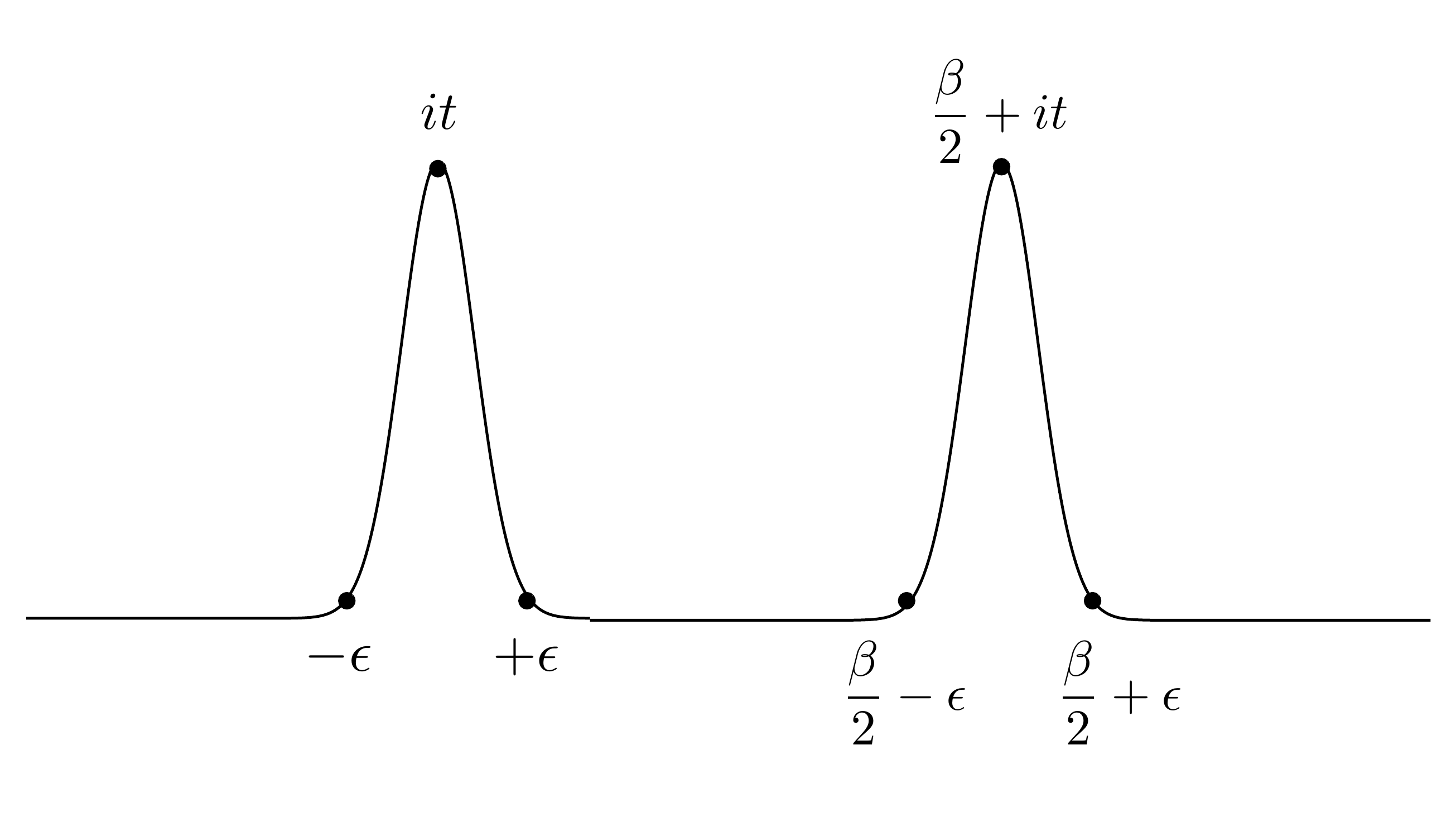}
\caption{\small{A complex time contour appropriate for computing the expectation value of the double commutator (\ref{bbc}).
Two operators are inserted at the points $it$ and $\beta/2+it$, and the other two at $\pm \varepsilon$ and $\beta/2\pm \varepsilon$ (with
some choices of the signs.  (To make a thermal trace, the left and right ends of the contour should be glued together.)}}\label{simpleone}
\end{center}
\end{figure}

To compute a correlator such as (\ref{dbc}), we choose a complex time contour that passes through all of the complex time points
at which one wishes to evaluate the correlation function.  It is convenient to pick a contour along which imaginary time is always increasing,
while real time may zig-zag or ``fold'' backwards and forwards, to pass through the desired complex time points.

For instance, a simple contour to compute the double commutator (\ref{dbc}) is shown in fig. \ref{simpleone}.  The contour hugs the imaginary
time axis except for real time excursions that are needed to ensure that the contour passes through the desired points that are off of that
axis.   In this example, the relevant points are $\tau=it$  and $\tau=\beta/2+it$.  

Feynman diagrams on such a complex time contour can be computed in a fairly normal way.  Feynman vertices that are derived from
a term in the action, for example an interaction $\int d\tau \,\U(\phi(\tau))$, are simply integrated over the chosen contour.  Since we have
chosen a contour along which imaginary time is everywhere increasing, for the propagator we can simply take the analytic continuation to
complex time of the Euclidean propagator.  

In \cite{Stanford:2015owe}, the large $t$ behavior of a four-point function in a weakly coupled scalar field theory was evaluated in the configuration of fig. \ref{simpleone}.
It was argued that the dominant contributions came from certain ladder diagrams (in which one includes self-energy corrections to the propagators).
The vertices of the ladder diagrams were arranged on the contour in a way that we will describe momentarily.

We will consider the case of the SYK model, taking for $V$ one of the elementary fields\footnote{Since we are taking
$V$ and $W$ to be fermionic,  the commutators in $\la [V,W]^2\ra$ must be replaced by anticommutators.}
 $\psi_i$ of this model and for $W$ another
of the elementary fields $\psi_j$.   The large $N$ disorder averaged four-point function of the SYK model is given by ladder diagrams
(with self-energy corrections) even before we take the limit relevant to chaos.  So we can skip that part of the argument from \cite{Stanford:2015owe}.
However, we will explain how the ladder diagrams should be arranged along the contour.

{\it A priori}, a vertex in a Feynman diagram might be attached to any part of the contour in fig. \ref{simpleone} -- either one of the horizontal
segments, or one of the vertical ``rails'' -- the left rail that connects to $\tau=it$, and the right rail that connects to $\tau=\beta/2+it$.  The rails
have left and right sides, and an operator attached to a rail might be attached on either the left or right side of the rail.   Two contributions differing
only by which side of a rail a given interaction vertex  is inserted on will often tend to cancel, simply because $d\tau$ is positive imaginary on one side
and negative imaginary on the other.

\begin{figure}[ht]
\begin{center}
\includegraphics[width=.25\textwidth]{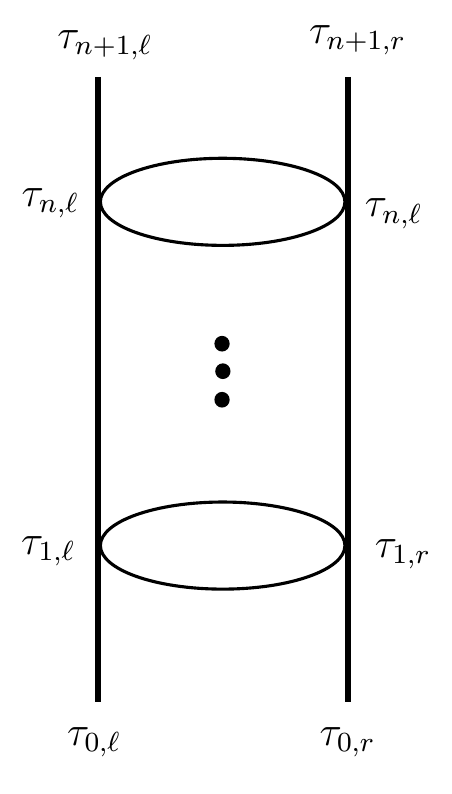}
\caption{\small{A ladder diagram with $n$ rungs. Vertices on the left and right sides of the ladder are labeled by complex time parameters $\tau_{i,l }$ and
$\tau_{i,r}$, $i=0,\dots,n+1$.   To construct a contribution to the double commutator of eqn.~(\ref{bbc}), we take $\tau_{0,l }=\pm\varepsilon$,
$\tau_{n+1,l }=it_1$, $\tau_{0,r}=\beta/2\pm \varepsilon$,  and $\tau_{n+1,r}=\beta/2+it_2$.    As explained in the text, to get a nonzero contribution, all
 the $\tau_{i,l }$ and $\tau_{i,r}$ must then be on the left or right ``rail'' of the complex time contour of fig. \ref{simpleone}, arranged
 in increasing order of real time.}}
 \label{imagineladder}
\end{center}
\end{figure}

In fig. \ref{imagineladder}, we consider a ladder diagram contributing in this situation to the double commutator
\be\label{bbc}W(t_1,t_2)= \bigl\la \bigl(\psi_i(\veps)-\psi_i(-\veps)\bigr)\;\psi_j(it_1)\;\bigl(\psi_i(\beta/2+\veps)-\psi_i(\beta/2-\veps)\bigr)\psi_j(\beta/2+it_2)\bigr\ra.~~t_1,t_2>0.\ee
In the diagram, the operator $\psi_i$ at the bottom of the left rail, inserted at $\tau_{0,l }=\pm\veps$,
is connected by a propagator to precisely one vertex
at complex time $\tau_{1,l }$.  Here $\tau_{1,l }$ must be located on the left rail, or we will get 0 as $\veps\to 0$, because of a cancellation
between the two contributions from $\tau_{0,l }=\pm\varepsilon$.  If $\tau_{1,l }$ is on the left rail, and inevitably, given the form of fig. \ref{simpleone},
at imaginary time between $\veps $ and $-\veps$, the cancellation is avoided because the two contributions involve two-point functions
$\la \psi_i(\tau_{1,l })\psi_i(\tau_{0,l })\ra$ with different operator orderings.  Likewise, $\tau_{1,r}$ must be on the right rail.

Now we have to consider the fact that $\tau_{1,l }$ could be on either side of the left rail.  For the same reason as in the last paragraph, this
will lead to a cancellation in the small $\veps$ limit unless $\tau_{2,l }$ is also inserted on the left rail, and at a value of real time greater than 
that of $\tau_{1,l }$.  Likewise, $\tau_{2,r}$ must be inserted on the right rail, and at a real time greater than that of $\tau_{1,r}$.   Continuing
in this way, we find that all the $\tau_{i,l }$ are on the left rail, at increasing values of $t$, and likewise the $\tau_{i,r}$ are similarly
arranged on the right rail.  Thus the ladder diagram of fig. \ref{imagineladder} is naturally mapped to spacetime, with the left and right sides of the ladder
mapping respectively to the left and right rails of the spacetime contour, and with higher rungs of the ladder mapping (on both left and right) to
later values of real time.  

What propagators should we use in this situation?  It is always correct to simply use the analytic continuation of the Euclidean propagator,
and this is what we do for the propagator $G_{l  r}$ from a point on the left rail to a point on the right rail.  However, for propagation between
two points on the same rail, say $\tau=it $ and $\tau'=it'$, with $t'>t$, it is convenient to define a ``retarded'' propagator $G_\ret$ in which 
we sum over which side of the rail $\psi_i(t)$ is inserted on. (Otherwise we would have to incorporate this sum in the Feynman vertices.)
We will make this more explicit in section \ref{sykchaos}.

\begin{figure}[ht]
\begin{center}
\includegraphics[width=.55\textwidth]{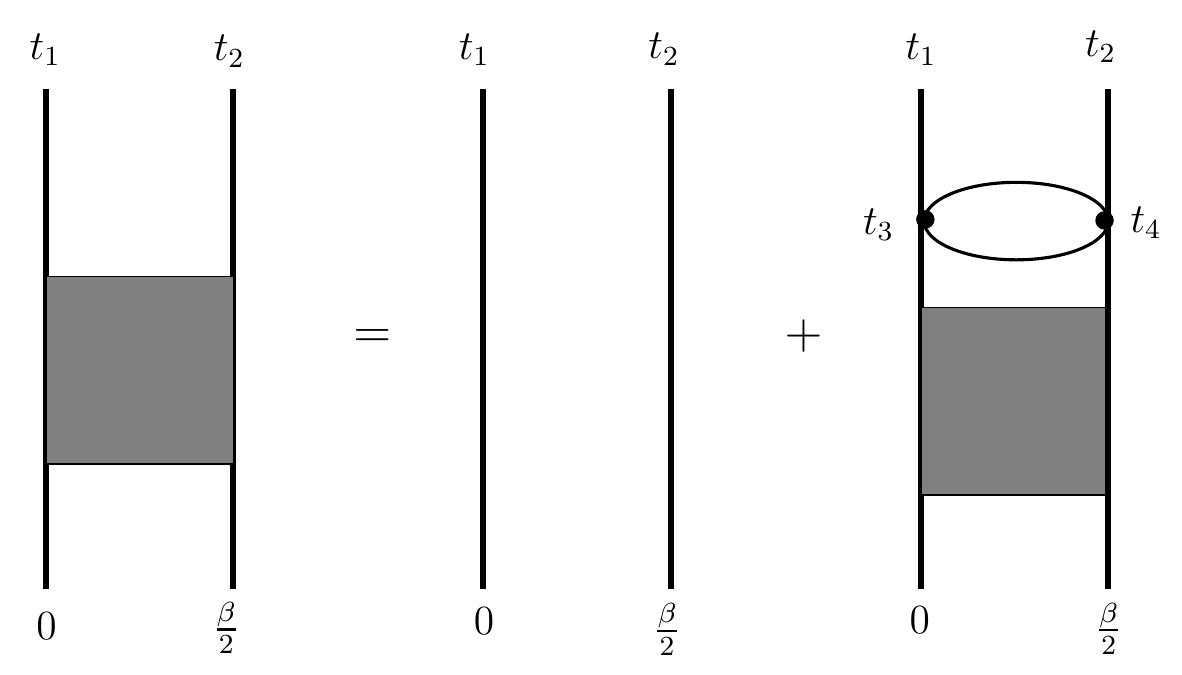}
\caption{\small{Integral equation for the sum of ladder diagrams.}}\label{intequation}
\end{center}
\end{figure}
Now we come to the question of how to determine the large time behavior of the sum of ladder diagrams.  In general, the OTO correlator
$W(t_1,t_2)$ of eqn.~(\ref{bbc}) obeys an integral equation (fig. \ref{intequation}) that can be written
\be\label{inteq}W(t_1,t_2)=W_0(t_1,t_2)+\int_0^{t_1}\!\!\!\!dt_3\int_0^{t_2}\!\!\!\!dt_4 \,\,K_\ret(t_1,t_2,t_3,t_4) W(t_3,t_4),  \ee
where $W_0$ is the free-field value of this correlator, and the retarded kernel is defined by adding a single rung to the ladder.  Thus $K_\ret$
is very similar to the conformal kernel constructed earlier in this paper, but constructed using the appropriate real time propagators.  

Quantum chaos means that $W(t_1,t_2)$ grows exponentially for large $t_1,t_2$.  If this is true, then at large times the free-field term $W_0$
is unimportant in eqn.~(\ref{inteq}), since it vanishes exponentially rather than growing exponentially.  Suppose that it is true that at late
times, $W(t_1,t_2)$ grows exponentially, say $W(t_1,t_2)\sim W_+(t_1,t_2)$ with
\be\label{nteq}W_+(t_1,t_2)=\exp\biggl(\frac{\lambda_L}{2}(t_1+t_2)\biggr) f(t_1-t_2), \ee
with some $\lambda_L>0$ and some function $f(t_1-t_2)$.  This form is natural because of the time-translation symmetry of the problem.   If $W$
does have this behavior, then eqn.~(\ref{inteq}) will reduce at late times to the same  equation for $W_+$ with the free field term dropped:
\be\label{teq} W_+(t_1,t_2)=\int_{-\infty}^{t_1}\!\!\!\!d t_3\int_{-\infty}^{t_2}\!\!\!\!dt_4 \,\,K_\ret(t_1,t_2,t_3,t_4) W_+(t_3,t_4). \ee
Note that in writing this equation, we have changed the lower limit on the $t_3,t_4$ integrals from 0 to $-\infty$.   This is permissible because,
if it is true that $W(t_1,t_2)$ grows exponentially with time, then for large $t_1,t_2$, the integral over $t_3,t_4$ in eqn.~(\ref{inteq}) is dominated
by $t_3,t_4$ near $t_1,t_2$, and hence it does not matter if the lower limit on the integral is 0 or $-\infty$.  

Eqn.~(\ref{teq}) says that $W_+$ is an eigenfunction of the retarded kernel $K_\ret$ with eigenvalue 1.  Thus eigenfunctions of $K_\ret$ with
eigenvalue 1 control the chaos region of the four-point function, somewhat analogously to the fact that eigenfunctions of the conformal
kernel $K$ with eigenvalue 1 control the operator product expansion.

Next we will review how this program was implemented for the 1d SYK model in \cite{kitaevfirsttalk, Maldacena:2016hyu}.

\subsection{Chaos In The SYK Model}\label{sykchaos}

Conformal invariance of the SYK model in the IR limit
means that there is no essential loss of generality to consider the case of inverse temperature $\beta=2\pi$.
Thus we parametrize imaginary time by an angle $\phi$.  The disorder-averaged Euclidean two-point function of the SYK model
is\footnote{In defining $\sgn(\phi_1-\phi_2)$, we subtract a multiple of $2\pi$ so that $\phi_1-\phi_2$ lies
 in a specified interval, for instance $[\alpha,\alpha-2\pi]$ for
some constant $\alpha$.  The discontinuity in sign when $\phi_1-\phi_2$ passes through $\alpha$ reflects the fact that in the thermal
ensemble, the elementary fermions change sign in going around the circle.}
\be\label{disav} G(\phi_1,\phi_2)=\frac{b\,\sgn(\phi_1-\phi_2)}{|2\sin\frac{1}{2}(\phi_1-\phi_2)|^{2\Delta}},\ee
with
\be\label{nisav}J^2b^q\pi
 =\bigl(\frac{1}{2}-\Delta\bigr)\tan\pi\Delta,~~\Delta=
1/q. \ee
In more detail,
\be\label{isav}G(\phi_1,\phi_2)=\begin{cases} \frac{b}{(2\sin(\frac{1}{2}(\phi_1-\phi_2))^{2\Delta}}, & \phi_1-\phi_2>0\cr
                                                                          \frac{-b}{(-2\sin\frac{1}{2}(\phi_1-\phi_2))^{2\Delta}}, & \phi_1-\phi_2<0.\cr\end{cases}\ee
                                                                          
Now it is straightforward to analytically continue to complex variables $\tau_i=\phi_i+it_i$.    The retarded propagator is 
\be\label{sav}G_\ret(t,t')=\theta(t-t')\left(G(it,-\veps+it')-G(it,\veps+it')\right)=\theta(t-t') \frac{2b\cos\pi\Delta}{(2\sinh\frac{1}{2}(t-t'))^{2\Delta}}. \ee
And the left-right propagator is
\be\label{av}G_{l  r}(t,t')=G(it,\pi+it')=\frac{b}{(2\cosh\frac{1}{2}(t-t'))^{2\Delta}}. \ee 
The retarded kernel is
\be\label{wav}K_R(t_1,\dots,t_4)=J^2(q-1)G_R(t_{13})G_R(t_{24})G_{l  r}(t_{34})^{q-2}. \ee
The sign here stands in contrast to the conformal kernel of the SYK model, which according to eqn.~(3.49) of \cite{Maldacena:2016hyu} is
\be\label{plav}K(\phi_1,\dots,\phi_4)=-J^2(q-1) G(\phi_{13})G(\phi_{24})G(\phi_{34})^{q-2}. \ee
The reason for the sign difference is simply that when we act on a wavefunction with $K$, we perform an integral $d\phi_3 d\phi_4$,
while the action of the real time kernel $K_R$ involves an integral $d t_3 d t_4$.  But $\tau=\phi+it$, so in passing to real time,
$d\phi_3 d\phi_4$ becomes $-d t_3 d t_4$.  (Equivalently, in Lorentz signature, an interaction vertex comes with a factor of $i$.)   It turns out that it is convenient to redefine the wavefunction that $K_R$ acts on
by 
\be\label{conjwav}W_+(t_1,t_2)=\exp(-\Delta(t_1+t_2))W'_+(t_1,t_2).\ee
This replaces $K_R$ with 
\be\label{wavprime}K_R'(t_1,\dots, t_4)=\exp\left(\Delta(t_1+t_2)-\Delta(t_3+t_4)\right)K_R(t_1,\dots,t_4).\ee 

We could now write down the integral equation (\ref{teq}), but it is more transparent to do this after changing variables to 
$z=e^{i\tau}=e^{i\phi-t}$.  On the left rail, $z=e^{-t}$, and on the right rail,  $z=-e^{-t}$.   Also, since we use $K_R'$ in conjunction
with integration over $t_3$ and $t_4$, what we really want to transform to the $z$ coordinates is $K_R'(t_1,\dots,t_4)d t _3 dt_4$.   A short computation
reveals that 
\begin{align}\label{plavprime}K_R'(t_1,\dots,t_4)dt_3dt_4=& (2\cos \pi\Delta)^2\theta(z_3-z_1)\theta(z_2-z_4)\cdot\left( \frac{J^2b^{q}(q-1)}
{ |z_{13}|^{2\Delta}|z_{24}|^{2\Delta}|z_{34}|^{2\Delta(q-2)}  }dz_3dz_4 \right)\end{align}
This has the following useful property.  Consider the conformal kernel of the SYK model formulated on the real $z$ axis.
It is constructed from the same Feynman diagrams but with different propagators:
\be\label{lav}K(z_1,z_2,z_3,z_4)= -J^2b^{q} \frac{\sgn(z_{13})\sgn(z_{24})}{  |z_{13}|^{2\Delta}|z_{24}|^{2\Delta}|z_{34}|^{2\Delta(q-2)}  }.\ee
However, in the region $z_3>z_1$, $z_4>z_2$ where $K_R'$ is nonzero, the different propagators are constant multiples of each other
and hence $K_R'$ is a constant multiple of $K$.   Therefore, $K_R'$ has some of the symmetry of $K$.  $K$ is invariant under
$SL(2,\R)$ acting on the real $z$ axis plus a point at infinity; $K_R'$ is invariant under the subgroup $z\to az+b$ that leaves fixed the
point at infinity (the $\theta$ functions in $K_R'$ are not invariant under the rest of $SL(2,\R)$).  This gives
$K_R'$ a non-obvious symmetry.
 The scaling symmetry $z\to az$
(which corresponds in terms of $t$ to time-translation symmetry) is a manifest symmetry of $K_R'$, but $z\to z+b$ is not a manifest symmetry
and is most apparent because of the relation to $K$.

We recall that $K$ has for eigenfunctions the three-point functions $\la \psi_i(z_1)\psi_i(z_2)V_h(y)\ra$, where $\psi_i$ is a conformal
primary of  dimension $\Delta$,
$V_h$ is a primary
of dimension $h$ and $y$ is arbitrary.  Taking $y\to\infty$ and rescaling, the eigenfunction reduces to $\text{sgn}(z_{12})/|z_1-z_2|^{2\Delta-h}$.    This particular
eigenfunction is also an eigenfunction for $z\to az+b$, and can be uniquely characterized by the way it transforms under that group (it is invariant
under $z\to z+b$, and scales with an $h$-dependent weight under $z\to az$).  It therefore  is also an eigenfunction of $K_R'$.  ($K$ has more eigenfunctions,
with $y\not=\infty$, and $K_R'$ also has more eigenfunctions, namely the ones that are not invariant under $z\to z+b$.)

With $z_1=e^{-t_1}$, $z_2=-e^{-t_2}$, the eigenfunction $W_+'=1/|z_1-z_2|^{2\Delta-h}$ is
\be\label{eif} W_+'(t_1,t_2) =\frac{\exp\left((
\Delta-h/2)(t_1+t_2)\right)}{\left(2\cosh\frac{1}{2}(t_1-t_2)\right)^{2\Delta-h}}.\ee
But remembering eqn.~(\ref{conjwav}), the eigenfunction $W_+(t_1,t_2)$ that appears in the four-point function is
\be\label{nelf} W_+(t_1,t_2)=\frac{\exp\left(-\frac{h}{2}(t_1+t_2)\right)}{\left(2\cosh\frac{1}{2}(t_1-t_2)\right)^{2\Delta-h}}.\ee
Thus we will get exponential growth in the chaos region if $\mathrm{Re}\,h<0$, and the fastest exponential growth will occur for the most negative
value of $\mathrm{Re}\,h$ such that $K_R'$ has eigenvalue 1.   

The eigenvalues $k(h)$ and $k_R(h)$ of $K$ and $K_R$ can be computed by setting $z_1=1$, $z_2=0$,
so that $W_+'(z_1,z_2)=1$.  Accordingly, the eigenvalue equations for $W_+'$ reduce to $k(h)=KW_+'(1,0)$, $k_R(h)=K_R'W_+'(1,0)$.  So as in 
\cite{KitaevTalks}, \cite{Maldacena:2016hyu}, 
\be\label{kcbose}k(h)= -J^2b^q(q-1)\int_{-\infty}^\infty \!\!\!\!d z_3 \int_{-\infty}^\infty \!\!\!\!dz_4\,\, \frac{\sgn(1-z_3)\sgn(-z_4)\sgn(z_{34})} { |1-z_3|^{2\Delta}|z_4|^{2\Delta}|z_{34}|^{2-2\Delta-h} } \ee
and
\be\label{krbose}k_R(h)=(2\cos\pi\Delta)^2 J^2b^q(q-1)\int_{1}^\infty \!\!\!\!d z_3 \int_{-\infty}^{0}\!\!\!\!dz_4 \,\,\frac{1} { |1-z_3|^{2\Delta}|z_4|^{2\Delta}|z_{34}|^{2-2\Delta-h }} .\ee

Let us first describe some qualitative properties of $k_R(h)$ that do not depend on doing the integral.   First of all, the integral for $k_R(h)$
converges for $\Re\, h<2\Delta$, and the integrand is positive-definite.  
  On the real $h$ axis for $h<2\Delta$, the integrand, for fixed $z_3,z_4$, is a monotonically increasing function of $h$, and therefore
 $k_R(h)$ likewise is monotonically increasing.   
 For $h\to -\infty$, $k_R(h)\to 0$ because the integrand vanishes pointwise, but for $h\to 2\Delta$, the integral ceases to converge and
 $k_R(h)\to +\infty$.  Along with the monotonicity, this implies that there is a unique real
$h_0$ for $h_0<2\Delta$ such that $k_R(h_0)=1$.  Finally, consider varying $h$ keeping fixed its real part.   This gives the integrand a nontrivial
phase, leaving its absolute value constant.  Therefore, along a line of fixed $\Re\,h<2\Delta$, the maximum absolute value of $k_R(h)$ is attained for
${\mathrm{Im}}\,h=0$.   It follows that there are no solutions to $k_R(h)=1$ with $\Re\,h<h_0$ or $\Re\,h=h_0$, ${\mathrm{Im}}\,h\not=0$.  Accordingly, if $h_0<0$, the 
four-point function grows exponentially in the chaos region and this growth is dominated by the eigenfunction with $h=h_0$.  

Evaluating the integral gives \cite{Maldacena:2016hyu}  
\be\label{evint}k_R(h)=\frac{\Gamma(3-2\Delta)\Gamma(2\Delta-h)}{\Gamma(1+2\Delta)\Gamma(2-2\Delta-h)},\ee
and therefore $h_0=-1$, a value that is in accord with expectations from black hole physics and that saturates a general bound on chaos \cite{Maldacena:2015waa}.
Moreover, evaluation of the integrals gives a relation between $k_R(h)$ and $k(h)$:
\be\label{wevint}\frac{k_R(1-h)}{k(h)}=\frac{\cos\pi(\Delta-h/2)}{\cos\pi(\Delta+h/2)}.\ee
(Because $k(h)=k(1-h)$, this can also be written as a relation between $k_R(h)$ and $k(h)$.)
This implies that $k_R(1-h)=k(h)$ if $h$ is an even integer.
It was shown in \cite{Maldacena:2016hyu} that this, along with the qualitative information about $k_R(h)$ found in the last paragraph, is very
useful in understanding the analytic continuation of the four-point function from the OPE region to the chaos region.  We will see in section \ref{sec:anacon}
that in the 2d models studied in the present paper, an analogous relation plays a similar role.

Unfortunately, we do not know a very direct way to obtain the relation (\ref{wevint}).  The integrals for $k_R(h)$ and $k(h)$ can be evaluated
by a method described in section 3.2.3 of \cite{Maldacena:2016hyu}, or by the method we followed in discussing eqn.~(\ref{kcbthree}).  Either
way, one sees without detailed calculation that the ratio $k_R(1-h)/k(h)$, or equivalently $k_R(h)/k(h)$, is a trigonometric function.
But to identify this function or even to deduce the important conclusion that $k_R(1-h)=k(h)$ when $h$ is an even integer seems
to require a detailed calculation.  In Appendix \ref{reta}, we give an explanation of sorts for an analogous relation in two dimensions.

\subsection{Extension To Two Dimensions}\label{twodext}
\newcommand{\barh}{\widetilde{h}}
In two dimensions, the logic behind the retarded kernel is identical to the one-dimensional case considered above. However, because we have an extra $x$ coordinate to integrate over, we will find a somewhat richer set of eigenfunctions. We will begin by discussing the formal bosonic model, before moving on to the better-defined supersymmetric model in the next section.

In order to form the retarded kernel, we need to know the propagators $G_R$ and $G_{lr}$. These follow from analytic continuation of the Euclidean correlator, which we write using Euclidean time $\tau$ on the thermal circle (still of circumference $2\pi$) and spatial position $x$ on an infinite line:
\be
G(\tau_1,x_1;\tau_2,x_2) = \frac{b}{(2\sinh\frac{x_{12}+i\tau_{12}}{2})^{\Delta}(2\sinh\frac{x_{12}-i\tau_{12}}{2})^{\Delta}}.
\ee
The retarded propagator is
\begin{align}
G_R(t_1,x_1;t_2,x_2) &= \theta(t_{12})\left[G(it_1,x_1;-\veps+it_2,x_2) - G(it_1,x_2;\veps + it_2,x_2)\right]\\
&=-2ib\sin(\pi\Delta)\frac{\theta(t_{12}-|x_{12}|)}{(2\sinh\frac{t_{12}-x_{12}}{2})^\Delta(2\sinh\frac{t_{12}+x_{12}}{2})^\Delta},         
\end{align}
and the left-right propagator is
\be
G_{lr}(t_1,x_2;t_2,x_2) = G(it_1,x_1;it_2+\pi,x_2) = \frac{b}{(2\cosh\frac{t_{12}-x_{12}}{2})^\Delta(2\cosh\frac{t_{12}+x_{12}}{2})^\Delta}.
\ee
The retarded kernel is then
\be
K_R(1,2,3,4) = -J^2(q-1)G_R(1,3)G_R(2,4)G_{lr}(3,4)^{q-2},
\ee
where we abbreviated the arguments as e.g. $t_1,x_1\rightarrow 1$. Note that the sign here is again opposite to that of the conformal kernel in two dimensions. As before, this sign comes from converting two Euclidean time integrals to Lorentzian time, or equivalently from two factors of $i$ coming from the Lorentz-signature interaction vertex.

The $x,t$ coordinates are helpful for thinking about the retarded kernel as a finite temperature problem, but for guessing eigenfunctions and doing the integrals, it is better to use coordinates $z = e^{x +i\tau}$ and $\bar{z} = e^{x-i\tau} = e^{x+t}$. In Lorentz signature, $z$ and $\bar{z}$ become real light cone coordinates. However, the following derivation is more symmetrical if we further change variables to $u = z$ and $v = \bar{z}^{-1}$. With $\tau=it$ or $\pi+it$ on the left or right rail, this gives explicitly
\begin{align}
u &= e^{x-t}, \hspace{29pt} v = e^{-x-t}, \hspace{30pt} (\text{left rail})\\
u &= -e^{x-t}, \hspace{20pt} v = -e^{-x-t}, \hspace{20pt} (\text{right rail}).
\end{align}
In particular, $u,v>0$ on the left rail, and $u,v<0$ on the right rail.
It is also convenient to conjugate the kernel by factors of $u,v$:
\be\label{conjKR}
K_R'(1,2,3,4) = (u_1v_1u_2v_2)^{-\frac{\Delta}{2}}K_R(1,2,3,4)(u_3v_3u_4v_4)^{\frac{\Delta}{2}}.
\ee
One finds
\be
K_R'(1,2,3,4)dt_3dx_3dt_4dx_4 = (2\sin(\pi\Delta))^2J^2b^{q}(q{-}1)\frac{\theta(u_{31})\theta(u_{24})}{u_{31}^\Delta u_{24}^\Delta u_{34}^{2-2\Delta}}\frac{\theta(v_{31})\theta(v_{24})}{v_{13}^\Delta v_{42}^{\Delta}v_{34}^{2-2\Delta}} \frac{du_3dv_3du_4dv_4}{4}.
\ee
A nice feature of this kernel is that it exactly factorizes into a product of one-dimensional kernels of the form (\ref{plavprime}) for left- and right-moving modes. We can find eigenfunctions by taking\footnote{As in the 1d discussion, we could motivate this by observing that $K_R'$ has some of the conformal symmetry of two-dimensional
Minkowski spacetime.  However, it is quicker and more or less equivalent to just use the way that $K_R'$ factors as a product of one-dimensional kernels.}  conformal three-point functions where the time argument of the third point has been taken to minus infinity, which corresponds to $u,v = \infty$. The candidate eigenfunctions are then
\begin{align}\label{findeig}
W_+'(u_1,v_1;u_2,v_2) &= \frac{1}{u_{12}^{\Delta-h}}\cdot\frac{1}{v_{12}^{\Delta-\barh}}\\
W_+(t_1,x_2;t_2,x_2) &= \frac{\exp\left(-\frac{h+\barh}{2}(t_1+t_2) + \frac{h-\barh}{2}(x_1+x_2)\right)}{(2\cosh\frac{x_{12}-t_{12}}{2})^{\Delta-h}(2\cosh\frac{x_{12}+t_{12}}{2})^{\Delta-\barh}}.
\end{align}

In the application of these functions to the chaos region of the four-point function, we look for eigenfunctions with eigenvalue 1; see eqn.~(\ref{teq}). We expect to have functions that are growing exponentially in time, but we demand normalizability in space, so that $h - \barh$ should be purely imaginary, i.e. the real parts of $h,\barh$ will be equal. The dominant growth will be the eigenfunction with the largest value of $h+\barh$, subject to the constraint that the eigenvalue of $K_R$ is 1. Because the kernel is a product of two one-dimensional kernels, we can use the logic described below (\ref{krbose}) to conclude that the maximum growth will be the solution to $K_R = 1$ with $h = \barh$.

To find exactly what the allowed growth rates are, we evaluate the integrals. One finds that (\ref{findeig}) does indeed give an eigenfunction of $K'_R$, with eigenvalue
\begin{align}
k_R(h,\barh) &= [2\sin(\pi\Delta)]^2J^2 b^q(q{-}1)\frac{1}{4}\frac{\Gamma(1-\Delta)^2\Gamma(\Delta-h)}{\Gamma(2-\Delta-h)}\frac{\Gamma(1-\Delta)^2\Gamma(\Delta-\barh)}{\Gamma(2-\Delta-\barh)}.
\end{align}
After substituting in for $q$ and $J^2b^q$ using (\ref{solution}), we find a simple relation between the eigenvalues of the retarded kernel and the regular Euclidean kernel in (\ref{bosker})
\be\label{retconf}
k_R(h,\barh) = k(1-h,\barh).
\ee
This relation looks a little strange, because  the integrals that define the retarded kernel make sense for arbitrary $h,\barh$, whereas for the Euclidean kernel we require the difference of the arguments (in this case $1-h-\barh$) to be an integer. The more precise statement is that (\ref{retconf}) holds for all values where the Euclidean kernel makes sense. However, it turns out that for the particular analytic expression we wrote for $k$ in (\ref{bosker}), eqn.  (\ref{retconf}) is true for all $h,\barh$. The fact that there is an analytic expression for $k$ such that (\ref{retconf}) holds will be essential for getting consistency with the method of direct analytic continuation.   A somewhat more direct explanation of
this relationship is presented in Appendix \ref{reta}.

We will now briefly discuss the solutions to $k_R = 1$. We argued above that the physically allowable solutions have $h -\barh$ purely imaginary. It is convenient to parametrize $h,\barh$ with 
\be\label{chaoslk}
h = -\frac{\lambda_L}{2} + \frac{ip}{2}, \hspace{20pt} \barh = -\frac{\lambda_L}{2}-\frac{ip}{2},
\ee
where in general $\lambda_L$ could be complex but $p$ is real. The associated eigenfunction is, for the simple case $t_1 = t_2 = t$ and $x_1 = x_2=x$, 
\be
W_+(t,x;t,x) = \exp\left(\lambda_L t +ip x\right).
\ee
The constraint $k_R = 1$ can be used to determine a function $\lambda_L(p)$. See figure \ref{fig:superlambda} for a plot of this function in the supersymmetric theory (the bosonic case is similar). For a reason noted above, the maximum value of the real part of $\lambda_L$ is at $p=0$. 
The behavior of $\lambda_L(0)$ as a function of $\Delta$ is quite interesting. It does not saturate the chaos bound $\lambda_L=1$, unlike in the one-dimensional theory. Instead, we get a function of $\Delta$ that vanishes as $\Delta$ approaches zero, consistent with the idea that the large $q$ theory is free. In the bosonic model, the function approaches a maximum of around $0.77$ when $\Delta = 1$, but this value corresponds to $q = 2$, which should not be a chaotic theory. We expect that in fact this value of $\Delta$ has to be treated separately, and not as a limit of $\Delta <1$. We should emphasize that the retarded kernel computation for the bosonic model is formal because the theory is unstable. However, it is a useful warmup for the supersymmetric version, which we discuss next.

\subsection{Supersymmetric Model In Two Dimensions}\label{supertwod}
The propagators for the supersymmetric retarded kernel can be obtained from analytic continuation of the Euclidean correlator
\be\label{eucsuper}
G(\tau_1,x_2,\theta_1,\bar{\theta}_1;\tau_2,x_2,\theta_2,\bar{\theta}_2) =\frac{b}{(2\sinh\frac{x_{12}+i\tau_{12}}{2}-\theta_1\theta_2)^{\Delta}(2\sinh\frac{x_{12}-i\tau_{12}}{2}-\bar{\theta}_1\bar{\theta}_2)^{\Delta}}.
\ee
As before, for discussing the eigenfunctions and eigenvalues of the retarded kernel, it is convenient to change variables to $z = e^{x + i\tau}$ and $\bar{z} = e^{x-i\tau}$ and further to $u = z$ and $v = \bar{z}^{-1}$. In the supersymmetric case, we also define new $\theta$ variables by
\begin{align}
\theta^u &= e^{\frac{x-t}{2}}\theta, \hspace{29pt}\theta^v = ie^{\frac{-x-t}{2}}\bar{\theta}, \hspace{25pt} (\text{left rail})\\
\theta^u &= ie^{\frac{x-t}{2}}\theta, \hspace{24pt}\theta^v = -e^{\frac{-x-t}{2}}\bar{\theta}, \hspace{20pt} (\text{right rail}).
\end{align}
This is simply saying that the $\theta$ variables transform with weight one half under conformal transformations. After continuing (\ref{eucsuper}) to find $G_R$ and $G_{lr}$, making the above coordinate transformations, and defining a conjugated kernel as in (\ref{conjKR}), one finds
\begin{align}\label{krsuper}
&K'_R(1,2,3,4)dt_3dx_3dt_4dx_4d\theta_3d\bar{\theta}_3 d\theta_4d\bar{\theta}_4 =\\
 &-(2\sin(\pi\Delta))^2J^2b^{\hat{q}}(\hat{q}{-}1)\frac{\Theta(u_{31})\Theta(u_{24})}{\langle 31\rangle_u^\Delta\langle 24\rangle_u^\Delta \langle 34\rangle_u^{1-2\Delta}}\frac{\Theta(v_{31})\Theta(v_{24})}{\langle 31\rangle_v^\Delta\langle 24\rangle_v^\Delta\langle 34\rangle_v^{1-2\Delta}} \frac{du_3dv_3du_4dv_4 d\theta_3^ud\theta_3^vd\theta_4^ud\theta_4^v}{4}\notag
\end{align}
where we are using the notation $\langle ij\rangle_u = u_i-u_j - \theta_i^u\theta_j^u$ and $\langle ij\rangle_v = v_i-v_j - \theta_i^v\theta_i^v$, and we have written the step functions with capital $\Theta$ to distinguish them from the Grassman $\theta$ variables.

As in the bosonic case, the kernel factorizes into a product of one-dimensional kernels. Candidate eigenfunctions are again given by three-point functions with a third operator that we take to $t = -\infty$. In principle, we have four choices for the third operator, since it could have either a bosonic or a fermionic primary in the $u$ and $v$ sides. All four options lead to eigenfunctions of $K_R$, but it is only the case with a bosonic primary on both sides that leads to a growing solution to $K_R=1$, so we will focus on this important case first. The eigenfunctions are
\begin{align}\label{findsupereig}
W_{+BB}'(z_1,\bar{z}_1;z_2,\bar{z}_2) &= \frac{1}{\langle 12\rangle_u^{\Delta-h}}\cdot\frac{1}{\langle 12\rangle_v^{\Delta-\barh}}\\
W_{+BB}(t_1,x_2;t_2,x_2) &= \frac{\exp\left(-\frac{h+\barh}{2}(t_1+t_2) + \frac{h-\barh}{2}(x_1+x_2)\right)}{(2\cosh\frac{x_{12}-t_{12}}{2}-i\theta_1\theta_2)^{\Delta-h}(2\cosh\frac{x_{12}+t_{12}}{2}+i\bar{\theta}_1\bar{\theta}_2)^{\Delta-\barh}}.
\end{align}
In these expressions, the $1$ coordinates are assumed to be on the left rail, and the $2$ coordinates on the right. We compute the eigenvalues by doing the integrals, e.g.
\begin{align}
\int_{u_1}^\infty \!\!\!\!du_3 \int_{-\infty}^{u_2} \!\!\!\!du_4 \int d\theta_3^ud\theta_4^u\frac{1}{\langle 31\rangle_u^{\Delta}\langle 24\rangle_u^{\Delta}\langle 34\rangle_u^{1-\Delta-h}} = -\frac{1}{\langle 12\rangle_u^{\Delta-h}}\frac{\Gamma(1-\Delta)^2\Gamma(\Delta-h)}{\Gamma(1-\Delta-h)}
\end{align}
Together with a similar integral for the $v$ coordinates, this leads to
\begin{align}
k_{R}^{BB}(h,\barh) &= \sin^2(\pi\Delta)J^2b^{\hat{q}}(\hat{q}-1)\frac{\Gamma(1-\Delta)^2\Gamma(\Delta-h)}{\Gamma(1-\Delta-h)}\frac{\Gamma(1-\Delta)^2\Gamma(\Delta-\barh)}{\Gamma(1-\Delta-\barh)}\\
&=-\frac{\Gamma(1-\Delta)^2}{\Gamma(\Delta+1)\Gamma(\Delta-1)}\frac{\Gamma(\Delta-h)\Gamma(\Delta-\barh)}{\Gamma(1-\Delta-h)\Gamma(1-\Delta-\barh)}.
\end{align}
Note that in the first line of this expression, the minus sign in (\ref{krsuper}) has canceled against another minus sign that we get when we exchange $d\theta^v_3$ and $d\theta_4^u$ in order to factorize the integral into separate $u$ and $v$ integrals.

A fact which will be important in comparing to the direct analytic continuation method is that this kernel is closely related to the Euclidean kernel $k^{FB}$. More precisely, when $1-h - \barh$ is an integer, we have that
\be\label{krkc}
k_{R}^{BB}(h,\barh) = k^{FB}(\tfrac{1}{2}-h,\barh).
\ee
These are the values for which our previous computation of $k^{FB}(1/2-h,\th)$ makes sense.\footnote{In our previous work, $k^{FB}(h, \th)$ is defined for $h-\th$ a half-integer,
and hence $k^{FB}(1/2-h,\th)$ is defined for $1-h-\th$ an integer.} However, for the particular analytic $k^{FB}$ that we wrote in (\ref{kFB}), this relation is not true away from these values. Of course, we could equally well use $k_{R}^{BB}$ to define an equivalent new $k^{FB}$ such that (\ref{krkc}) holds for all $h,\barh$.
\begin{figure}[t]
\begin{center}
\includegraphics[width=.4\textwidth]{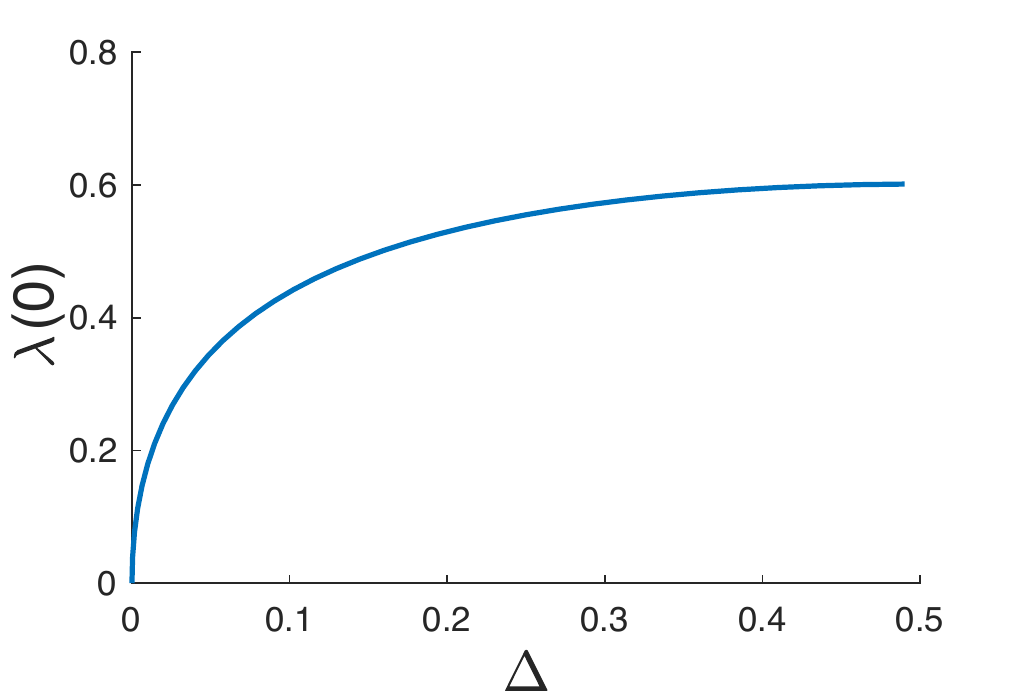}
\includegraphics[width=.4\textwidth]{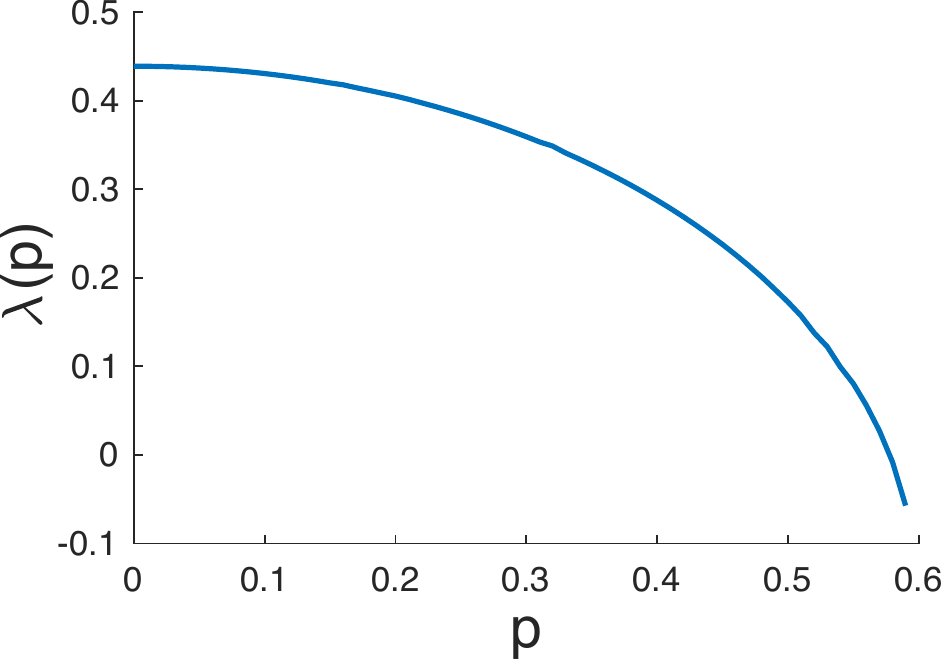}
\caption{\small{{\bf Left}: the chaos exponent $\lambda_L(0)$ as a function of $\Delta$ for the two-dimensional supersymmetric theory. {\bf Middle}: the function $\lambda_L(p)$ for $\Delta = 0.1$. The function is symmetric in $p\rightarrow -p$ and it becomes complex just beyond where the curve ends. 
}}\label{fig:superlambda}
\end{center}
\end{figure}

As in the bosonic model, we define $\lambda_L(p)$, the chaos exponent as a function of $p$, by solving the equation $k_{R}^{BB}(h,\barh) = 1$ with $h,\barh$ related to $\lambda_L$ and $p$ as in (\ref{chaoslk}). In figure \ref{fig:superlambda} we plot $\lambda_L(0)$ as a function of $\Delta$ and we also plot $\lambda_L(p)$ for a particular value of $\Delta$. The largest physical value of $\lambda_L(0)$ is for $\h q = 3$, where we find $\lambda_L(0) \approx 0.5824$.

Although they do not lead to growing solutions, for completeness we will also discuss the eigenvectors of the kernel where the third operator has a fermionic primary on one or both of the holomorphic and antiholomorphic sides. The kernel is a product, so we will discuss the holomorphic side. Including a factor as in eqn.~(\ref{onormal}) and specializing to the case where the time argument of the third operator goes to infinity, we see that the holomorphic factor of the eigenfunction will be
\be
\frac{1}{\langle 12\rangle_u^{\Delta-h}}\frac{\theta_1^u-\theta_2^u}{u_{12}^{1/2}} = \frac{\theta_1^u-\theta_2^u}{u_{12}^{1/2+\Delta-h}}.
\ee
The eigenvalue can be worked out using
\be
\int_{u_1}^\infty \!\!\!\!du_3 \int_{-\infty}^{u_2} \!\!\!\!du_4 \int d\theta_3^ud\theta_4^u\frac{\theta_3^u-\theta_4^u}{\langle 31\rangle_u^{\Delta}\langle 24\rangle_u^{\Delta}u_{34}^{\frac{3}{2}-\Delta-h}} = \frac{\theta_1^u-\theta_2^u}{u_{12}^{\frac{1}{2}+\Delta-h}}\frac{\Gamma(1-\Delta)^2\Gamma(\frac{1}{2}+\Delta-h)}{\Gamma(\frac{3}{2}-\Delta-h)}.
\ee
Combining with similar factors from the $v$ integrals, one finds
\begin{align}
k_{R}^{FB}(h,\barh) &= -k^{BB}_R(h-\tfrac{1}{2},\th) = k^{BB}(\tfrac{1}{2}-h,\th)\\
k_R^{FF}(h,\th) &= k^{BB}_R(h-\tfrac{1}{2},\th-\tfrac{1}{2}) = k^{BF}(\tfrac{1}{2}-h,\th)\\
k_R^{BF}(h,\th) &= -k_R^{BB}(h,\th-\tfrac{1}{2}) = k^{FF}(\tfrac{1}{2}-h,\th).
\end{align}
In these expressions, the relations to the Euclidean kernels are valid only for values of $h,\th$ such that the Euclidean kernels are well defined. In all cases, the relationship between the retarded and Euclidean kernels involves switching the $B$ or $F$ index and taking $h\rightarrow \tfrac{1}{2}$ on the holomorphic side. (Because applying this operation to both sides is a symmetry of the kernel, we could equivalently do it just on the antiholomorphic side.)

\section{Analytic Continuation To The Regge/Chaos Region}\label{sec:anacon}
So far, we have computed the chaos exponent using the retarded kernel. But given that we have already derived an expression for the four-point function, we can also simply continue that expression to the chaos region \cite{Roberts:2014ifa}. In our 2d setting, this is the same as the Regge limit studied more generally in \cite{Cornalba:2007fs,Banks:2009bj,Costa:2012cb} and recently in \cite{Afkhami-Jeddi:2016ntf,Caron-Huot:2017vep,Alday:2017gde,Kulaxizi:2017ixa,Li:2017lmh}. Whereas the retarded kernel approach only gives the exponents $\lambda_L(p)$, this direct continuation approach will give the actual OTO correlator, including the coefficients of the contributions with different values of $p$. This section also might be useful as an example where the Regge limit can be worked out explicitly.

\subsection{Bosonic Model In Two Dimensions}
To be concrete, we will focus on the OTO correlator in the specific configuration given by
\be\label{OTO}
F(t,x) = \frac{\langle \Phi_i(\frac{3\beta}{4}+it,x)\Phi_j(\frac{\beta}{2},0)\Phi_i(\frac{\beta}{4}+it,x)\Phi_j(0,0)\rangle}{\langle \Phi_i(\frac{3\beta}{4},x)\Phi_i(\frac{\beta}{4},x)\rangle\langle\Phi_j(\frac{\beta}{2},0)\Phi_j(0,0)\rangle}.
\ee
where $\langle \cdot\rangle$ denotes the thermal expectation value. In our 2d setting with an infinite spatial line, we can relate this correlator to a zero-temperature vacuum correlator using the conformal mapping $z = e^{\frac{2\pi}{\beta}(x + i\tau)}$. From now on we will set $\beta = 2\pi$, but it can be restored by dimensional analysis. Because we have divided by the two-point functions in defining $F(t,x)$, it is conformally-invariant and therefore only a function of the two cross ratios $\chi,\bar{\chi}$. To work out the cross ratios for the configuration in (\ref{OTO}), we evaluate the $z$ coordinates of each of the four points and then compute $\chi = \frac{z_{12}z_{34}}{z_{13}z_{24}}$. One finds
\be\label{crossexact}
\chi = \frac{2}{1+i \sinh(t-x)}, \hspace{20pt} \bar{\chi} = \frac{2}{1+i \sinh(t+x)}.
\ee
We would like to understand what kinematic limit of the four-point function large $t$ corresponds to. It is helpful to think about starting from $t = 0$, where we have a purely Euclidean correlator, and slowly increasing $t$. (Let us suppose for the rest of the discussion that $x>0$ fixed.) As we do this, $\chi$ and $\bar{\chi}$ stop being complex conjugates, so the four-point function has branch points. In particular, there is an important branch cut running from $\chi = 1$ to $\chi = \infty$. As $t$ increases, $\chi$ goes around the branch point at $\chi = 1$ but $\bar{\chi}$ does not \cite{Roberts:2014ifa}, see figure \ref{fig:chichibar}. Ultimately, for large $t$, both $\chi$ and $\bar{\chi}$ become small, 
\be\label{crosschaos}
\chi \rightarrow -4i e^{-t+x},\hspace{20pt} \bar{\chi}\rightarrow -4ie^{-t-x}.
\ee
Since both cross ratios are small, one's first instinct is that this should be efficiently described by an OPE. In fact, because we have taken one of the cross ratios around the branch point, this is not true. Instead, we are probing an intrinsically Lorentzian regime of the correlator known as the Regge limit \cite{Cornalba:2007fs,Banks:2009bj,Costa:2012cb}. (See also the recent works \cite{Afkhami-Jeddi:2016ntf,Alday:2017gde,Kulaxizi:2017ixa,Li:2017lmh} for discussion of the Regge limit in CFT.) In other words, for 2d CFTs the Regge limit is the same as the chaos region. We will elaborate on this in section \ref{sec:reggevschaos} below, and in particular on why it does not hold in higher dimensions.
\begin{figure}
\begin{center}
\includegraphics[width=0.8\textwidth]{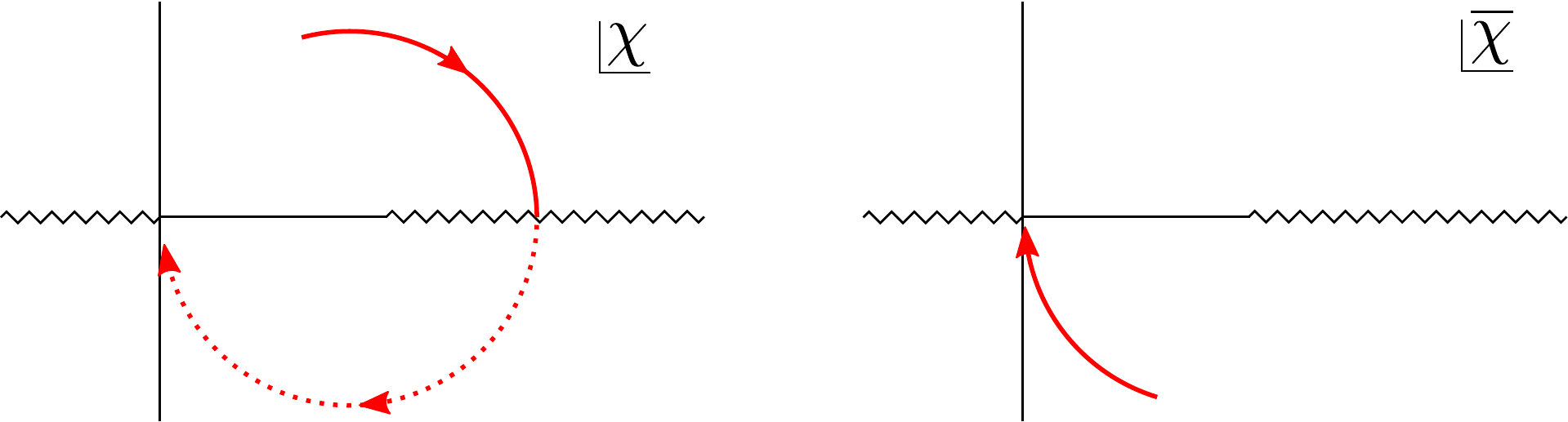}
\caption{\small{The paths taken by the cross ratios $\chi,\overline{\chi}$ as we continue from $t = 0$ to larger values of $t$. The dotted line indicates that $\chi$ has passed under the branch cut onto the second sheet.}}\label{fig:chichibar}
\end{center}
\end{figure}

To understand the difference between the Regge/chaos region and a normal OPE limit, one can try doing the continuation (\ref{crossexact}) to a given conformal block that contributes to the correlation function in the $\chi\rightarrow 0$ channel. After passing around $\chi = 1$ and then taking $\chi$ small, one finds the leading behavior (assuming $h>1/2$)
\be\label{aroundone}
\frac{\Gamma(h)^2}{\Gamma(2h)}F_h(\chi) \rightarrow 2\pi i\frac{\Gamma(2h-1)}{\Gamma(h)^2}\chi^{1-h} + ...
\ee
For large positive values of $h$, this function will be growing as we take $\chi$ small. The power law growth in $\chi$ translates to exponential growth as a function of time, see (\ref{crosschaos}). Combining with the antiholomorphic block, which was not continued around one and therefore has the standard dependence $\bar{\chi}^{\widetilde{h}}$, we find
\be\label{block2nd}
F_h(\chi) F_{\widetilde{h}}(\bar{\chi}) \propto \chi^{1-h}\bar{\chi}^{\widetilde{h}} \propto e^{t(h-\widetilde{h}-1) - x(h+\widetilde{h}-1)}.
\ee
In the Regge/chaos limit, we are interested in the behavior for fixed $x$ and large $t$, and we see that the contribution is exponentially growing with the spin $\ell$, as $e^{t(\ell-1)}$. So far we have considered a single block, but if we try to do this term by term inside the OPE in the $\chi\rightarrow 0$ channel, we will find that the contributions from high spin operators will be uncontrolled, and the expansion will fail. Similarly, the expression (\ref{nadsum}) as an integral over the principal series and a sum over spins will also fail, because the sum involves arbitrarily large positive spins.

This does not mean that $F(t,x)$ is divergent. It only means that we should not try to organize the answer in terms of an OPE expansion in the $\chi\rightarrow 0$ channel. In order to get a useful formula for the large $t$ behavior of $F(t,x)$, we will have to massage our expression (\ref{nadsum}) for the four-point function in the bosonic SYK model  into a more suitable form. The correct way to do this, as explained in \cite{Cornalba:2007fs,Costa:2012cb} and recently discussed in \cite{Kulaxizi:2017ixa}, is a procedure similar to a Sommerfeld-Watson resummation, where the sum over spins is replaced by a contour integral, and the contour is then deformed to a region where the real part of the spin is small, so we can safely continue to the Regge/chaos limit. To represent the sum as a contour integral, we take\footnote{We recall that in this model, because of bose statistics,
the single-sum operators have even $\ell$.}
\be\label{offint}
\sum_{\ell = \text{even}>0}\rightarrow -\frac{\pi}{2}\int_{\mathcal{C}}\frac{d\ell}{2\pi i}\frac{1+e^{-i\pi \ell}}{\sin (\pi \ell)}
\ee
where $\mathcal{C}$ is a contour that circles clockwise around the individual poles at $\ell = 2,4,6,...$. (Naively, one could use a different trigonometric function with the same poles and residues, but in fact later steps will require that we use this particular factor.) After making the substitution (\ref{offint}), we then pull the contour $\mathcal{C}$ to the left, until it is running parallel to the imaginary $\ell$ axis, and with a real part less than one. At this point it becomes safe to do the continuation to the Regge/chaos limit. Since the real part of the spin $\ell$ is less than one, we will not find any growing contribution as the time $t$ becomes large.

Based on this procedure, it sounds like we will prove something too strong: that there is no growing contribution in the Regge/chaos region. However, as we deform the contour, we might pick up poles with real part of $\ell$ greater than one. The locations of these poles determine the chaos behavior. We will find that they are also consistent with the computation of the chaos exponent from the retarded kernel.

The main subtle point in the analysis is the following: the expression for the four-point function (\ref{nadsum}) that we derived was valid for the case where the spin $\ell$ is an integer. In general, there are many different analytic functions of spin with the same values at the integers, and for the continuation procedure described above to work, it will be important to pick the right one.

We will now go through this procedure in more detail, starting from the expression for the four-point function in (\ref{nadsum})
\be
\mathcal{F} = \frac{\lambda}{2\pi^3}\sum_{\ell \text{ even}}\int_{-\infty}^\infty \frac{ds}{2\pi}(\ell^2+4s^2)\frac{\sin(\pi h)}{\cos(\pi\widetilde{h})}\frac{\Gamma(h)^2}{\Gamma(2h)}\frac{\Gamma(\widetilde{h})^2}{\Gamma(2\widetilde{h})}F_{s,\ell}(\chi,\bar{\chi})\frac{k(\ell,s)}{1-k(\ell,s)}.
\ee
In this equation, and in several others below, we will use $h,\widetilde{h}$ and $\ell,s$ interchangeably, with the understanding that
\be\label{hhtells}
h = \frac{1+\ell}{2}+is, \hspace{20pt} \widetilde{h} = \frac{1-\ell}{2}+is.
\ee
We would like to use (\ref{offint}) to write the sum over $\ell$ as a contour integral. In order for this to be helpful, it is important to pick the right continuation away from even integer $\ell$. In principle, we are free to insert ratios of trigonometric functions in various places, as long as they are equal to one one when $\ell$ is an integer. For the bosonic model, it turns out that the original expression for $k(\ell,s)$ in (\ref{bosker}) is already in a convenient form, but (as we will see in a moment) it is important to multiply the whole integrand by $\sin(\pi \widetilde{h})/\sin(\pi h)$, which is equal to one for even spin. We then continue away from the even integers, writing
\be
\mathcal{F} = -\frac{\lambda}{4\pi^2}\int_{-\infty}^\infty \frac{ds}{2\pi}\int_{\mathcal{C}}\frac{d\ell}{2\pi i}\frac{1 + e^{-i\pi\ell}}{\sin(\pi \ell)}(\ell^2+4s^2)\tan(\pi\widetilde{h})\frac{\Gamma(h)^2}{\Gamma(2h)}\frac{\Gamma(\widetilde{h})^2}{\Gamma(2\widetilde{h})}F_{s,\ell}(\chi,\bar{\chi})\frac{k(\ell,s)}{1-k(\ell,s)}.
\ee
Here the contour $\mathcal{C}$ circles clockwise around all even integers. So far this is just a trivial rewriting of the Euclidean correlation function. Our goal is to deform the $\mathcal{C}$ contour in such a way that we can safely continue to the Regge/chaos region. Recall from (\ref{block2nd}) that the growing terms after taking the chaos limit are contributions with $\text{Re}(\ell)>1$, so the goal will be accomplished if we can deform the $\mathcal{C}$ contour into a region with $\text{Re}(\ell)<1$. Let us imagine that $s$ is fixed, and try deforming the contour. We can leave the parts of $\mathcal{C}$ that circle negative integer $\ell$ as they are, but for the terms with positive $\ell$, we pull the contour back so that it is running parallel to the imaginary $\ell$ axis, and with $\text{Re}(\ell) <1$. 

Does the contour cross poles during this deformation? Notice that the real part of $h$ will be positive, so we do not need to worry about poles in the holomorphic factors in the conformal block $F_{s,\ell}$ or the expression $\Gamma(h)^2/\Gamma(2h)$. Notice also that the real part of $\widetilde{h}$ will be negative, so the expression $\tan(\pi\widetilde{h})\Gamma(\widetilde{h})^2/\Gamma(2\widetilde{h})$ does not have poles either. (This was the reason that it was important to multiply by $\sin(\pi \widetilde{h})/\sin(\pi h)$ before continuing in $\ell$.) However, the block $F_{s,\ell}$ has poles at negative half-integer values of $\widetilde{h}$, and we do encounter such values when we deform the $\mathcal{C}$ contour. Fortunately, these contributions are not important in the Regge/chaos limit, because (see section \ref{absdisc}) the residues of these poles are proportional to different blocks, of the form $F_h(\chi)F_{1-\widetilde{h}}(\bar{\chi})$.  After continuing $\chi$ around the branch point at one to get to the chaos region, we find schematically $F_h(\chi) F_{1-\widetilde{h}}(\bar{\chi}) \rightarrow \chi^{1-h}\bar{\chi}^{1-\widetilde{h}}\propto e^{(2is-1)t - \ell x}$ which is not growing in time, and is in fact suppressed for large $\ell$. The upshot of all of this is that as we deform the $\mathcal{C}$ contour, there are no ``kinematic'' poles coming from the blocks or the measure factor that lead to growing terms in the Regge/chaos region. 

Of course, we can also have poles coming from solutions to $k(\ell,s) = 1$, and these do lead to interesting behavior. The contribution of such a pole is
\be\label{partofF}
\mathcal{F} \supset \frac{\lambda}{4\pi^2}\int \frac{ds}{2\pi}\frac{1 + e^{-i\pi\ell}}{\sin(\pi\ell)}(\ell^2+4s^2)\tan(\pi\widetilde{h})\frac{\Gamma(h)^2}{\Gamma(2h)}\frac{\Gamma(\widetilde{h})^2}{\Gamma(2\widetilde{h})}F_{s,\ell}(\chi,\bar{\chi})\frac{1}{\partial_\ell k(\ell,s)}\Bigg|_{k(\ell,s) = 1}
\ee
where for each value of $s$ in the integral, we evaluate at $\ell$ such that $k(\ell,s)=1$. In general there will be some set of values of $s$ such that the real part of the corresponding $\ell$ is greater than one, and to evaluate the growing part of the correlation function we only need to integrate over these values of $s$. In principle, there could have been several solutions to $k(\ell,s)$ with $\text{Re}(\ell)>1$ for each value of $s$, but in practice we find at most one.

The point of the expression (\ref{partofF}) is that this is the only part of the four-point function that is growing in the chaos region, and it is now expressed in a form that can be safely continued. We continue the block $F_{s,\ell}(\chi,\bar{\chi}) = F_{h}(\chi)F_{\widetilde{h}}(\bar{\chi})$ to the chaos region using (\ref{aroundone}). Keeping only the leading term in the behavior of the conformal blocks, we then find an expression of the form
\be\label{ofform}
\mathcal{F} \supset \int ds f(s) \chi^{1-h}\bar{\chi}^{\widetilde{h}},
\ee
where $h,\widetilde{h}$ are determined in terms of $s$ by the conditions $k(h,\widetilde{h}) =1$ and $h + \widetilde{h} = 1 + 2is$, and the coefficient function $f$ results from combining the measure factors in (\ref{partofF}), the derivative of $k(\ell,s)$ and the factor from the transformation around one in (\ref{aroundone}). At this point we can also explain why it was important to use the particular trigonometric function in (\ref{offint}). The reason is that at an intermediate stage in the continuation of $\chi$ around one, the blocks become exponentially growing for values of $\ell$ with large negative imaginary part. (This is visible already in the leading term in the product of the blocks, with spin dependence $(\chi/\bar{\chi})^\ell$.) The trigonometric factor that we chose provides convergence in this region.

We can now write an expression for the OTO correlator (\ref{OTO}), by substituting the formula (\ref{crosschaos}) for the cross ratios in terms of $x,t$  into our expression for the growing parts of the four-point function (\ref{ofform}). In order to compare to the retarded kernel discussion, it is convenient to write the answer in terms of
\be\label{plambda}
p = -2s,\hspace{20pt} \lambda_L = \ell-1.
\ee
instead of $\ell,s$ or $h,\widetilde{h}$. We find that the connected part of the OTO correlator $F(t,x)$ is
\be\label{finalOTO}
F(t,x) = (\text{non-growing}) - \frac{1}{N}\int \frac{dp}{2\pi}g(p)\exp\big(\lambda_L(p)t + i p x\big)
\ee
where $\lambda_L(p)$ is determined by the combination of (\ref{plambda}) and $k(\ell,s) = 1$. The integral over $p$ is over the range where $\lambda_L(p)>0$. Explicitly, the function $g(p)$ is given by
\be
g(p) = \frac{\lambda}{\pi2^{2\ell}}(\ell^2+4s^2)\frac{\tan(\pi\widetilde{h})}{\sin(\frac{\pi\ell}{2})}\frac{\Gamma(2h-1)}{\Gamma(h)^2}\frac{\Gamma(\widetilde{h})^2}{\Gamma(2\widetilde{h})}\frac{1}{-\partial_\ell k(\ell,s)}\Bigg|_{k(\ell,s) = 1}.
\ee
The factor of $1/\sin(\pi\ell/2)$ came from combining the trigonometric factor in (\ref{offint}) with a factor of $e^{i\pi \ell/2}$ that arises when we write $\chi^{1-h}\bar{\chi}^{\t h}$ in terms of $x,t$. All $h,\t h,\ell,s$ variables on the RHS should be considered functions of $p = -2s$ using $k(\ell,s) = 1$ and the relation between $h,\widetilde{h}$ and $\ell,s$ in (\ref{hhtells}).

We have intentionally used the same notation $\lambda_L(p)$ as in the discussion of the retarded kernel. One can check from (\ref{retconf}) that the equation that defines $\lambda_L(p)$ is the same in both cases. 
 More simply, we can explain the agreement as follows. Suppose we start with a Euclidean contribution with weights $h,\widetilde{h}$ satisfying $k(h,\widetilde{h})=1$. After continuing to the Regge/chaos region, we find a growing piece with weights $1{-}h,\widetilde{h}$, as we saw in (\ref{ofform}). For consistency with the retarded kernel approach, we need this growing behavior to be a solution to $k_R = 1$, which means $k_R(1{-}h,\widetilde{h}) = 1$. The key point is that this is guaranteed by $k(h,\widetilde{h}) = 1$ and eqn.~(\ref{retconf}), which we can write as $k(h,\widetilde{h}) = k_R(1{-}h,\widetilde{h})$.

Let us now look at the behavior of the function (\ref{finalOTO}). For fixed $x$ and large $t$, the integral over $p$ gets concentrated near $p = 0$ where $\lambda_L(p)$ is largest, see figure \ref{fig:superlambda}. We can do the Gaussian integral around $p = 0$ and we get a function with the $t$ and $x$ dependence
\be
F(t,x) \propto -\exp\left[\lambda_L(0)t - \frac{x^2}{2t \lambda_L''(0)}\right].
\ee
In particular, the spatial profile becomes smooth on scales of $x \sim \sqrt{t}$. In theories dual to string theory, this is attributed to ``transverse string spreading'' but it appears to be a more general feature of the Regge limit \cite{Li:2017lmh}.

It is also interesting to consider the case where we take $x$ and $t$ large simultaneously. This is outside the Regge limit, and takes us closer to the light cone limit of the four-point function. As we make $x$ large, the $s$ integral is dominated by the closest non-analyticity as a function of $s$. We then get exponential decay as a function of $x$, rather than the Gaussian decay just described. We expect the closest non-analyticity to the real $p$ axis to be from the zero of the $\sin(\pi \ell/2)$ function at $\ell = 2$. When we take the residue of this pole, we find the contribution of the Regge/chaos continuation of the stress tensor. This is in keeping with expectations that the stress tensor/graviton should dominate in the light cone limit \cite{Penedones:2007ns,Fitzpatrick:2012yx,Komargodski:2012ek,Hartman:2015lfa}.

\subsection{Supersymmetric Model In Two Dimensions}
We will now briefly comment on the supersymmetric case. For simplicity, we set $\zeta = \b\zeta = 0$, see (\ref{fourpt}). This gives a formula very similar to the bosonic model, except that we have to sum over different contributions from the kernels in the four channels. It turns out that the only growing contribution comes from the $FB$ channel, which can be analyzed following the general approach described above. Following the logic described above, to find the growing part of the OTO correlator, we set 
\be
h = 1 + \frac{\lambda_L}{2}-\frac{ip}{2}, \hspace{20pt} \widetilde{h} = -\frac{\lambda_L}{2}-\frac{ip}{2}
\ee
and look for solutions to $k^{FB}(h-\frac{1}{2},\th) =1$ with Re$(\lambda_L)>0$. If we use the expression (\ref{kFB}) and continue naively in spin, we find a problem: there are an infinite number of solutions with positive $\lambda_L$ for each value of $s$. To make progress we have to find another analytic function $\hat{k}^{FB}$ that is equal for integer spin\footnote{Here, we mean the spin of the bosonic descendant, which is $h+\frac{1}{2}-\widetilde{h}$ if we consider $k^{FB}(h,\widetilde{h})$, or simply $h - \widetilde{h}$ if we consider $k^{FB}(h-\frac{1}{2},\widetilde{h})$.} and that leads to only finitely many growing solutions. An obvious choice is to define this function by requiring that the relationship with the retarded kernel (\ref{krkc}) should hold for all values of $h,\widetilde{h}$. Explicitly, this leads to
\be
\hat{k}^{FB}(h,\widetilde{h}) = -\frac{\Gamma(1-\Delta)^2}{\Gamma(\Delta-1)\Gamma(\Delta+1)}\frac{\Gamma(h+\Delta-\frac{1}{2})}{\Gamma(h - \Delta+\frac{1}{2})}\frac{\Gamma(-\widetilde{h}+\Delta)}{\Gamma(1-\widetilde{h}-\Delta)}.
\ee
This function agrees with $k^{FB}(h,\widetilde{h})$ for the values that we sum over in (\ref{fourpt}), so it leads to the same four-point function. But it is an improvement, in the sense that $\hat{k}^{FB}(h-\frac{1}{2},\widetilde{h})=1$ leads to at most a single solution with positive Re$(\lambda_L)$. In fact, because it is related to the retarded kernel $k_R^{BB}$ by 
\be
\hat{k}^{FB}(h-\tfrac{1}{2},\widetilde{h})= k_R^{BB}(1-h,\widetilde{h}),
\ee
we find the same function $\lambda_L(p)$ that we found using the retarded kernel, and that we plotted in figure \ref{fig:superlambda}. Our analytic continuation method gives somewhat more information than the retarded kernel approach, because we can substitute $\lambda_L(p)$ (and a small modification of $g(p)$ appropriate for the supersymmetric case) into (\ref{finalOTO}) to get the full answer for the growing terms in the four-point function at order $1/N$.
\begin{figure}[ht]
\begin{center}
\includegraphics[width = .5\textwidth]{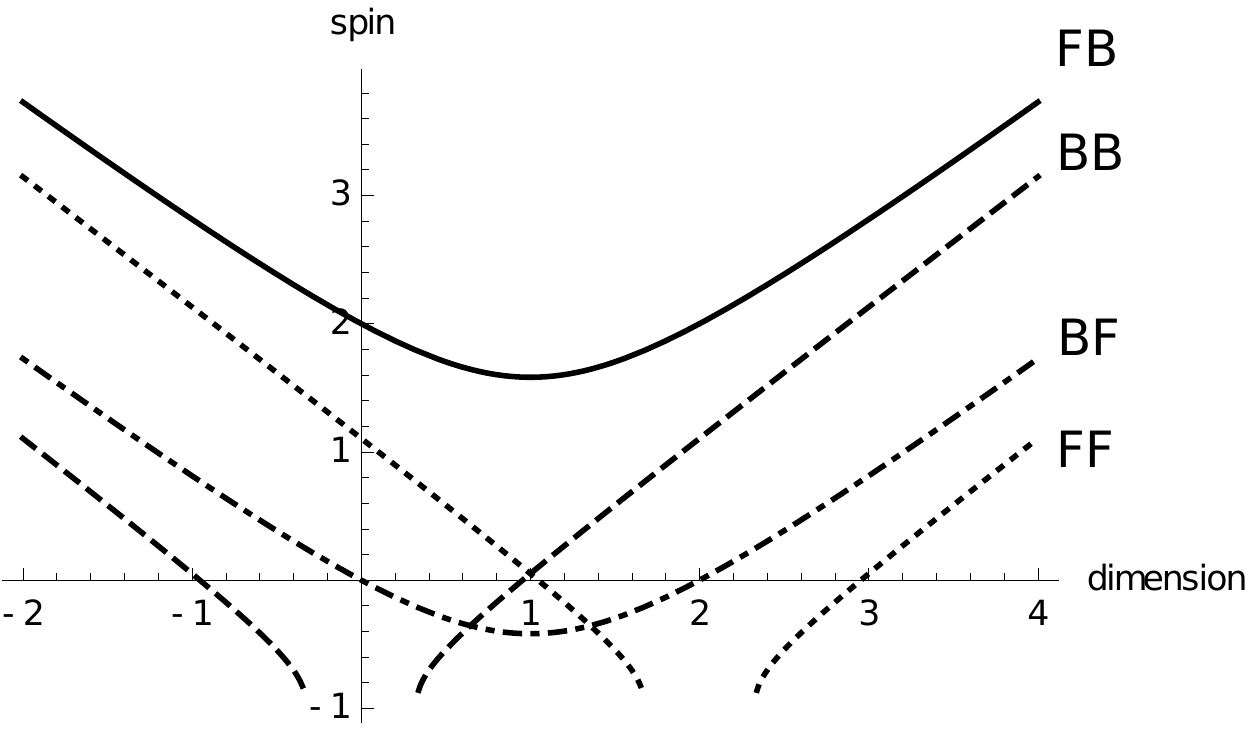}
\caption{The spin as a function of dimension $J(E)$ for the leading Regge trajectory appearing in the correlation function of primaries ($\zeta = \bar{\zeta} = 0$) in various channels, for $\q =3$. The FB and BF curves differ by two units of spin, and the BB and FF curves differ by two units of dimension. In between the gaps in the BB and FF curves, the spin is complex.}\label{reggeplot}
\end{center}
\end{figure}

We have checked that the other channels do not lead to growing solutions. For  this it is convenient to parametrize $h = \frac{1+J}{2}+is, \th = \frac{1-J}{2}+is$ with $\text{Im}(s) = 1/2$ for the $BB$ channel and $\text{Im}(s) = -1/2$ for the $FF$ channel. One can then numerically search for solutions to $k^{BB}(h,\th) = 1$ or $k^{FF}(h{-}\frac{1}{2},\th{-}\frac{1}{2})=1$ with the largest real parts of $J$. This defines $J(s)$ for each channel. One finds that for all values on the contour, the real part of $J$ is negative in these channels. The chaos exponent $\lambda_L = J-1$, is even more negative, so we do not find any growing contributions.

In this discussion, a central role has been played by the spin as a function of conformal dimension with fixed real part and varying imaginary part. Although it is not directly relevant for the studying the Regge limit, it is also sometimes useful to consider the spin as a function of purely real dimension, $J(E)$. We plot this for each of the channels in figure \ref{reggeplot}.

\subsection{Regge vs. Chaos}\label{sec:reggevschaos}
The limit of the four-point function that was relevant in the above discussion is known as the Regge limit. This can be defined for a CFT in a general dimension by taking $\chi$ around the branch point at one, and then taking $\chi,\bar{\chi}$ small with a fixed ratio. This type of limit was discussed in \cite{Cornalba:2007fs,Costa:2012cb}. One reason that it is called the Regge limit is that in theories with holographic duals, this region of the four-point function probes the S-matrix of the bulk theory in the regime of fixed impact parameter and high energy \cite{Cornalba:2006xk}. This is conventionally called the Regge limit in the context of scattering theory.

For two-dimensional conformal field theories, it was shown in \cite{Roberts:2014ifa} that this same kinematical limit is relevant for the computation of OTO correlators at finite temperature. This was the content of figure \ref{fig:chichibar}.

We can ask what happens in higher dimensions. The Regge limit is defined in a similar way in arbitrary dimensions. One begins with a four-point function in the vaccuum, applies a conformal transformation so that all four points lie in a 
2d plane, and then takes the limit of the two-dimensional kinematics just discussed. By contrast, the chaos region is defined for thermal correlation functions, which are not simply related to vaccuum correlation functions in higher than two dimensions. So the Regge and chaos limits are in general different.

More precisely, what we should say is that the Regge limit is different from the chaos limit for a $d>2$ theory in flat space. In fact, vacuum correlation functions in a conformal field theory {\it can} be understood as thermal correlation functions, but for a theory in a hyperbolic space with radius of curvature equal to the thermal length $\beta$. So if we like, we can think of the Regge limit as the chaos limit for a special thermal system defined in hyperbolic space.

This raises an important conceptual point. For a generic theory in flat space, even if it is weakly coupled, one expects the chaos exponent to be positive. However, one does not always expect a growing contribution in the Regge limit for $d>2$. This is consistent with the fact that these two limits are different. But we have just argued that we can think of Regge as chaos in hyperbolic space; does not this mean that we should always get growth in the Regge limit? The essential point is that for weakly interacting scalar theories in hyperbolic space, excitations can get diluted in the space fast enough that the butterfly effect fails, and the chaos exponent is negative. A simple model for the growth of many-body chaos is a self-reproducing random walk, or the spread of an epidemic \cite{Stanford:2015owe}. In any space this type of process leads to an exponentially growing number of random walkers or infected agents. A more interesting question, and one more closely related to the chaos exponent, is whether the {\it density} grows. In flat space, it always does, because the power-law decay from diffusion is overcompensated by the exponential growth from self-reproduction. But in hyperbolic space diffusion by itself leads to an exponential decay, so if we have slow self-reproduction the density can actually exponentially decrease in time.\footnote{The situation is different for weakly-coupled theories with vector degrees of freedom. There the BFKL analysis shows that one gets growth in the Regge region even if the interactions are weak \cite{Kuraev:1977fs,Balitsky:1978ic}.} In this situation, we expect a negative chaos exponent, or equivalently a Regge intercept with spin less than one.

\section{Discussion}
In this paper we studied SYK-like field theories in $d = 2$ spacetime dimensions. In the first few sections, we studied the general representation of a four-point function in terms of conformal or super-conformal eigenfunctions. We then applied this to SYK-like theories in one and two dimensions. In two dimensions, the bosonic theories we studied did not lead to a reliable SYK-like IR phase, but the supersymmetric theory did. We computed the four-point function at order $1/N$ and discussed some aspects of the out-of-time-ordered correlators that diagnose chaos. One simple takeaway from these models is that they are solvable large $N$ CFTs with interaction strength of order one. This can be diagnosed either by the twist of lightest ``single-sum'' spin-four operator, for which $E-J = 0.2888$ (in the $\hat{q}=3$ model) or the chaos exponent, which is $\lambda_L = 0.5824\frac{\beta}{2\pi}$, again for $\hat{q}=3$.

It would be nice to understand what happens to the models at higher orders in $1/N$, or even at finite $N$. An expectation is the following. As we back off from infinite $N$, different realizations of the disordered model will become somewhat different from each other. The operators that we have to adjust to move between theories are of the form $\int d^2\theta \,\uphi_{i_1}\uphi_{i_2}\cdots \uphi_{i_{\hat{q}}}$. If no indices are the same, then at large $N$ this operator corresponds to a marginal deformation, since its dimension is $1+\hat{q}\Delta = 2$. However, perturbatively in $1/N$ we expect the dimension to be modified. So, beyond the large $N$ limit, one expects that the theories are not connected by exactly marginal deformations. Instead, the SYK-like models should define a slow RG flow in a landscape of theories. It is possible that this landscape contains a large number of true fixed points.

At the level that we have discussed the theory, our results would apply also to the tensor models that realize SYK without disorder \cite{Witten:2012bg,KT}. We can think of these models as specific choices of the disorder couplings that are simple enough to have a large global symmetry, but arranged in such a way as to give the same results as random couplings at order one and order $1/N$. An important point, however, is that the exact global symmetry of such models will lead to the same problem that we enountered in section \ref{modone} for models with a $U(1)$ symmetry. In particular, we will find a formally divergent contribution to a four-point function in the antisymmetric channel.

This difficulty is reminiscent of what happens in the the one-dimensional SYK model. There, one finds a naive divergence in the IR theory from an integral over reparametrization zero modes. The correct treatment of these modes breaks conformal symmetry. Also, in models with a continuous global symmetry, one expects further zero modes from local group transformations \cite{Sachdev:2015efa,Fu:2016vas,KT}. In $d = 2$, we also have an emergent reparametrization symmetry, but unlike the situation in one dimension, the reparametrizations are not normalizable so they do not lead to a divergence; they can be treated consistently in the strict IR limit. However, local internal symmetry transformations remain normalizable in 2d, corresponding to modes with $(h,\widetilde{h}) = (1,0)$ that are part of the principal series. The integral over such modes can lead to a divergence that invalidates the naive IR fixed point.

This suggests one way to avoid the problem, which is to go one dimension higher and study a similar model in $d = 3$. The supersymmetric interaction with $\hat{q} = 3$ is relevant, so an SYK-like fixed point is possible. Now a local symmetry transformation corresponds to a mode with dimension $E = 2$ and spin $J = 1$, which is not part of the normalizable set of conformal eigenfunctions. So we do not expect three-dimensional models with symmetry to suffer from the problem of section \ref{modone}. Instead, it seems that non-disordered tensor model CFTs should be possible.

Finally, one would like to understand the holographic dual of the models considered in this paper. Of course, since they do not saturate the chaos bound, and instead have a tower of reasonably light higher-spin operators, we do not expect an approximately local gravitational dual. One possibility would be some kind of string theory with tension of order the $AdS$ scale. However, this would predict a Hagedorn-like spectrum of single sum operators that we do not find, so it does not seem like the bulk theory could be a conventional string theory. Instead, we could imagine a bulk theory in which the ``single-sum'' operators are some kind of composite state of two particles, but where the potential between the particles does not arise from a dynamical object like a string with many degrees of freedom of its own. This question obviously deserves further attention.

\vskip 2cm\noindent
{\it Acknowledgements}  We thank K. Bulycheva, F. Gesztesy, P. Goddard, A. Jevicki, I. Klebanov, J. Maldacena,  D. Simmons-Duffin, and B. Simon for discussions.  The work of JM was supported in part by NSF grant Phy-1606531 at the Institute for Advanced Study and NRF grant GUN 87667 at the University of Cape Town.
DS is supported by  Simons Foundation grant 385600.    Research of EW is supported in part by NSF Grant Phy-1606531.
\appendix  

\section{Normalization Of Bound State Wavefunctions}\label{normo}

The derivation of the operator product expansion of the SYK model in  \cite{Maldacena:2016hyu} depended on an identity relating the normalization of bound
state wavefunctions to scattering data of continuum states.   Essentially the same identity came into play when we deformed the contour for the $s$ integral in
eqn.~(\ref{morehelpful}).  In this appendix, we give a direct explanation of this identity, which actually is an example of a general relationship between bound
state wavefunctions and scattering data for Schrodinger-like operators.\footnote{See for example eqn.~(12.52) of   \cite{Newton},
which is used in section 12.1.5 of that book in explicitly showing the completeness of the bound state plus scattering state spectrum of the Schrodinger equation. This
explicit proof of completeness involves a contour deformation analogous to the one used in deriving the OPE for the SYK model.}

The eigenvalue equation for the wavefunction $\Psi_h(\chi)$ can be written
\be\label{eiprob}\partial_\chi(1-\chi)\partial_\chi\Psi_h =\frac{h(h-1)}{\chi^2}\Psi_h. \ee
For some other dimension $h'$, we likewise have
\be\label{neiprob}\partial_\chi(1-\chi)\partial_\chi\Psi_{h'} =\frac{h'(h'-1)}{\chi^2}\Psi_{h'}. \ee
We now look at the Wronskian:
\be\label{ipbrob}\partial_\chi\bigl(\Psi_{h'}(1-\chi)\partial_\chi \Psi_h-(1-\chi)\partial_\chi\Psi_{h'}\;\Psi_h\bigr)=\frac{h(h-1)-h'(h'-1)}{\chi^2}\Psi_{h'}\Psi_h. \ee
Therefore
\be\label{lipbrop}\bigl[ \Psi_{h'}(1-\chi)\partial_\chi \Psi_h-(1-\chi)\partial_\chi\Psi_{h'}\cdot\Psi_h \bigr]_0^2=\bigl(h(h-1)-h'(h'-1)\bigr)\int_0^2\frac{d\chi}{\chi^2}\Psi_{h'}\Psi_h. \ee

On the left, there is no surface term at $\chi=2$, because of the boundary condition $\left.\partial_\chi\Psi_h\right|_{\chi=2}=\left.\partial_\chi\Psi_{h'}\right|_{\chi=2}=0$
that is satisfied by the wavefunctions.  However, we have to look at the surface term at $\chi=0$.  For this, we make the usual expansion
\be\label{pbrob}\Psi_h(\chi)=\t A(h)F_h(\chi)+\t B(h)F_{1-h}(\chi).\ee
A bound state occurs at a value of $h$ with $\Re\,h>1/2$ at which $\t B(h)=0$.  We set $h$ to such a value, and we take $h'=h+\varepsilon$.  Thus
\be\label{hibor}\Psi_h(\chi)=\t A(h) F_h(\chi),\ee
while to   first order in $\varepsilon$,
\be\label{libor}\Psi_{h'}(\chi)=\t A(h) F_h(\chi)+\varepsilon \t B'(h) F_{1-h}(\chi)+\cdots. \ee
In eqn.~(\ref{libor}), we omitted some terms (involving the derivative of $\t A(h)F_h(\chi)$ with respect to $h$) that are not singular enough for small $\chi$ to be relevant in
the computation we are about to perform.  

To first order in $\varepsilon$, the right hand side of eqn.~(\ref{lipbrop}) is $-\varepsilon(2h-1)\int_0^2\frac{d\chi}{\chi^2}\Psi_h^2$.  Since $F_h(\chi)\sim \chi^h$,
$F_{1-h}(\chi)\sim \chi^{1-h}$ for small $\chi$, the surface term at $\chi=0$ on the left hand side of eqn.~(\ref{lipbrop}) to first order in $\varepsilon$
is $-\varepsilon (2h-1) \t A(h)\t B'(h)$.  Comparing, we learn that at the values of $h$ that correspond to bound states, 
\be\label{nilop}\t A(h)\t B'(h) =\int_0^2\frac{d\chi}{\chi^2}\Psi_h^2=\left(\Psi_h|\Psi_h\right). \ee
This is the identity that is important in the derivation of the OPE.

\section{The KLT Integral}\label{eval}

A certain integral that is useful in relating tree-level amplitudes for open and closed strings \cite{KLT} has frequently arisen in the present paper. This integral and generalizations of it were also studied earlier by other authors, see e.g. \cite{Symanzik:1972wj}. In this appendix, we describe a method to evaluate this integral and some related ones.

We start with the integral
\be\label{lotag}
\int \frac{d^2\x}{|\x|^{2a}}e^{i\p\cdot \x} = c(a)|\p|^{2a - 2}, \hspace{20pt} c(a)\equiv \frac{\pi}{2^{2a-2}}\frac{\Gamma(1-a)}{\Gamma(a)}.
\ee
 Here $\x=(\x_1,\x_2)$ and $\p=(\p_1,\p_2)$ are Cartesian two-vectors, and the coefficient $c(a)$ can be derived by doing a Gaussian integral over $\p$ on both sides of the equation. 
Now introduce complex variables with $x=\x_1+i \x_2$, $p=\p_1+i\p_2$.  Thus for example  $|\x|^2=x\b x$, $\p\cdot \x =\frac{1}{2}(p\b x+\b p x)$.
By taking 
 derivatives of eqn.~(\ref{lotag}) with respect to $p$, we find that
\be\label{hotag}
\int \frac{d^2 \x}{|\x|^{2a}}\left(\frac{i}{2}\bar{x}\right)^n e^{ip\cdot x} = c(a,n) p^{a-1-n}\bar{p}{}^{a-1}, \hspace{20pt} c(a,n) \equiv \frac{\pi}{2^{2a-2}}\frac{\Gamma(1-a)}{\Gamma(a-n)}.
\ee
Relabeling this equation, we find the Fourier representation 
\be\label{fourier}
x^{b+n}\bar{x}^{b} = \frac{(i/2)^n}{c(b+1+n,n)}\int \frac{d^2\p \,e^{i\p\cdot \x}}{p^{b+1}\bar{p}^{b+1+n}}.
\ee

\def\1{{\mathbf 1}}
We can use this to evaluate the desired integral:
\begin{align}\label{KLT}\notag
\int d^2 \x \,x^{a+n}\bar{x}^a (1-x)^{b+m}(1-\bar{x})^b &= \int \frac{d^2\p\, e^{i\p\cdot \1}}{p^{a+b+2}\bar{p}^{a+b+2+n+m}}\frac{(2\pi)^2(i/2)^{n+m}}{c(b+1+m,m)c(a+1+n,n)}\\ \notag
&=(2\pi)^2\frac{c(a+b+2+n+m,n+m)}{c(b+1+m,m)c(a+1+n,n)}\\
&=\pi\frac{\Gamma(1+a)\Gamma(1+b)\Gamma(-1-a-b-m-n)}{\Gamma(2+a+b)\Gamma(-b-m)\Gamma(-a-n)}. 
\end{align}
In the first step, we used (\ref{fourier}) on both  factors $x^{a+n}\b x^a$ and $(1-x)^{b+m}(1-x)^b$, and we then did the integral over $\x$.  Note that $x=1=\b x$
corresponds to $(\x_1,\x_2)=(1,0)$; this two-vector has been denoted as $\1$, so $\p\cdot \1=\p_1$.

All of these integrals have analogs in $d$ dimensions that can be proved with similar methods.  The generalization of eqn.~(\ref{lotag}) is 
\be\label{fourierd}
\int \frac{d^d \x}{|\x|^{2a}}e^{i\p\cdot \x} = \c_d(a)|\p|^{2a-d}, \hspace{20pt} \c_d(a) \equiv \frac{\pi^{d/2}}{2^{2a-d}}\frac{\Gamma(\frac{d}{2}-a)}{\Gamma(a)}.
\ee
Using this and repeating the above derivation, one gets an analog of the KLT integral (\ref{KLT}) for the special case $m=n=0$:
\be\label{kltan}
\int d^d\x |\x|^{2a}|\1-\x|^{2b} = \pi^{\frac{d}{2}}\frac{\Gamma(-a-b-\frac{d}{2})\Gamma(a+\frac{d}{2})\Gamma(b + \frac{d}{2})}{\Gamma(-a)\Gamma(-b)\Gamma(a+b+d)}.
\ee  Here $\1$ is the $d$-vector $(1,0,\cdots,0)$.
For a more complete analog of the KLT integral, we  
start with eqn.~(\ref{fourierd}) and differentiate with respect to $\p$.  If $A_{i_1i_2\cdots i_s}$ is an arbitrary symmetric traceless tensor of rank $s$,
then
\be\label{fourierderiv}
i^s\int \frac{d^d \x}{|\x|^{2a}}e^{i\p\cdot \x}A_{i_1i_2\cdots i_s}\x_{i_1}\x_{i_2}\dots \x_{i_s} =2^s\frac{\Gamma(a-d/2+1)}{\Gamma(a-d/2-s+1)}
 \c_d(a)|\p|^{2a-2s-d}A_{i_1i_2\cdots i_s}\p_{i_1}\p_{i_2}\dots \p_{i_s} .
\ee
Now we can imitate the derivation of (\ref{KLT}) to get
\begin{align}\label{KLTtwo}&\int d^d\x |\x|^{2a}|\1-\x|^{2b
}A_{i_1\cdots i_s}\x_{i_1}\cdots \x_{i_s}\cr &=\pi^{d/2}A_{11\cdots 1} \frac{\Gamma(a+1)\Gamma(a+b+s+d/2+1)\Gamma(-a-b-s-d/2)
\Gamma(a+d/2+s)\Gamma(b+d/2)}{\Gamma(a+1+s)\Gamma(a+b+d/2+1)\Gamma(a+b+d+s)\Gamma(-a-s)\Gamma(-b)}\cr
&=\pi^{d/2}A_{11\cdots 1}\frac{\sin\pi(a+s)}{\sin\pi(a+b+s+d/2)}\frac{\Gamma(a+1)\Gamma(b+d/2)\Gamma(a+d/2+s)}{\Gamma(a+b+d/2+1)\Gamma(a+b+d+s)\Gamma(-b)}.\end{align}
Here $A_{11\cdots 1}=A_{i_1i_2\cdots i_s}\1_{i_1}\1_{i_2}\cdots \1_{i_s}$ is the component of $A_{i_1i_2\cdots i_s}$ with all indices set to 1.
(By a simple rotation and scaling, one can generalize eqn.~(\ref{KLTtwo}) with $\1$ replaced by any $d$-vector.)

From these formulas, one can deduce other useful ones.  For example 
\be\label{itin}
\int \frac{d^d\x \,d^d\y}{|\1-\x|^{2\Delta}|\y|^{2\Delta}|\x-\y|^{2d-2\Delta - 2h}} = \pi^d\frac{\Gamma(\frac{d}{2}-\Delta)^2\Gamma(\Delta-h)\Gamma(\Delta+h-\frac{d}{2})}{\Gamma(\Delta)^2\Gamma(d-h-\Delta)\Gamma(\frac{d}{2}+h-\Delta)}.
\ee
 eqn.~(\ref{itin}) can be proved by using eqn.~(\ref{kltan}) to integrate
over $\y$ (after a rotation and rescaling of $\x$) and then using the same formula to integrate over $\x$.
With a similar use of eqn.~(\ref{KLTtwo}), we get
\begin{align}\label{symmtrac}
\int \frac{d^d\x_3 d^d \x_4}{|\x_{13}|^{2\Delta}|\x_{24}|^{2\Delta}|\x_{34}|^{2d-4\Delta}}\cdot \frac{A_{i_1i_2\cdots i_s}\x_{34}^{i_1}\cdots\x_{34}^{i_s}}{|\x_{34}|^{2\Delta-2a}} = \lambda\frac{A_{i_1i_2\cdots i_s}\x_{12}^{i_1} \cdots
\x_{12}^{i_s}}{|\x_{12}|^{2\Delta-2a}}\end{align}
with
\begin{align}
\lambda = \pi^d\frac{\Gamma(\frac{d}{2}-\Delta)^2}{\Gamma(\Delta)^2}\frac{\Gamma(\Delta-a)}{\Gamma(d-a-\Delta)}\frac{\Gamma(s+a+\Delta-\frac{d}{2})}{\Gamma(s +a - \Delta+\frac{d}{2})}.
\end{align}

\section{More Details On The Central Charge Computation}\label{app:c}

\def\O{{O}}
In this appendix we show how the central charges of charge of our 2-dimensional models is computed by evaluating the contribution of the stress tensor to the 4-point function. To this end, consider a 2d CFT with a holomorphic stress tensor $T$ that satisfies
\be\label{sts} \langle T(x)T(y)\rangle =\frac{c}{2(x-y)^4}. \ee
Suppose the theory also has a spinless operator $\O$ of dimension $(h, \t h)=(\Delta/2,\Delta/2)$, and a corresponding two-point function
\be\label{tpf} \langle \O(x,\b x) \O(0,0)\rangle = \frac{b}{|x|^{2\Delta}}\ee
for some $b$.  The singular part of the $T \cdot \O$ operator product expansion is as usual
\be\label{singp} T(y) \O(0,0)\sim \frac{h}{y^2} \O(0,0)+\frac{1}{y}\partial_y\O(0,0).\ee

\noindent
Conformal invariance then determines the $\langle \O \O T\rangle$ three-point function:
\be\label{threp}\langle \O(x_1,\b x_1)\O(x_2,\b x_2) T(y)\rangle =\frac{b}{|x_1-x_2|^{2\Delta}}\frac{h(x_1-x_2)^2}{(y-x_1)^2(y-x_2)^2}.\ee
Concerning the derivation of this formula, we note the following.  By inspection, the right hand side of eqn.~(\ref{threp}) has the double poles
at $y=x_1$ and $y=x_2$ that one would expect from the leading part of the OPE (\ref{singp}).  One can verify that the single poles are also as expected,
but it is actually not necessary to check this: the right hand side of eqn.~(\ref{threp}) is the unique formula that has the right double poles at $y=x_1,x_2$, is holomorphic
elsewhere,
and vanishes as $y^{-4}$ at infinity (so that the quadratic differential $\langle \O(x_1)\O(x_2) T(y)\rangle (dy)^2$ has no pole at infinity). The $\O\cdot \O$ OPE therefore reads in part
\be\label{tops}\O(x_1,\b x_1)\O(x_2,\b x_2)\sim \frac{b}{|x_1-x_2|^{2\Delta}}\left(1+\frac{2h}{c}(x_1-x_2)^2 T(x_2)+\dots\right) \ee
The leading term is a restatement of eqn.~(\ref{tpf}), and the term proportional to $T$ follows by comparing eqns.~(\ref{threp}) and (\ref{sts}).  

\noindent
Now suppose that this theory has a second operator $\O'$ with the same dimension and two-point function as $\O$, and that we are interested in the normalized
four-point function
\be\label{normpt}W(x_1,\b x_1;\dots; x_4,\b x_4)=\frac{\langle\O(x_1,\b x_1)\O(x_2,\b x_2)\O'(x_3,\b x_3)\O'(x_4,\b x_4)\rangle}{
          \langle\O(x_1,\b x_1)\O(x_2,\b x_2)\rangle\langle\O'(x_3,\b x_3)\O'(x_4,\b x_4)\rangle} .  \ee
This function can be expressed as a sum over operators propagating in the 1-2 channel. In particular, the stress tensor will give a contribution that is proportional to $(x_1-x_2)^2(x_3-x_4)^2/(x_2-x_4)^4$ in the limit that $x_1\to x_2$ and $x_3\to x_4$.  This contribution can be determined from
the OPE (\ref{tops}) and the two-point function (\ref{sts}):
\be\label{ormpt}W\sim \cdots +   \frac{2h^2}{c}\frac{(x_1-x_2)^2(x_3-x_4)^2}{(x_2-x_4)^4}.   \ee    
To leading order, the right hand side is the same as $(2h^2/c)\chi^2$, where $\chi=\langle 1,2\rangle \langle 3,4\rangle/\langle 1,3\rangle \langle 2,4\rangle$ is the usual cross ratio. So
the stress tensor contribution is
\be\label{strcont}W_T=\frac{2h^2\chi^2}{c}=\frac{\Delta^2\chi^2}{2c}. \ee
To apply this result to SYK-like models, we have to remember that the four-point function $\F$ of such models is usually defined with an extra factor of $N$
(eqn.\eqref{normalfn}).  Thus the corresponding result for $\F$ is
\be\label{strcontt}\F_T=\frac{2h^2N\chi^2}{c}=\frac{N\Delta^2\chi^2}{2c}. \ee
Let us consider now the naive bosonic model of section \ref{firsttry}, starting with  eqn.\eqref{withcoeff}, which we reproduce here:
\begin{eqnarray}\label{pll}
 \mathcal{F} =  \frac{2\Delta}{\pi(2{-}\Delta)(1{-}\Delta)^2} \sum_{\ell=\text{ even}}\int_{-\infty}^\infty \frac{ds}{2\pi} (\ell^2+4s^2)\A(s,\ell)F_{s,\ell}(\chi,\bar\chi) \frac{k(\ell,s)}{1-k(\ell,s)}.
\end{eqnarray}
In the spin two sector, the integrand has a simple pole at $h,\th = 2,0$ or equivalently $\ell,s =2,-\tfrac{i}{2}$ coming from a zero of $1-k(l,s)$ in the denominator. Expanding the kernel near this point, one finds
\begin{eqnarray}
  k(\ell,s) = k(2,s) = 1 + \frac{2-2\Delta}{\Delta(2-\Delta)}(is-\tfrac{1}{2})+ O((is-\tfrac{1}{2})^2)\,,
\end{eqnarray}
This implies that
\begin{eqnarray}
  \frac{k(2,s)}{1-k(2,s)} = -\frac{\Delta(2-\Delta)}{2(1-\Delta)(is-\tfrac{1}{2})} + \cdots\,.
\end{eqnarray}
where the dots are regular at $s = -\frac{i}{2}$. The contribution of the stress tensor comes from the residue of this pole. To evaluate this residue, we use that when $\ell,s = 2,-\tfrac{i}{2}$, we have $\mathcal{A}(s,\ell) = \frac{\pi}{6}$,  $\ell^2 + 4s^2 = 3$, and $F_{s,\ell} = \chi^2+\dots$. Integrating clockwise in (\ref{pll}) around the pole in $s$, we find the contribution
\be
\mathcal{F}_T = \frac{\Delta^{2}}{2(1-\Delta)^{3}} \chi^2.
\ee
Finally, comparing this with \eqref{strcontt} gives
\begin{eqnarray}
  c = \left(1 - \Delta\right)^3 N = \left(1-\frac{2}{q}\right)^{3}N\,.
\end{eqnarray}
Even though the naive bosonic model was not well-defined, this formula for the central charge has some reasonable properties. It decreases as we decrease $q$. This is important because the $c$ theorem requires $c$ to decrease along RG flows, and one can flow from larger $q$ to smaller $q$ by perturbing with the more relevant smaller-$q$ interaction. Also, we get the free field answer in the $q\to\infty$ limit, and zero in the limit $q\rightarrow 2$, much as in the better-defined supersymmetric case.

\def\veps{\varepsilon}
\section{Wick Rotation And The Kernel}\label{reta}
\def\O{{O}}
In this appendix we would like to elaborate on the relationship between the conformal and retarded kernels.
The Euclidean formula \eqref{zolt} for the kernel of the naive bosonic model in two dimensions can be written
\be\label{euclidean} k(h,\th) =C\int d^2\x d^2\x' \,\,\frac{(\bar x-\bar x')^{-J}}{|1-x|^{2\Delta}|x'|^{2\Delta} |x-x'|^{4-2\Delta-2h}},~~~~C=\frac{(q-1)(1-\Delta)^2}{\pi^2}.\ee
In writing the formula this way, we assume that the integer $J=h-\th$ is $\leq 0$; otherwise, in the following, one should use a similar formula with the roles of $h$ and $\th$ and of $x$ and $\bar x$ exchanged.

\noindent
We want to Wick rotate this formula to Lorentz signature.  For this, we first write $x$ and $x'$ in terms of their real and imaginary parts:
$x=s+i\tau$, $x'=s'+i\tau'$. Then we rotate $\tau$ and $\tau'$ in the complex plane, setting $\tau=e^{i\alpha}t$, $\tau'=e^{i\alpha}t'$, with
real $t$, $t'$.  It is important that $\tau$ and $\tau'$ are  rotated by the same angle $\alpha$.   This ensures that, for $0\leq \alpha<\pi/2$,
 the integral remains convergent and that it is invariant under this deformation.
Now set $\alpha=\pi/2-\epsilon$ and consider the behavior for small $\epsilon>0$.  We  have, for example, $|x|^2=s^2-t^2(1-2i\epsilon)+{\mathcal O}(\epsilon^2)$.
For $\epsilon\to 0^+$, we can here drop the $\epsilon^2$ term and replace $2t^2\epsilon$ by $\epsilon$, and so $|x|^2$ becomes $s^2-t^2+i\epsilon$.    
In terms of light cone coordinates $u=s-t$, $v=s+t$, this is $uv+i\epsilon$.  For the measure, we observe that $d^2\x=ids \,dt=\frac{i}{2} du\,dv$.  
In these coordinates, the integral that computes the kernel is  then
\be\label{lorentzian} k(h,\th) =-\frac{C}{4}\int du\,du'\,dv\,dv'  \frac{(v-v')^{-J}}{((u-1)(v-1)+i\epsilon)^{\Delta}(u'v'+i\epsilon)^{\Delta} ((u-u')(v-v')+i\epsilon)^{2-\Delta-h}}.\ee
There is no need for an $i\epsilon$ in the numerator, since $-J$ is a nonnegative integer. Eqn.\eqref{lorentzian} is simply the formula that we would have
written down to begin with if we had carried out the whole analysis in Lorentz signature.

If we could naively ignore the $i\epsilon$'s in the denominator in eqn.\eqref{lorentzian}, the integrand would factor as a function of $u$ times a function of $v$, and the integral
would likewise factor.  Of course, this is not correct.  Instead, we proceed as follows. Keeping $v$ fixed, the integrand as a function of $u$ has singularities
at $u=1-i\epsilon/(v-1)$ and at $u=u'-i\epsilon/(v-v')$.    If $1-v$ and $v-v'$ have opposite signs, then both singularities are in the upper or lower half-plane.
When this is so, by closing the contour for the $u$ integral in the half-plane that has no singularities, one finds that the $u$ integral vanishes.  Thus the $u$ integral is nonzero only if $1-v$ and $v-v'$ have the same sign.
Similarly the $u'$ integral is nonzero only if $v'$ and $v-v'$ have the same sign.  Since $(1-v)+(v-v')+v'=1$, the signs must all be positive, so we can restrict the integration
region to  $1\geq v\geq v'\geq 0$.

At this point, the integral actually can be factorized as the product of an integral over $u,u'$, and an integral over $v,v'$:
\begin{align}\label{factorized}k(h,\th)=&-\frac{C}{4}\int_{-\infty}^\infty \!\!\!\!du\,du' \,\,\frac{1}{(1-u+i\epsilon)^{\Delta}(u'+i\epsilon)^{\Delta}(u-u'+i\epsilon)^{2-\Delta-h}   }\cr &\times\int_0^1 
\!\!\!\!dv \int_0^v \!\!\!\!dv' \,\,\frac{1}
{(1-v)^\Delta (v')^\Delta (v-v')^{2-\Delta -\th}}. \end{align}
The $u$ integral runs along the real axis.  The integrand has branch points at $u=1+i\epsilon$ and $u=u'-i\epsilon$, 
with cuts as shown in fig.(\ref{deformation}a).  The integral over the real $u$ axis can be replaced with an integral on a contour that, as shown in the figure, starts at $u=+\infty$,
loops counterclockwise around the cut at $u=1+i\epsilon$, and returns to $+\infty$.  
 Now we can take the limit $\epsilon\to 0$ and replace the $u$ integral with
\be\label{zold}\left(e^{-i\pi\Delta}-e^{i\pi\Delta}\right)\int_1^\infty \!\!\!\!du \,\,\frac{1}{(u-1)^\Delta (u-u')^{2-\Delta-h}}.\ee
Here the  parts of the integral below and above the cut contribute respectively $e^{-i\pi\Delta}$ and $-e^{i\pi\Delta}$.    

\begin{figure}[ht]
 \begin{center}
   \includegraphics[width=6.5in]{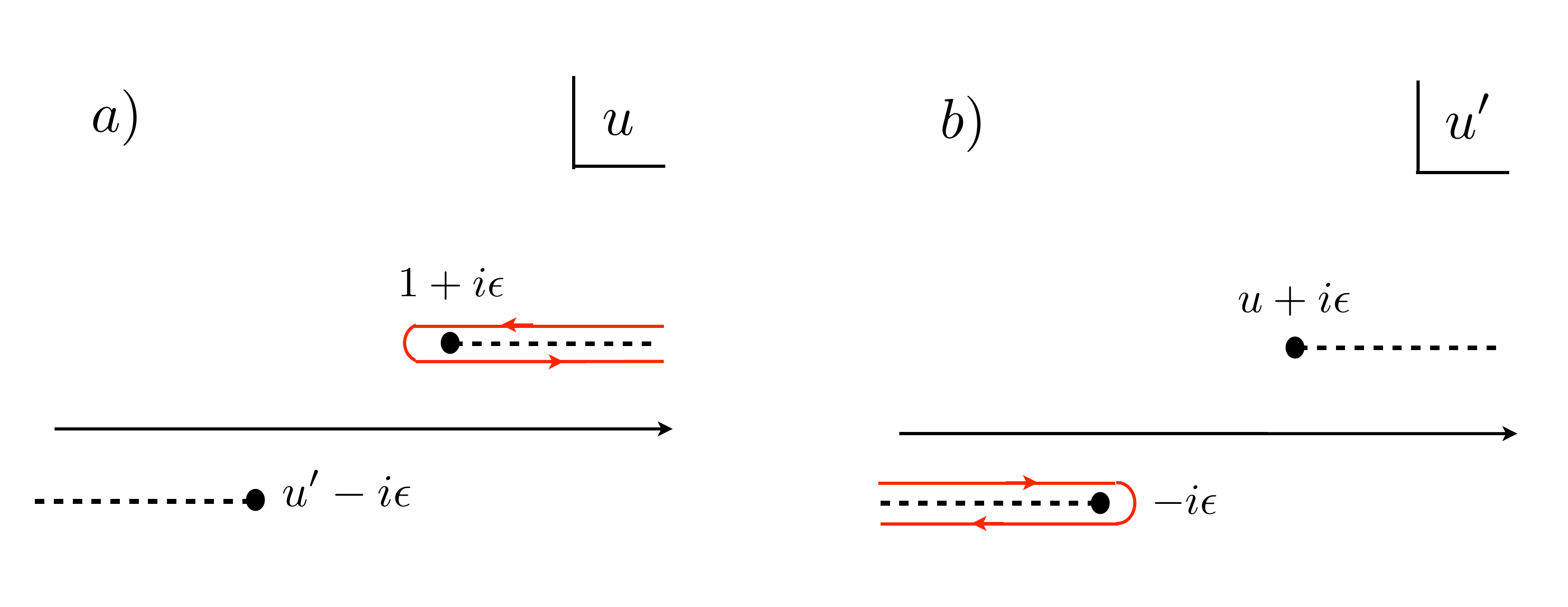}
 \end{center}
\caption{\small (a)  The $u$-plane with the indicated cuts starting   at $u=1+i\epsilon$ and at $u=u'-i\epsilon$ and running parallel to the real axis.  
The integral on the real $u$ axis can be deformed
to an integral on the indicated contour. (b) The picture in the $u'$ plane is a sort of mirror image, and the integration contour can be deformed  from the real axis
to the one shown.  \label{deformation}}
\end{figure}
\noindent
The picture in the $u'$ plane is a sort of mirror image of this (fig.(\ref{deformation}b)) and now the integral over the real $u'$ axis can be replaced
with an integral over a contour that starts at $-\infty$, loops clockwise around the cut at $u'=-i\epsilon$, and returns to $-\infty$.

The $u$ and $u'$ integral then combine, in the limit $\epsilon\to 0$, to
\be\label{preffo} -4\sin^2\pi\Delta \int_1^\infty \!\!\!\!du \int_{-\infty}^0 \!\!\!\!du' \frac{1}{(u-1)^\Delta|u'|^\Delta (u-u')^{2-\Delta -h}}.\ee
This matches the integral over the $u$ variables in the retarded kernel (see section \ref{twodext}).  So it is natural to try to likewise compare the integral over $v$ and $v'$
in eqn.\eqref{factorized} to the corresponding integral in the retarded kernel.  It turns out that this can be done, although it requires a somewhat mysterious change of
variables.   We set
\be\label{reffo} v=\frac{a}{a-b},~~~~~v'= \frac{a-1}{a-b}.\ee
The conditions $1\geq v\geq v'\geq 0$ give $a>1$, $b<0$, and the integral over $v$ and $v'$ is converted to
\be\label{leff}\int_1^\infty \!\!\!\!da \int_{-\infty}^0 \!\!\!\!db\,\, \frac{1}{ (a-1)^\Delta |b|^\Delta(a-b)^{1-\Delta+\th}}. \ee
Putting the ingredients together, we have
\be\label{zedd}k(h,\th)=C\sin^2\pi\Delta \int_1^\infty \!\!\!\!du \int_{-\infty}^0 \!\!\!\!\!du'\, \frac{1}{(u-1)^\Delta |u'|^\Delta (u-u')^{2-\Delta -h}}  \int_1^\infty \!\!\!\!da \int_{-\infty}^0 \!\!\!\!\!db\, \frac{1}{(a-1)^\Delta 
|b|^\Delta(a-b)^{1-\Delta+\th}}. \ee
Comparing to the formula for the retarded kernel, we see finally
\be\label{edd}k(h,\th)=k_R(h, 1-\th). \ee
From this relation and the fact that $k(h,\th) = k(1-h,1-\th)$, it follows that $k_R(h,\th) = k(1-h,\th)$.

\bibliography{references}

\bibliographystyle{utphysmodb}


\end{document}